\documentclass[prb,frontmatterverbose,twocolumn,showpacs,amsmath,amssymb,floatfix,superscriptaddress]{revtex4-2}

\usepackage{graphicx}%
\usepackage{dcolumn}%
\usepackage{bm}%
\usepackage{comment}
\usepackage{hyperref}
\usepackage{cleveref}
\usepackage{placeins}
\usepackage{xcolor}
\usepackage{booktabs}
\usepackage[commandnameprefix=ifneeded]{changes}

\def\doubleunderline#1{\underline{\underline{#1}}}

\begin{document}

\title{Ab initio framework for deciphering trade-off relationships in multi-component alloys}

\author{Franco Moitzi}
\affiliation{Materials Center Leoben Forschung GmbH, Roseggerstra{\ss}e 12, A-8700 Leoben, Austria}

\author{Lorenz Romaner}
\affiliation{Chair of Physical Metallurgy and Metallic Materials, Department of Materials Science, 
             University of Leoben, Roseggerstra{\ss}e 12, A-8700 Leoben, Austria}

\author{Andrei V. Ruban}
\affiliation{Materials Center Leoben Forschung GmbH, Roseggerstra{\ss}e 12, A-8700 Leoben, Austria}
\affiliation{Department of Materials Science and Engineering, Royal
         Institute of Technology, 10044 Stockholm, Sweden}

\author{Max Hodapp}
\affiliation{Materials Center Leoben Forschung GmbH, Roseggerstra{\ss}e 12, A-8700 Leoben, Austria}
         
\author{Oleg E. Peil}
\affiliation{Materials Center Leoben Forschung GmbH, Roseggerstra{\ss}e 12, A-8700 Leoben, Austria}

\date{\today}

\newcommand{\oep}[1]{\textcolor{blue}{#1}}
\newcommand{\maxho}[1]{\textcolor{orange}{#1}}
\newcommand{\fmo}[1]{\textcolor{green}{#1}}

\newcommand{\remark}[1]{\textcolor{red}{#1}}

\newcommand{\taus}{\tau_y}

\begin{abstract}

While first-principles methods have been successfully applied to characterize individual properties
of multi-principal element alloys (MPEA), their use to search 
for optimal trade-offs between competing properties is hampered by high computational demands.
In this work, we present a novel framework to explore Pareto-optimal compositions by integrating 
advanced ab initio-based techniques into
a Bayesian multi-objective optimization workflow complemented with
a simple analytical model providing straightforward analysis of trends.
We benchmark the framework by applying it to solid solution strengthening and ductility of
refractory MPEAs,
with the parameters of the strengthening and ductility models being efficiently computed using
a combination of the coherent-potential approximation method, accounting for finite-temperature effects, 
and actively-learned moment-tensor potentials parameterized with ab initio data.
Properties obtained from ab initio calculations are subsequently used to
extend predictions of all relevant material properties to a large class of refractory alloys
with the help of the analytical model validated by the data and
relying on a few element-specific parameters and universal functions that describe bonding between elements. 
Our findings offer new crucial insights into the traditional 
strength-vs-ductility dilemma of refractory MPEAs. The proposed framework is versatile and
can be extended to other materials and properties of interest, enabling a predictive and 
tractable high-throughput screening of Pareto-optimal MPEAs over the entire composition space.

\end{abstract}

\maketitle

\section{Introduction}

Refractory multi-principal element alloys (MPEAs) have gained significant interest for high-temperature applications due to their excellent mechanical properties. These 
alloys consist of several high-melting-point elements such as tungsten (W), 
tantalum (Ta), molybdenum (Mo), zirconium (Zr), hafnium (Hf), and vanadium (V), that 
form body-centered cubic (bcc) solid solutions. Despite their high yield strength at 
ambient temperatures and remarkable strength retention at high temperatures above 
1000\,K~\cite{Senkov2011,Senkov2016,Wei2020,Gao2015,Dixit2022}, refractory 
MPEAs are limited by their low tensile ductility and exhibit 
a sharp ductile-to-brittle transition~\cite{Lo2019,Xie2022,Juan2015,Miracle2017,Zou2014}, which 
restricts their use to high-temperature applications, such as gas turbine engines~\cite{Dixit2022,Sheikh2016}.

However, certain alloy compositions have been identified to have a more 
favorable combination of strength and ductility than others~\cite{Li2022}. For instance, 
the ductilisation effect of rhenium (Re) and iridium (Ir) in tungsten is a well-known 
example that keeps the overall strength of the material virtually 
unchanged~\cite{Luo1991,Romaner2010,Geach1955}. Consequently, the quest for 
discovering new compositions of refractory MPEAs with improved mechanical properties 
remains an important challenge in materials science.
The conventional trial-and-error method of 
materials design is too laborious for complex MPEAs. This is where 
computer-guided design opens new opportunities for efficient exploration of the 
vast composition space of these materials.

In recent years, many works have introduced computational methods to identify 
and characterize alloys with tailored properties \cite{Khatamsaz2022,Gao2023,Debnath2020,Rickman2019,Ferrari2021,Ouyang2023,Roy2023,Singh2023},
such as the solid solution strengthening, hardness, ductility index, and thermodynamic stability.
Moreover, some of these methods have been employed in the context of the optimization of 
several target properties \cite{Khatamsaz2022,Khatamsaz2023a,Khatamsaz2023b,Solomou2018}, with
the relevant material properties being obtained from various modeling 
approaches using experimental and theoretical input.
For example, methods based on machine learning (ML) use various sets of descriptors 
partly derived from ab initio calculations
in order to fit experimental data \cite{Hu2021,Hart2021,Zhang2020}. This approach, 
however, often requires large data sets and, most crucially, it is intrinsically
interpolative and cannot give predictions outside the domain covered by an existing data set.
A more transferable methodology is to use properties of elements, or 
simple compounds, for extrapolation of target properties to multi-component properties, mainly
using the rule-of-mixture \cite{Shaikh2020,Gao2017,Elder2023a}.
Despite its appeal due to its simplicity, this approach cannot capture the full
complexity of a system in the case of intricate interactions (e.g., due to abrupt 
changes in the near-Fermi level band structure \cite{Singh2018}).

A typical method of choice for truly predictive calculations of alloy properties is density functional theory (DFT).
However, explicit evaluation of fracture or yield strength in alloys is mostly out of reach for
available DFT implementations.
Although this may become possible with general-purpose interatomic potentials \cite{li_complex_2020,Song2023},
a fairly robust description can be obtained by using appropriate models,
which have been shown to accurately predict strengthening \cite{Maresca2020_a,Maresca2020_b,Lee2021} and ductility \cite{Novikov2022,Rice1974,Mak2021,Li2020_a}.
These models combine linear elasticity and microscopic principles,
and the only required input parameters are certain material properties.
Specifically, apart from basic equilibrium properties, such as the molar volume and elastic constants,
the strengthening model requires misfit volumes related to concentration derivatives of the alloy volume,
while the ductility model relies on surface and unstable stacking fault (USF) energies computed for specific
orientations. Importantly, all of these parameters are accessible through simulations using first-principles methods, yet, such simulations are still challenging for disordered MPEAs.

The most common approach to ab initio alloy modeling is to represent a disordered alloy 
using special quasi-random structures (SQS) or similar supercell representations
that try to mimic perfect randomness~\cite{avdw:atat2,avdw:maps,Singh2021}. However, this method has several 
inherent limitations for MPEAs. First, the computational effort grows exponentially as the number of alloy 
components increases. Moreover, properties influenced by global symmetry or local chemical 
environment~\cite{Tasnadi2015,Holec2014}, such as surface energies or elastic constants, 
present significant challenges for convergence and necessitate the use of even larger cells,
limiting the number of components in the best case to three or four. Calculating 
properties for arbitrary compositions,
especially when components are present in low concentrations,
is even more cumbersome.

One alternative methodology is the coherent potential approximation
(CPA) \cite{Soven1967,Velicky1968}, whose advantages are its extremely small computational 
expenses, irrespective of the number of components,
as well as its ability to treat arbitrary alloy concentrations, providing a straightforward way for accurate
evaluation of concentration derivatives. Unfortunately, the single-site nature of CPA renders it
impossible to take into account effects of local atomic relaxations that are crucial for a correct
characterization of defects, such as USFs, or inhomogeneous structures, e.g., surfaces.

Another alternative to traditional supercell methods are machine-learning interatomic potentials (MLIPs) that are trained on ab initio data~\cite{behler_generalized_2007,Bartok2015,thompson_spectral_2015,Shapeev2016}. 
The size of supercells is not limited with MLIPs, which is especially important when simulating extended defects like interfaces, where proper configurational averaging becomes an issue with MPEAs. 
In contrast to the previously proposed way of passively training
general-purpose interatomic potentials {on a pre-defined fixed training set}
(e.g., \cite{li_complex_2020,Byggmastar2021,Song2023,Lopanitsyna2023}),
{whose accuracy is not sufficient for our goals, especially for } parts of the configurational space not covered by the training set, the use of active learning enables
automatic generation of {MLIP} training sets {specifically tailored to
a given class of systems and a given set of properties}. This allows one
to predict any material property of interest with DFT accuracy over the entire composition space of an alloy \cite{Hodapp2021,Novikov2022} with a minimal amount of DFT calculations.

In this work, we combine the two approaches---CPA and actively learned Moment Tensor Potentials (MTP), a class of MLIPs---to characterize solid solution strengthening and ductility of refractory MPEAs. In particular, we use CPA to compute the elastic constants and misfit volumes, needed for computing
the critical resolved shear stress (CRSS) within the model of Maresca and Curtin~\cite{Maresca2020_a,Maresca2020_b}. MTPs are used to simulate large supercells, necessary to
calculate surface and USF energies which are needed to parameterize a model for intrinsic ductility based on the Rice-Thomson theory \cite{Rice1974}.

The crucial point of our approach is to enhance the simulation efficiency
by employing the most suitable method for different sets of properties.
Specifically, CPA's quantum mechanical treatment of solid solutions
and its analytical nature eliminates finite-size cell effects
and sampling errors related to random atomic configurations,
making it excel in calculations of the concentration dependence of properties
that are relatively insensitive to local relaxations,
such as elastic constants and misfit volumes.
In comparison to supercell calculations with MLIPs, CPA is especially
superior in accuracy when dealing with abrupt changes of the Fermi-surface topology,
which is often encountered in bcc alloys and can have a significant impact on
the concentration dependence of properties.
On the other hand, ab initio based MTP methodology surpasses CPA in dealing with
defects and special geometries, such as surfaces, due to its capability to
incorporate local atomic relaxations.

{Another important part of our framework} is a simple model that is only slightly more
complicated than the rule of mixtures but can accurately capture the 
concentration dependence of all the quantities relevant for strengthening and
ductility over the entire composition space {of a large class of refractory alloys}.
The model is based on simple element-specific descriptors
and universal functions of the $d$-electron valence, with a few adjustable parameters that 
can be easily fitted to data from simple compounds.
Importantly, the model is based on fundamental properties of bonding in transition metals,
which makes it capable of extrapolative predictions, rendering it as a potentially useful tool for
exploring the \emph{entire} space of refractory MPEAs.
Moreover, the analysis of higher-order contributions beyond the first order
can provide an insight into interactions between individual alloy components.

As a proof of concept for the flexibility of such an
approach, we perform multi-objective optimization (MOO) of strength and ductility for
three- and four-component alloys.
{Then, we use these results to validate the model and extend our
predictions to a larger class of alloys}. By analyzing the results, we reveal several potential pitfalls
of conventional methods and show how they are easily overcome within our approach.

\section{Results}

\subsection{Multi-objective optimization of strength and ductility}

\begin{figure*}
   \centering
   \includegraphics[width=0.8\textwidth]{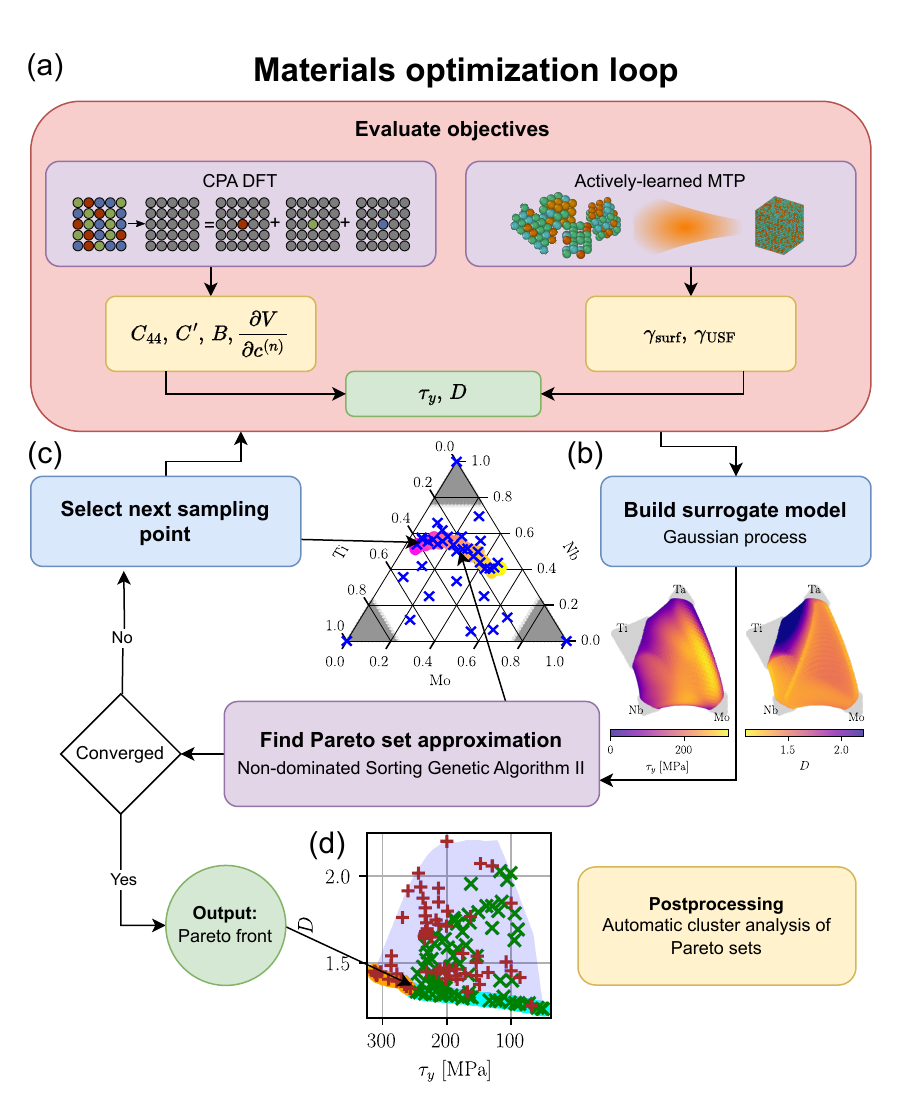}
   \caption{Flowchart of the materials design loop using Bayesian multi-objective optimization:
   (a) Advanced ab initio techniques such as the coherent potential approximation (CPA) and actively-learned moment tensor potentials (MTPs) to calculate the strength, $\taus$, and ductility, $D$, objectives.
   Quantities obtained from ab initio are: Shear, $C_{44}$, $C'$, and bulk, $B$, moduli,
   specific volume (per atom), $V$, and its concentration derivatives, $\partial V / \partial c^{(n)}$, all evaluated at
   room temperature; the surface energy, $\gamma_{\mathrm{surf}}$, and the USF energy, $\gamma_{\mathrm{USF}}$.
   (b) Employing Gaussian process regression, we construct a surrogate model, and a genetic algorithm is applied to the search of Pareto-optimal compositions. 
   (c) The sampling points are chosen from the Pareto front approximation until convergence is reached. 
   (d) Pareto-optimal alloys are the key outcome of the process.}
   \label{fig:fig_01}
\end{figure*}

We consider homogeneous alloys whose strength can be characterized by the CRSS due to impeded mobility
of dislocations interacting with the alloy components that act as a random field of solutes.
Within the Maresca-Curtin model \cite{Maresca2020_b}, the CRSS is given by $\taus \equiv \taus(V, \{\Delta V_n\}, C_{ij})$,
where $V$ is the specific volume, $\Delta V_n$ are the misfit volumes for each component $n$,
and $C_{ij}$ are the elastic constants (see Methods~\ref{method_sss} for details).
If the material parameters are obtained from first principles, the model allows for parameter-free predictions of yield strength without relying on experimental input.

The strength of a solid solution is known to compete with its ductility.
The ductility of a refractory MPEA is mainly determined by the intrinsic fracture behavior of single-crystalline grains.
The ability of a material to undergo plastic deformation affects the way unavoidable defects grow to cracks and propagate through the material.
If a crack is blunting (propagating) upon external loading, the material is said to be intrinsically brittle.
If, instead, a dislocation is emitted from the crack tip, the material is said to be intrinsically ductile.
This competition between blunting and dislocation emission on active slip systems in bcc alloys can be quantified using the Rice-Thomson model~\cite{Rice1974,Rice1992}.
Within the Rice-Thomson model, ductility is described by means of a ductility index, $D$, which depends on the ratio of the unstable stacking fault energy $\gamma_{\mathrm{USF}}$ and the surface energy $\gamma_{\mathrm{surf}}$, times
a prefactor depending on the crack orientation and the elastic constants.
In this theory, lower $D$ values indicate increased alloy ductility, and $D < 1$ signifies the absence of intrinsic brittleness. A recent application of the ductility index  by Mak \textit{et al.}~\cite{Mak2021} has led to the development of an RT ductility criterion for refractory metal alloys. Their study showcased the correlation of $D$ with measured fracture strain and its predictive capability for ductility trends.

Improved ductility is expected for alloys with lower stacking fault energies, which facilitates dislocation emission.
At the same time, the opposite is true for improved strength, and to investigate this competition, we leverage a Bayesian multi-objective optimization framework to uncover
sets of Pareto-optimal compositions, with the objective functions obtained from first principles.
The flowchart in Fig.~\ref{fig:fig_01} presents an overview of the workflow developed in this work.
The two main ingredients of the workflow are 1) the calculations of the objectives, the CRSS, $\taus$, and the ductility index, $D$
(Figure~\ref{fig:fig_01}a), and 2) a Bayesian optimization loop based on the non-dominated sorting genetic algorithm (NSGA-II) \cite{Deb2022}
for finding the Pareto-set approximation (Figure~\ref{fig:fig_01}b,c); see Methods~\ref{method_moo} for details.
The key feature here is that NSGA-II utilizes an intermediate surrogate model (Gaussian process in our case), significantly reducing the need for direct evaluations of the objective functions.

As shown in Fig.~\ref{fig:fig_01}a, the quantities obtained from first-principles calculations are the elastic constants, $C_{44}$,
$C'$, and $B$, the concentration derivatives of the specific volume, and the planar defect energies: the surface energies,
$\gamma_{\mathrm{surf}}$ (\{100\}, \{110\}), and the unstable stacking fault energy, $\gamma_{\mathrm{USF}}$ (\{110\}, \{112\}), evaluated for relevant orientations $\{\bullet\}$ \cite{Mak2021}.

We use CPA to calculate the elastic constants and volumes because these quantities are insensitive to local atomic relaxations, and evaluation of concentration derivatives is much more efficient with this approach compared to SQS calculations.
Furthermore, a similar methodology has already been successfully applied to computing CRSS of iron-group MPEAs~\cite{Moitzi2022,Biermair2023}.

On the other hand, the planar defects are strongly affected by local lattice relaxations, and
we, therefore, resort to a supercell approach for these quantities.
However, carrying out high-throughput calculations
with the conventional SQS is computationally unwieldy,
especially for properties requiring non-trivial structure relaxations (e.g., surfaces).
Therefore, we use actively-learned MTPs that are able to predict material properties with near-DFT accuracy \cite{Novikov2022,Hodapp2021},
requiring only a small fraction of the computational expenses of a direct DFT calculation.
This allows us to perform structure relaxation on supercells with hundreds of thousands of atoms at a cost that is only slightly higher 
than that of classical interatomic potentials, such as the embedded atom method.
In comparison to previously published works \cite{Mak2021, Hu2021, Zheng2022, Xu2020}, our approach consistently produces comparable USF energies
for specific alloy compositions (see Results and Discussion) but we can also treat any other composition
of the alloy once the MLIPs are trained. Moreover, we anticipate greater accuracy
for systems affected by Fermi surface nesting (\cite{Kormann2017}) or for alloys experiencing local lattice reconstructions (\cite{Treglia1999})
because our methodology does not impose constraints on cell size or lattice relaxations.
Indeed, in case of surface energies, the differences between MTP-based and SQS calculations become much more pronounced
(sometimes, more than 200 mJ/m$^2$, which is more than 10\% of the typical value of $\gamma_{\mathrm{surf}}$),
mainly due to a considerably slower convergence of the surface energy with the cell 
size, as will be discussed in more details later on (see Section~\ref{section:sqs}).

\subsection{Search for Pareto-optimal alloys}

\begin{figure}
    \centering
    \includegraphics{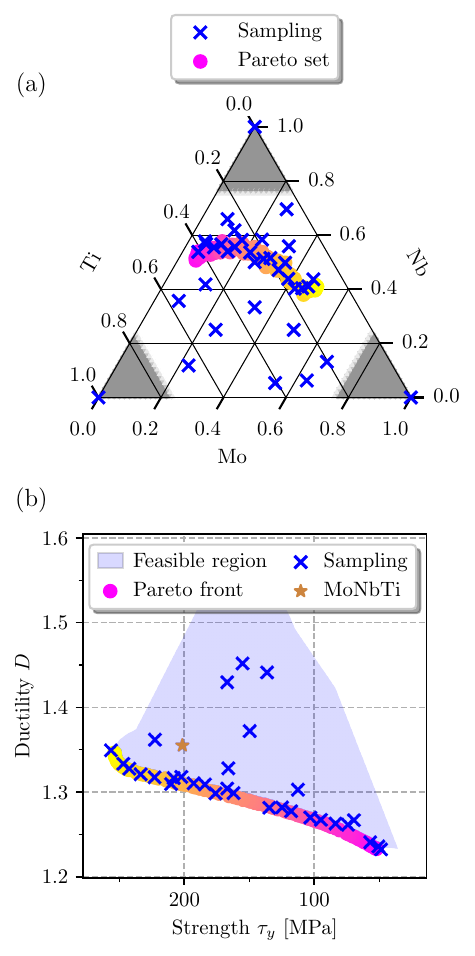}
    \caption{Optimization results for the MoNbTi system: (a) Pareto set approximation (PSA) alongside with sampling points in the search space. (b) Pareto front approximation (PFA), sampling points and feasible region in objective space. PFA was gradually color-coded from yellow, which corresponds to high strength, to pink, which corresponds to high ductility. PSA uses the same color coding.}
    \label{fig:fig_02}
\end{figure}

\begin{figure*}
    \centering
    \includegraphics{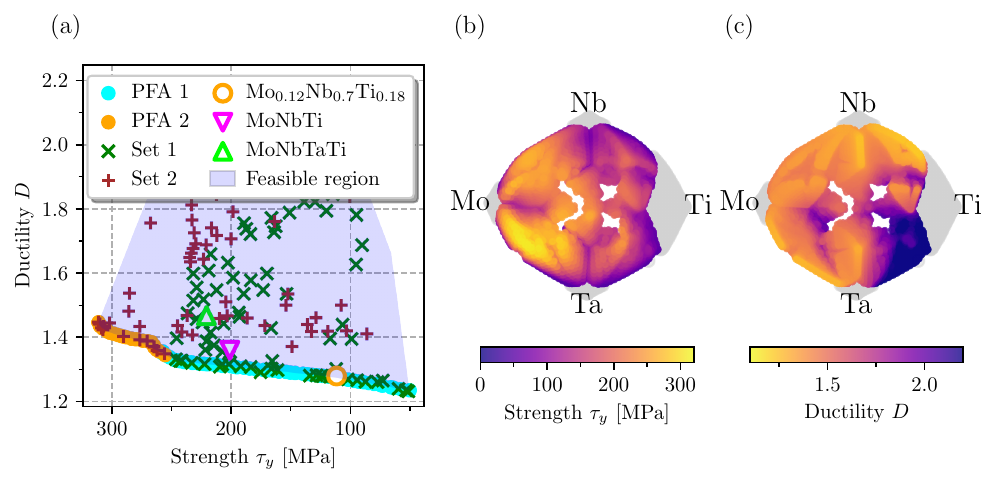}
    \caption{Optimization results for the MoNbTiTa system:
    (a) Visualization of the PFA and sampling points. The Pareto set comprises two distinct regions in the concentration space derived from automated clustering. These regions define the corresponding sections (PFA1) and (PFA2) on the Pareto front. The sampling points are color- and marker-coded based on their proximity to the respective PFA regions.
    Open symbols mark specific alloys mechanically tested in Refs.~\cite{Coury2019,Senkov2019c}.
    t-SNE projections of (b) ductility index $D$ and (c) strengthening $\taus$ for the whole search space. Grey areas are beyond the search space. Sets 1 and 2 mark sampling points that are closer in the search space to the points of PFA1 and PFA2, respectively}
    \label{fig:fig_03}
\end{figure*}

We now combine the calculations of the CRSS and the ductility index, $D$, and perform a search for Pareto-optimal alloys in the composition space, according to the workflow shown in Fig.~\ref{fig:fig_01}.
We first analyze the MoNbTi alloy, whose search (design) space can be easily visualized using ternary diagrams.
We constrain the search space to a concentration range of $0.0 \leq c_{i} \leq 0.75$ for each component $i$. This is motivated by the fact that pure Ti is stable in the hcp structure ($\alpha$-Ti) under ambient conditions and it has limited miscibility with the other two elements, Mo~\cite{Barzilai2017} and Nb~\cite{Boenisch2020} {(for more detailed analysis of phase stability, see 
Supplemental Materials -- SM~\ref{sup_stab_ss})}.
Moreover, dilute alloys offer minimal potential for solid solution strengthening.

The constraints are handled by penalty functions within NSGA-II, which penalize undesirable solutions by reducing their fitness values in proportion to the degree of constraint violation. In principle, this approach could be extended to define more complex regions in the search space that should be avoided, e.g., regions of thermodynamic instability or other unfavourable properties.

Fig.~\ref{fig:fig_02}a shows the sampling points in the search space of MoNbTi,
alongside with the Pareto set approximation (PSA), which is a set of compositions that are non-dominated by each other
but are superior to the other compositions in the Pareto sense.
This can be seen more clearly in Fig.~\ref{fig:fig_02}b, where we show the Pareto front approximation (PFA) at the bottom
of the feasible region, resulting from the sampling in the objective space.
There is a continuous mapping between PFA and PSA. The PFA's left portion corresponds to PSA's right portion, and vice versa. 
On the right side of PSA and the left side of PFA are stronger but less ductile alloys. 
In both PFA and PSA, the color-coded scheme designates yellow for high strength and pink for high ductility.
Generally, we observe that relative changes in the ductility index across the concentration space are not as severe as changes in the strength. The lower limit of the strength for Pareto-optimal alloys within the constraints given above is about 50 MPa, while the strongest alloys have
$\taus \approx 260$ MPa. On the other hand, the corresponding ductility index, $D$, changes from 1.23 to 1.35
{(with the error of $D$ estimated to be less than 0.015, see SM~\ref{sup_error_propagation})}.
It is also worth noting that the Pareto set avoids the region around the equimolar composition.

Regions of higher strength can be found close to the equimolar MoNb alloy with small additions of Ti. Reducing the Mo content in the alloy enhances its ductility and reduces the likelihood of brittle fracture. 
On the other hand, the alloy must contain a certain amount of Mo to ensure a reasonable strengthening effect.
More ductile alloys can be found in a region close to the Nb\textsubscript{0.6}Ti\textsubscript{0.4} binary.
The improved ductility comes at a price of the strength that
is almost five times smaller than the maximum value.

Pareto-optimal alloy compositions in the three-component MoNbTi alloy system exhibit a continuous and smooth transition across the entire design space, forming a connected Pareto set.
To investigate non-trivial alloying effects, we expand the composition space by introducing Ta, anticipating Fermi surface nesting effects caused by Ta alloying.
Similarly to the case of MoNbTi, we confine the search space to concentration values ranging
from $0.0$ to $0.75$ for each component, with the exception of Ti, which is restricted 
to a maximum atomic fraction of $0.55$. This limitation is motivated by the limited 
solubility of Ti, as described in Ref.~\cite{Barzilai2016}
{(see also SM~\ref{sup_stab_ss})}. The results of the 
optimization, represented by the PFA and sampling points, as illustrated in Fig.~\ref{fig:fig_03}a
reveal a more intricate concentration dependence of the objectives, which can also be seen in the 
t-SNE projections displayed in Figs.~\ref{fig:fig_03}b,c (see Methods~\ref{method_tsne} for details).

These plots indicate that, while the ductility index, $D$, is, again, minimal near binary NbTi, the strength, $\taus$,
is maximal in two regions: 1) close to binary MoNb, with small additions of Ti, as in the MoNbTi case;
2) the vicinity of binary MoTa, which exhibits considerably higher strength than the 3-component alloy.
This leads to two effectively disjoint Pareto subsets in the composition space of the MoNbTiTa alloy.

Indeed, clustering analysis has detected that parts of the Pareto front denoted by PFA1 and PFA2 in Fig.~\ref{fig:fig_03}a
can be associated with two separate regions in the concentration space. Despite a seemingly 
continuous crossover between these
two regions in the objective space, the regions are truly disjoint in the composition space.
The right, ductility-favoring portion of the Pareto front, PFA1,
corresponds to alloys similar to the three-component MoNbTi system with minor additions of Ta.
The optimal compositions in this region are characterized
by the Nb content ranging from 0.4 to 0.6, by the Ti content ranging from 0.2 to 0.4, and moderate Mo concentration maintaining the strengthening.
At the same time, the left portion of the Pareto front, PFA2, represents a separate region comprising alloys with high Mo ($0.4-0.6$) and moderate Ta ($0.2-0.35$) content, whereby Ta is responsible for a strength above 250 MPa, i.e., beyond
those reachable by the MoNbTi system. An important feature of the PFA2 is a small knee corresponding to $\taus \approx 270$ MPa.
Starting from this point, one can either improve ductility without reducing the strength much, or enhance the 
strength without compromising on ductility.

In Figs.~\ref{fig:fig_02}a and \ref{fig:fig_03}a, we highlight three alloys from Ref.~\cite{Coury2019,Senkov2019c} which were mechanically tested. 
Both the equimolar MoNbTiTa and MoNbTi exhibit similar strength in experimental measurements, with the four-component alloy showing slightly higher strength. The Nb-rich Mo\textsubscript{0.12}Nb\textsubscript{0.7}Ti\textsubscript{0.18} alloy demonstrates approximately 2/3 of the strength of the others. This behavior is also observed in our results. Additionally, ductility was indirectly assessed in these experimental works by measuring strain at fracture. The Nb-rich alloy exhibited the highest ductility, while the other two alloys showed lower ductility, yet remained non-brittle.
Our ductility index corroborates this finding.
All three alloys are located near the Pareto-front, with the Nb-rich alloy being located precisely on the ductile portion of the Pareto front. For MoNbTiTa, our results suggest that increasing the Mo and Nb content, while adjusting the Ta content, can improve the ductility of the equimolar MoNbTiTa alloy, without sacrificing the strength.

\subsection{ Virtual bond approximation model}

\begin{figure}
   \centering
   \includegraphics{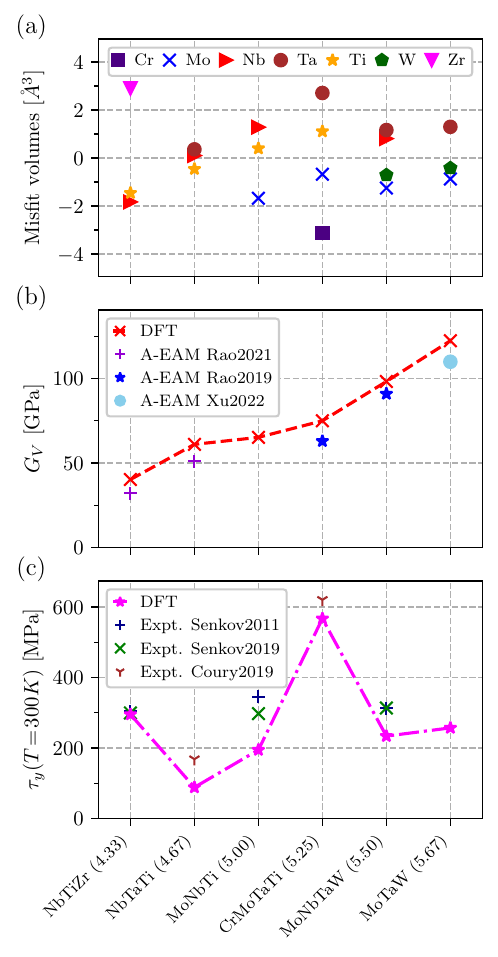}
   \caption{Comparison of misfit volumes $\Delta V_i$ (a), 
            the shear modulus, $G_V$, (b), and the solid solution strengthening, 
            $\taus$ (c) for various refractory alloys. The alloys 
            are sorted according to their valence electron concentration (VEC). 
            Experimental values for $\taus$ are from Refs. \cite{Senkov20211,Senkov2019,Coury2019}. Shear modulus is compared to results obtained with $A$-atom embedded-atom potential from Refs. \cite{Rao2019,Rao2021,Xu2022}.}
   \label{fig:paper_compare_alloy}
\end{figure}

{So far, the main idea of the workflow was to get accurate and predictive results with high computational efficiency.
Still, characterizing multiple alloy systems in this ways would require a lot of DFT calculations (MTP training and CPA).
Instead, we would like to use the results obtained thus far for MoNbTiTa to analyze the strengthening and ductility
behavior in a broader range of refractory alloys.
One way to achieve this would be} to define a set of ``good'' descriptors and train a black box ML model,
which can then be used to obtain properties at arbitrary compositions.
Such descriptors may encompass purely local structural characteristics~\cite{Lupopasini2022,Lupopasini2023,Nyshadham2019}, effective interatomic interactions~\cite{Liu2021} or 
local electronic structure-related properties like moments of the local density of states
and energies of simpler substructures~\cite{Hu2021,Huang2024}.
One significant disadvantage of such data-driven models is that they often lack interpretability
and they are also only interpolative, meaning that one has to have a large amount of data covering
enough of alloy systems to be able to predict properties of an alloy not included in the initial data set.
Physical models, on the other hand, are more transferable because they require fewer data points
for training and provide clear insights into system behavior, fostering a deeper understanding of cause-and-effect relationships.
Moreover, such models based on fundamental principles can be also considered as extrapolative, which prevents implausible outcomes, ensuring reliability.

One of the simplest examples of a physics-motivated model is a correlation between the valence electron count
(VEC) and the ductility.
In particular, a lower VEC is known to be associated with an enhanced intrinsic ductility due to 
a greater elastic softness \cite{Li2020_a, Yang2018, Mak2021, Sheikh2016}.
Brittle alloys are prevalent at a VEC of around 5 to 6, with an improved ductility beyond this range \cite{Yang2018}.
A similar correlation can be seen for the shear modulus, $G$, as shown in Fig.~\ref{fig:paper_compare_alloy}b. The shear modulus increases more significantly than the bulk modulus, resulting in a larger
shear-to-bulk modulus ratio ($G/B$) at higher values of VEC.
This trend is frequently associated with embrittlement.

\begin{figure}
   \centering
   \includegraphics{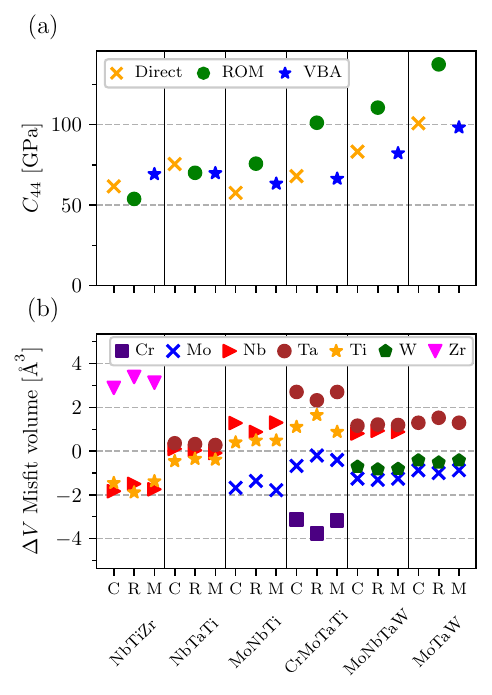}
   \caption{Comparison of (a) $C_{44}$ and (b) $\Delta V$ misfit volumes 
    obtained from the rule-of-mixture (R), 
    virtual bond approximation (M), 
    and direct calculations (C).
   }
   \label{fig:paper_misfit_elastic_combined}
\end{figure}

Despite its clear physical meaning as the bonding state filling factor, the VEC does not take into account
interactions between alloy components.
This, in particular, can explain why it fails to reveal any particular trend in misfit volumes
(Fig.~\ref{fig:paper_compare_alloy}a) and it is, thus,
not a useful descriptor for the CRSS, as can be seen in Fig.~\ref{fig:paper_compare_alloy}c.
Another popular model -- the rule of mixtures (ROM), also known 
as Vegard's law when applied to the lattice constant or volume -- shares the same drawback
(see, e.g., Fig.~\ref{fig:paper_misfit_elastic_combined}).

Our approach, novel in the context of strengthening and ductility, is based on a slightly modified version of the
virtual bond approximation (VBA) \cite{Ruban1998} that can be considered
as a generalization of classical bonding theory of transition metals\cite{Pettifor1995} to alloys.
The key idea of the VBA is to assign universal functions to local bonds between alloy components. These so-called \emph{virtual bond} functions depend on the average $d$-valence of elements forming the bond, as well as element-specific bandwidths (equivalently, second moments of the density of states) of the corresponding elemental compounds. To second order in inter-component interactions
{(possible extension to higher orders is straightforward)}, the VBA takes the form
\begin{align}
    \mathbf{\xi}(\{c_i \}) = & \sum_i c_i \, w_i \, \mathbf{\xi}^{(1)} \big( N_i \big) + \notag \\
       & + \sum_{i j} c_i \, c_j \, w_{ij} \, \mathbf{\xi}^{(2)} \big( N_{ij} \big),
\end{align}
where $\mathbf{\xi}(\{ c_i \})$ stands for $B$, $C'$, $C_{44}$, $V$, or $\Delta E$ (for USF and surface energies),
with $c_i$ being atomic fractions of components and $\mathbf{\xi}^{(k)}$ representing the corresponding virtual bond functions
of the average $d$-valences,
$N^{(k)} \equiv \{ N_i, N_{ij} \}$, and of bandwidth parameters $w^{(k)} \equiv \{ w_i, w_{ij} \}$ (see Methods~\ref{method_vba} for details).
Note that $N_i$ and $w_i$ are element-specific parameters, while the second-order parameters are defined as
$N_{ij} = (N_i + N_j) / 2$ and $w_{ij} = \sqrt{w_i w_j}$.

The unknowns $\mathbf{\xi}^{(k)}$ are obtained using linear regression over a set 
of values of the associated quantity 
computed for a series of systems including pure elements (Cr, Mo, Nb, Ta, Ti, W, Zr)
and equimolar binary alloys in the bcc structure.
Since the inclusion of Ta causes deviations and abrupt changes in material properties attributed to Fermi 
surface transitions (e.g., see Fig.~\ref{fig:appendix_5} in SM),
we also calculate several three- and four-component alloys 
containing Ta.
Significant deviations from both directly calculated and ROM values, 
serve as a reliable indicator of whether additional alloys should be included. This procedure was 
only deemed necessary for elastic properties. The fitted model is applicable to any alloy composed of the seven mentioned elements.

A comparable approach was presented by \citet{Ferrari2018,Ferrari2021} where
energies and material properties were fitted using polynomial functions of concentrations.
A notable advantage of our method lies in the fact that we do not 
require parameters to be fit individually for each alloying system. Our 
approach promotes the coefficients from system-specific tabulated values to transferable parameters that encapsulate local bonding characteristics.
For example, the VBA applied to iron-group MPEAs produced good
results despite major complications due to the complex magnetic behavior of this class of alloys~\cite{Moitzi2022}.

\begin{figure}
   \centering
   \includegraphics{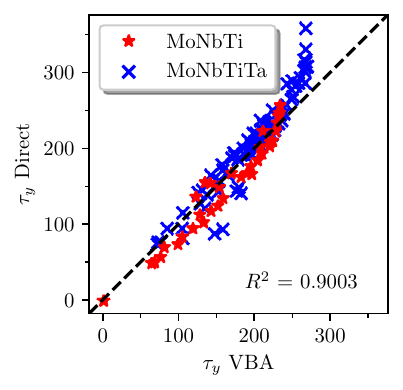}
   \includegraphics{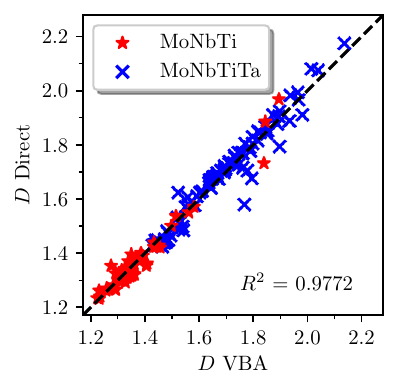}
   \caption{Correlation between directly calculated $D$ and $\taus$ and their counterparts $D_\text{VBA}$ and ${\taus}_\text{VBA}$ derived from the virtual bond approximation.
   }
   \label{fig:figure22}
\end{figure}

To demonstrate the performance of our VBA model, we compare direct and 
VBA calculations for the most important properties
for six previously considered equimolar alloys: NbTiZr, NbTaTi, MoNbTi, CrMoTaTi, MoNbTaW and MoTaW. 
The results are presented in Fig.~\ref{fig:paper_misfit_elastic_combined}, where we 
compare calculated (labeled as ``C''), VBA model (``M''), alongside ROM (``R'') values 
for the misfit volumes, $C_{44}$, and $\gamma_{\mathrm{surf}}$.
The misfit volumes for equimolar compositions (Fig.~\ref{fig:paper_misfit_elastic_combined}b) 
are reproduced well by both the VBA model and the ROM, with
the former giving systematically better values. However, the true advantage of using a model that includes inter-component interactions, becomes apparent when the composition is varied:
In this case, the ROM often fails to reproduce the concentration trends, while the VBA model
produces almost exact results corresponding to direct calculations
(e.g., see Fig.~\ref{fig:paper_misfit_volumes_changes} in SM).

Obtaining virtual bond functions, $\xi^{(k)}$, for surface and USF energies is 
more difficult due to relaxation effects. 
To take them into account, we apply an additional rescaling of the functions, as described in Methods~\ref{method_vba}.
{Test results can be found in SM~\ref{sup_vba} (Figs. ~\ref{fig:paper_surf_rescale} and ~\ref{fig:figure15}).}

{To validate our VBA model against our ab initio results},
we use the compositions previously generated during the
optimization process for the MoNbTi and MoNbTiTa alloy systems, compute $\taus$ and $D$ using the material parameters
obtained from the VBA model, and compare the results with direct CPA+MTP calculations.
As can be seen from Fig.~\ref{fig:figure22}, the agreement is 
remarkably good for such a simple model, with minor deviations due to
its inability to replicate strong non-linear effects in the 
properties upon changing the concentration. A more detailed analysis reveals
that most of the discrepancies (especially for $\taus$ in Fig.~\ref{fig:figure22}) are 
associated with Ta causing substantial non-linearity in the concentration dependence of the elastic properties (see SM~\ref{sup_valid_sss}).
Such a non-linear behavior is due to Fermi-surface effects and pose a challenge to any simple model.

The validation on MoNbTiTa has shown the ability of the VBA model to
to describe trends in multicomponent alloys based on the data from elemental and simple alloy compounds.
However, a much stricter test (also addressing an eventual overfitting issue) would be to check if 
the model can predict properties for alloys containing elements not included in the training set.
With this in mind, Hf and V (or any alloy containing them) have been excluded from the training set used
to fit parameters $\xi^{(k)}$. To evaluate properties for alloys involving these elements, the only input
needed is the corresponding bandwidth parameter (corrected for the row in the periodic table) and the number of $d$-electrons,
which are element-specific constants.

\begin{figure}
   \centering
   \includegraphics{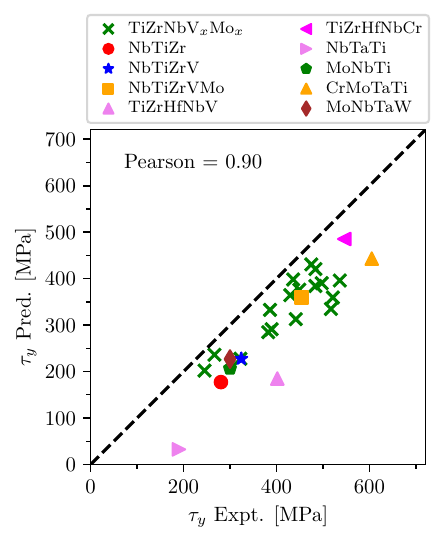}
   \caption{Parity plot of predicted and experimentally obtained CRSS, $\tau_y$, for various alloys,
   including alloys containing Hf and V.
   }
   \label{fig:appendix_parity_hfv}
\end{figure}

The model is then applied to a series of alloys containing nine $d$ elements from three groups: Ti, V, and Cr.
The results for the CRSS are shown in Fig.~\ref{fig:appendix_parity_hfv} where they are plotted against
experimentally measured $\taus$ \cite{Wu2014a,Wu2015a,Fazakas2014a,Xiong2023a}
(tabulated data can be found in Table S2 of SM~\ref{sup_tab_data}).
One can see that the model captures trends very well in a large range of V-containing alloys. Some alloys display underestimated predictions, but the overall strength ranking remains accurate, reflected in a high Pearson coefficient.
The overall underestimation is anticipated as the strengthening model gives the lower bound,
as discussed in more details in SM~\ref{sup_valid_ductilty} (the effect can also be seen in Fig.~\ref{fig:paper_compare_alloy}).

\begin{figure}
   \centering
   \includegraphics{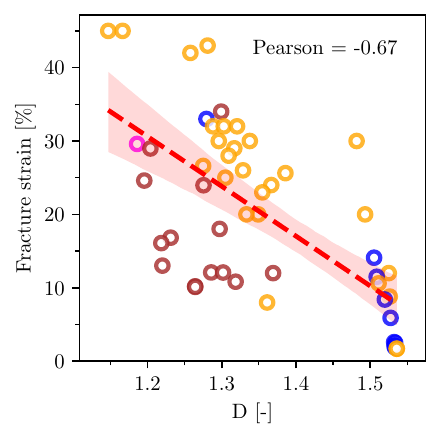}
   \caption{{Correlation plot of predicted $D$ and fracture strain for various alloys from Ref~\cite{Singh2023}. Alloys containing Hf or V are marked with orange and brown dots, respectively. Alloys containing both Hf and V are marked with pink dots.}
   }
   \label{fig:appendix_parity_d_82}
\end{figure}

The predicted ductility index is much more difficult to validate, as it is not a measurable quantity.
Therefore, we compare it against the fracture strain for a range of alloys from \citet{Singh2023}.
As one can see from the results in Fig.~\ref{fig:appendix_parity_d_82}
(tabulated data can be found in Table S1 of SM \ref{sup_tab_data}),
there is a significant
anti-correlation between the measured strain at fracture and $D$,
with the Pearson coefficient of -0.67. In critically assessing this outcome, one should keep in mind
that it captures not only errors of the VBA model itself but also the qualitative nature
of the ductility index, which is not directly related to the fracture strain, but rather, it is
an indicator for the tendency to embrittlement. Furthermore, the $D$-criterion does not apply to
intergranular fracture that can be facilitated by elements segregating at grain boundaries and
reducing their cohesive strength. Nevertheless, it is important that most of the alloys
on the brittle and ductile ends (having extreme values of $D$/fracture strain)
are ranked correctly. Note also, that our results are very similar to an analogous plot of
the fracture strain versus the ductility index in \citet{Singh2023}, if one takes into account
that the ductility index in the reference is related to the inverse of our $D$. Overall, the prediction results suggest that the VBA is sufficiently accurate to be used for a
qualitative assessment and analysis of ductility-strength trade-offs for refractory MPEAs,
as will be discussed in the next section in more details.

\section{Discussion}

\subsection{Comparison to SQS results}
\label{section:sqs}

Although our results on the CRSS and the ductility index are qualitatively consistent
with previously published calculations, there are quantitative differences, which we
now analyze by examining individual quantities involved in the models.

The key parameters for calculating the ductility index, $D$, are the energies of 
USF (for two orientations, $\{110\}$ and $\{112\}$)
and the surface ($\{100\}$ and $\{110\}$). In this context, we compare the results of equimolar MoNbTi
of our calculations with actively-learned MTPs to respective values from previous 
studies in Table~\ref{tab:comp}.

In particular, the USF are often used in theories predicting the 
mechanical properties of bcc metallic alloys, and its energy is therefore often studied.
Our MTP approach yields values of 730 $mJ/m^2$ and 853 $mJ/m^2$, for $\{110\}$ and $\{112\}$ respectively.
The values for both orientations converge quickly with cell size, which validates the use of smaller cells in SQS calculations.

The SQS-based references values of~\citet{Xu2020} and values from MTPs trained 
on the large data set of~\citet{Zheng2022} render similar values, with rather small 
difference up to 40 $mJ/m^2$. Errors on this scale can be attributed to the choice of a specific lattice parameter, 
statistical sampling errors and method-specific numerical errors
({Supplemental Materials~\ref{sup_conv_elements} contain a detailed analysis of errors for surface and USF energies)}. Only~\citet{Mak2021} shows 
consistently higher values for all orientations.
Notably, our analysis reveals the origin of the 
discrepancies among the reference values reported in the 
literature. \citet{Xu2020} and~\citet{Zheng2022} calculated USFs by keeping the 
cell cubic with the lattice parameter of the alloy. 
In contrast,~\citet{Mak2021} relaxed the out-of-plane lattice vector to zero stress, along with the ionic positions perpendicular to the stacking fault, while keeping the fault plane lattice vector fixed. 
Replicating this procedure with our MTP on similar cell sizes revealed a {systematic} increase of approximately 70-100 $mJ/m^2$ in $\gamma_{\mathrm{USF}}$ for various compositions. In comparison, alternative approaches, such as the surrogate model of~\citet{Hu2021}, give comparable, but overestimated, values (the last row in Table~\ref{tab:comp}).
Additionally, our calculations highlight the substantial contribution of local ionic relaxations to the USF energy, causing an energy change of about 25\%. Despite this, neglecting local relaxations does not affect overall trends and relative changes upon concentration variations, potentially justifying the use of CPA calculations for qualitative studies (more detailed analysis provided in SM~\ref{sup_valid_mat1}).

\begin{table}
\caption{Comparison of stacking fault and surface energies (in mJ/m$^2$) for MoNbTi obtained from actively-learned MTP (this work), MTP trained on large data set (\cite{Zheng2022}), SQS (\cite{Mak2021},\cite{Xu2020}) and surrogate model trained on SQS data (\cite{Hu2021}).}
    \begin{ruledtabular}
        \begin{tabular}{c|c|c|c|c}
              & $\gamma_{surf}$ $\{100\}$
              & $\gamma_{surf}$ $\{110\}$
              & $\gamma_{USF}$ $\{110\}$
              & $\gamma_{USF}$ $\{112\}$ \\
               \hline
  This work              &  1989     &  1832  &  730  &  853 \\
  Mak~\cite{Mak2021}     &  2317     &  2171  &  820  &  927 \\
  Zheng~\cite{Zheng2022} &    -      &    -   &  774  &   -  \\
  Xu~\cite{Xu2020}       &    -      &    -   &  765  &  865 \\
  Hu~\cite{Hu2021}       &    -      &  2189  &  817  &   - \\
        \end{tabular}
    \end{ruledtabular}
    \label{tab:comp}
\end{table}

\begin{figure}
   \centering
   \includegraphics{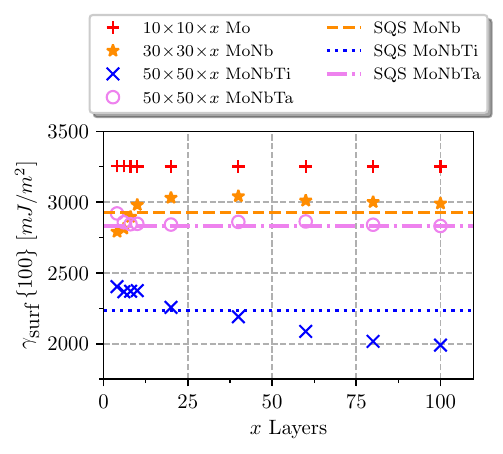}
   \caption{Convergence of surface energies with respect to the number of layers for pure Mo, equimolar MoNb, MoNbTi, and MoNbTa, along with corresponding results for an SQS cell sized 4x4x10. All values are obtained with MTPs.}
   \label{fig:figure5}
\end{figure}

In contrast to USF energies, surface energies for random alloys are much more challenging to compute, hence there are relatively few references available. One of the reasons is the presence of surface reconstructions, which can result in contracted or extended bond lengths near surfaces. Lattice spacing only gradually recovers to the bulk spacing, making it necessary to use relatively large cells for accurate calculations. In addition, finite size ordering effects on the surface must be avoided for random alloys.

According to~\citet{Mak2021}, the $\{100\}$ and $\{110\}$ surfaces of MoNbTi exhibit surface energies of 2317 $mJ/m^2$ and 2171 $mJ/m^2$, respectively. Moreover, the surrogate model of~\citet{Hu2021} has determined the surface energy for the $\{110\}$ surface to be 2189 $mJ/m^2$. Both results were obtained with the SQS approach. However, our methodology has produced lower values for these surface energies. Specifically, our calculations yield surface energies of 1989 $mJ/m^2$ and 1832 $mJ/m^2$ for the $\{100\}$ and $\{110\}$ surfaces, respectively.
To understand the source of such a considerable discrepancy, we performed the convergence analysis with respect
to the slab size (number of layers), which can be easily done with MTP.
The analysis, whose results are presented in Fig.~\ref{fig:figure5}, clearly shows that the convergence of the surface energies can be
rather slow. In the case of MoNbTi, up to 100 layers are required to produce the correct (converged) values.
A more detailed examination of this behavior suggests that it can be traced back to a strong effect of statistical
fluctuations of local atomic configurations on the relaxation contribution to the total energy (see SM~\ref{sup_valid_mat2} for details).
This emphasizes the importance of the analysis of supercell size convergence and also shows insufficiency of
typical SQS sizes for accurate surface energies of random alloys.

\subsection{Analysis of trade-offs}

\begin{figure*}
   \centering
\includegraphics{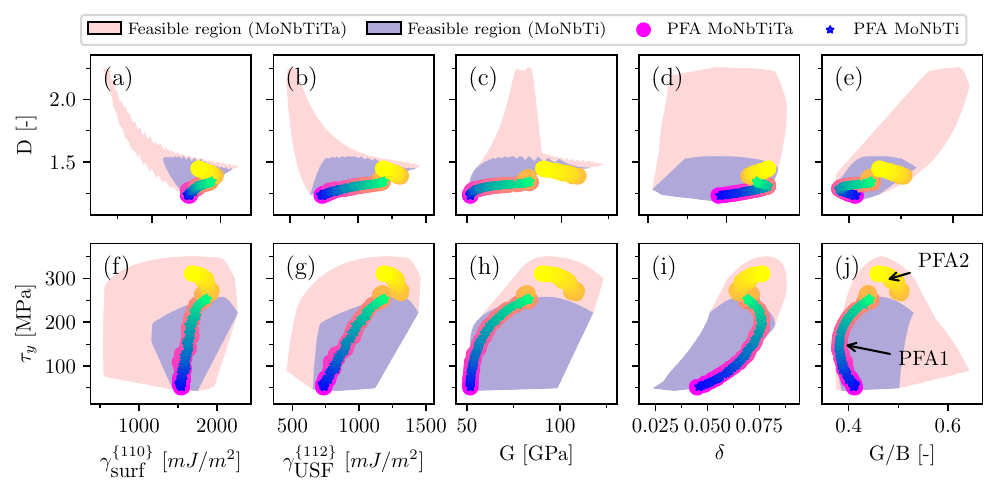}
   \caption{Feasible regions (light blue for MoNbTi, light red for MoNbTiTa) and correlations between materials parameters
   (surface energy, $\gamma_{\mathrm{surf}}^{\{110\}}$, USF energy, $\gamma_{\mathrm{USF}}^{\{112\}}$, (c) shear modulus, $G$,
   misfit parameter, $\delta$, and Pugh's ratio, $G/B$, where $B$ is the bulk modulus) and objectives
   (the ductility index, $D$, and CRSS, $\taus$) evaluated along the Pareto fronts (marked with blue-green for MoNbTi
   and with red-yellow for MoNbTiTa). MoNbTiTa's PFA has two segments: PFA2 is represented by the predominantly yellow tail, PFA1 overlaps with the PFA of MoNbTi.
   }
   \label{fig:figure19}
\end{figure*}

The VBA model is very efficient and accurately captures the strength and ductility parameters, enabling exhaustive scans across the entire concentration space. This allows us to track the evolution of material parameters as new alloy components are introduced
{and also quickly evaluate feasible regions with high resolution, which would otherwise imply significant
computational costs.
For example, by scanning over a dense mesh of concentrations for MoNbTi, MoNbTiTa, we can visualize the feasible regions of the two objectives ($D$ and $\taus$)} as functions of the most influential material {properties},
with the results displayed in Fig.~\ref{fig:figure19}.
For the planar defect energies, we use the dominating set of orientations ($\{110\}/\{112\}$), which determines the value
of $D$ for most of the Pareto-optimal compositions.

Examination of the feasible regions themselves already reveals some general trends. From the shapes of the regions one can infer
that, generally, the CRSS correlates with the misfit parameter, $\delta$, and the shear modulus, $G$, and only weakly with
the USF energy, $\gamma_{\mathrm{USF}}$, while possessing practically no correlation with $\gamma_{\mathrm{surf}}$.
Interestingly, despite the considerable correlation with the shear modulus,
$\taus$ exhibits no particular trends in terms of the Pugh's ratio, $G/B$.
At the same time, the ductility index, $D$, has clear correlations with the surface and USF energies (in the form of the reciprocal
relation) and a rather strong correlation with $G/B$, while the correlations with $G$ and $\delta$ are relatively weak.
An important conclusion to be drawn here is that, since $\taus$ and $D$ only weakly correlate with $\gamma_\text{surf}$
and $\delta$, respectively, the latter two material parameters offer a possibility for an independent fine-tuning of
the ductility and strength.

The trends inferred from the shapes of the feasible regions are, of course, not surprising.
However, the analysis of the Pareto-optimal compositions provides us with a more refined picture. 
Firstly, one portion of the PFA for MoNbTiTa is almost completely overlapping with the PFA for MoNbTi. We will refer to this portion as MoNbTi(Ta). This portion of the PFA of MoNbTiTa corresponds to PFA1, which was previously discussed. The other visible portion of MoNbTiTa corresponds to PFA2.

The most prominent observation, when examining the Pareto fronts, is the consistent 
change in trends between the two portions of the PFA of MoNbTiTa. If one 
Pareto front shows a positive correlation, another one
shows a negative correlation.
For instance, although high surface energies are generally beneficial for ductility,
$D$ in MoNbTi(Ta) increases (i.e., the material becomes less ductile) at higher values of $\gamma_\text{surf}$,
mainly because $\gamma_\text{USF}$ (and $G$) increases faster.
However, in the other portion of MoNbTiTa, this trend is reversed along the PFA, where the ductility improves ($D$ decreases) thanks to a comparable rate of increase
in both $\gamma_\text{surf}$ and $\gamma_\text{USF}$.

Notably, the behavior of the Pugh's ratio, $G/B$, is much more non-trivial than what one could previously deduce from
the plots of feasible regions.
While it effectively distinguishes between brittle and ductile
composition regions similar to $D$, the positive correlation
with $D$ is observed only along a more brittle part of the PFA for MoNbTi(Ta), while the more ductile
part as well as the PFA of MoNbTiTa reveal a clear negative correlation with the ductility index.
This suggests that blindly following the Pugh's criterion for improving ductility might result in very
unfavorable compromises on the strength.

Finally, another way to use Fig.~\ref{fig:figure19} is to rationalize the effects of the addition of Ta already visible
in the Pareto-front representation in Fig.~\ref{fig:fig_03}.
The observed increased strength of optimal compositions with Ta (along the PFA) can be primarily attributed to Ta enhancing the misfit parameter, $\delta$ (see also Fig.~\ref{fig:paper_compare_alloy}a), and, to a certain extent, enhancing the shear modulus $G$. However, although high 
strength alloys are associated with larger values for both, $G$ and $\delta$, the maximum of these
two quantities does not necessarily mean maximum strength, as is evident from the locations of the highest values of $\taus$ at the Pareto fronts.
Furthermore, the reduced ductility of Ta-containing alloys can be attributed to lower surface energies,
which is not compensated by the modest decrease in USF energies. This also highlights how the Pugh's criterion
can fail to predict the correct ductility trend when the alloy surface energy is ignored.

\begin{figure}
   \centering
\includegraphics{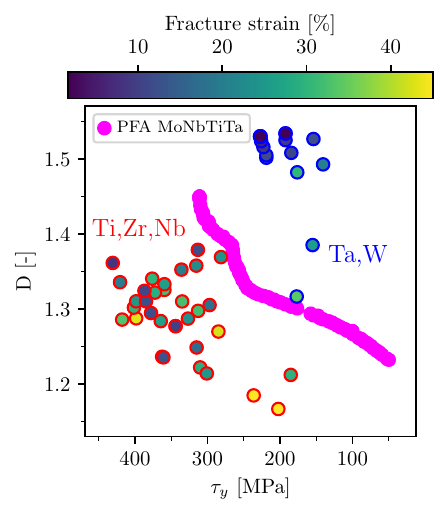}
   \caption{ Ductility index, $D$, and CRSS, $\taus$ evaluated for various alloys using the VBA model. Red circles indicate Ti, Zr, and Nb-rich alloys, while blue circles denote Ta and W-rich alloys, with the fill color (from dark blue to yellow) corresponding to the measured fracture strain (\cite{Singh2023}). The Pareto front of MoNbTiTa is plotted in magenta points.
   }
   \label{fig:appendix_parity_d_8}
\end{figure}

To check some of the above observations and to demonstrate the potential of the model for exploration of other systems,
we extend our investigation of trends in 
$D$ and $\taus$ to a wider range of alloys comprising elements Nb, Ta, Mo, W, Ti, V, Zr, and Hf.
Specifically, we consider the same alloy compositions as in Fig.~\ref{fig:appendix_parity_d_82}
and apply the VBA model to compute all necessary properties for the main objectives, $D$ and $\taus$, with the results shown in Fig.~\ref{fig:appendix_parity_d_8}. 
We also add values from the Pareto set of MoNbTiTa (mind the difference
in scales when comparing to Fig.~\ref{fig:fig_03}), which puts this alloy system into a broader context.

A remarkable feature of Fig.~\ref{fig:appendix_parity_d_8} is an evident separation 
of points into two distinct clusters.
These clusters can be categorized into specific alloy groups characterized by certain primary elements. 
One cluster, with points marked by red circles, consists of alloys rich in Ti, Zr, and Nb,
which span a range of CRSS values from 200 to 400 MPa and have moderate to high values of the fracture strain
(equivalently, moderate to low values of $D$). 
Another cluster, with points marked by blue circles, comprises alloys rich in Ta and W
and these alloys are characterized by smaller fracture strains and systematically high values of $D$. 
Another remarkable feature is that the Pareto front of MoNbTiTa almost perfectly separates these two
clusters, suggesting that the Pareto set represents the most optimal alloys in the class of Ta- or Mo-rich alloys
(and containing only moderate amount of ductilizers in the form of Ti).
A more detailed analysis shows that superior properties of most of the alloys in the ``red'' cluster
can be roughly rationalized along the lines of the conclusions we have made from Fig.~\ref{fig:figure19}.
The $D$-parameter of these alloys is within the range of values on the Pareto front of MoNbTiTa,
which also applies to the surface and USF energies. However, values of $\delta$ for the alloys
from the cluster exceed the misfit parameter of MoNbTi by a factor of 2-4,
making them much stronger and thus placing them far beyond the Pareto front of MoNbTiTa.

\subsection{Conclusion}

In conclusion, we have demonstrated how a combination of CPA and actively-learned MTPs can solve the problem of predicting the CRSS and intrinsic ductility of refractory MPEAs over the whole composition space with ab initio accuracy, a problem that is currently not tractable with conventional
ab initio-based methods.
In particular, by integrating our \textit{ab initio} automated workflow into a robust Bayesian multi-objective optimization framework
we have applied the developed approach to identifying high-performance Pareto-optimal compositions.

Furthermore, our supercell size convergence tests enabled by MTP-powered molecular statics have revealed that sizes of typical
supercell sizes used in conventional SQS calculations are often insufficient for accurate calculations of certain important materials properties, such as the surface energy, of multi-component alloys. Specifically, the surface energy of MoNbTi has been shown to exhibit a rather slow convergence with
the supercell size, which leads to errors of 10--20 \% for SQS supercells of about 100 atoms.

We have applied our multi-objective optimization framework to the four-component MoNbTiTa alloy system and exemplified its use by investigating how alloying with Ta affects the strength-vs-ductility trade-off.
Interestingly, we have found that, in this particular case, instead of advancing the Pareto front of the three-component system, the addition of Ta resulted in the emergence of a separate optimal region, extending the initial Pareto front of MoNbTi to higher 
strength values. Such a behavior represents a peculiar case of a disjoint Pareto sets in the composition space,
which might have important implications for alloy design.

Furthermore, we have used the results from the workflow to validate a
model based on the virtual bond approximation (VBA), which captures the bonding mechanism of transition metals.
Relying on only a couple of element-specific parameters and universal functions,
our VBA model has proved to provide
estimates of alloy properties over the entire concentration space that are in
very good agreement with direct CPA+MTP calculations.
We have, {then}, used the VBA model to examine the trade-off between $\taus$ and $D$ in terms of relevant material
parameters {for a large class of refractory alloys}.
The results suggest that an improved trade-off in alloy design can generally be achieved
if one fine-tunes the CRSS and ductility index by targeting the misfit parameter and the surface energy, respectively.
This analysis also demonstrates how the VBA model can be a useful tool for examining trade-offs
between competing properties in a large compositional space of refractory alloys.

Although our study focuses on strength and ductility, the proposed methodology can be readily applied to other mechanical properties, such as precipitation strengthening, which would involve
interface energies or anti-phase boundary energies.
Moreover, given the good accuracy of the proposed VBA model, 
it may also be used as a surrogate model in a multi-fidelity optimization framework, where ab initio
evaluations of the objectives are triggered only rarely, allowing for efficient on-the-fly optimization
of alloy properties.

Another prospective direction for development is the integration of recently developed multicomponent magnetic MTPs
\cite{kotykhov_constrained_2023} into the presented framework, opening the possibility to treat magnetic
alloys with complex magnetic configurations. This can be also complemented by the learning approach,
whereby the MLIPs are trained on the fly using fragments of the whole system \cite{hodapp_operando_2020},
which could broaden the range of applicability of our framework to systems that are, in principle, barely
tractable by conventional Kohn-Sham DFT, e.g., isolated dislocations.

\section{Methods}

\subsection{Analytical model for intrinsic ductility}
\label{method_ductility}

To predict the intrinsic ductility of a material, the 
Linear Elastic Fracture Mechanics (LEFM) based theory is employed. This theory, initially 
introduced by Rice and Thomson~\cite{Rice1992}, has recently found application in the description of 
ductility of alloys and has been enhanced by incorporating atomistic insights~\cite{Mak2021, Andric2019}. It 
aims to describe the behavior of ideal sharp cracks in elastic bodies and evaluates 
a material's propensity for failure due to crack propagation. This is done by balancing the stress intensity 
or strain energy release rate $K_\text{Ic}$ and $K_\text{Ie}$ of the governing processes. The competition between
brittle cleavage and ductile dislocation emission at the crack tip determines the material's 
behavior, which is reflected in the stress intensity. Plastic deformation at the 
crack front results in a reduction of stress intensity, which ultimately can lead to an 
increase the total fracture toughness.
We only consider in our investigation mode $I$ loading (stress orthogonal to the local plane of the crack surface), which 
generally limits the ductility the most. Crack geometries that dominate ductility show 
cleavage on the $\{110\}$ or $\{100\}$ planes and dislocation emission on the $\{110\}$ or $\{112\}$ planes based 
on Ref.~\cite{Mak2021,Li2020_a,Tyson1973,Vitek1968}. We will refer to these fracture systems as $\{\text{crack plane}\}/\{\text{emission plane}\}$. The ratio of the $K$-factors for a given crack geometry gives the ductility index 
\begin{equation}
    D = \dfrac{K_{Ie}}{K_{Ic}},
\end{equation}
where intrinsic ductility is achieved for $D < 1$ at $T = 0K$. The stress 
intensity factor $K_\text{Ic}$ for cleavage fracture is given by
\begin{equation}
    K_\text{Ic} = \sqrt{\dfrac{2 \gamma_\text{surf} }{ \lambda_{22}(\doubleunderline{C}) }}
\end{equation}
and stress intensity factor $K_\text{Ie}$ for dislocation emission is given by
\begin{equation}
    K_\text{Ie} = \dfrac{\sqrt{\gamma_\text{USF} 
    \,o(\doubleunderline{C}, \theta, \phi) }}
    {F_{12}(\doubleunderline{C},\theta) \cos(\phi) },
\end{equation}
where $\theta$ is the angle of the slip plane with respect to the crack front, $\phi$ is 
the angle of the Burgers vector $b$ inclined to the slip direction, 
$\gamma_{\text{USF}}$ is the unstable stacking fault energy of the emission plane, 
$\gamma_{\text{surf}}$ is the surface energy of the crack plane and 
$\doubleunderline{C}$ is the linear elastic tensor. $\lambda_{22}$ is Stroh's anisotropic elasticity parameter~\cite{Ting1996}. $F_{12}$ and $o$ are geometrical and anisotropic parameters~\cite{Andric2019}. 

\subsection{Analytical model for yield strength}
\label{method_sss}

The yield strength in bcc alloys with dominating solid solution strengthening can be evaluation with the model proposed by Maresca and Curtin~\cite{Maresca2020_a, Maresca2020_b}. Within their model, the temperature and strain-rate-depended yield stress is characterized by two thermal activation models specific to certain temperature ranges.
In the low-temperature/high stress ($\tau_y/\tau_{y0} > 0.5$) range, $\taus(T)$ is given by,
\begin{align}
\taus(T) = \tau_{y0} 
    \left( 1 
        - \left(
        \frac{k T}{\Delta E_{b}}
        \ln \frac{\dot{\varepsilon}_{0}}{\dot{\varepsilon}}
        \right)^{\frac{2}{3}}
    \right),
    \label{eq:tau_lowT}
\end{align}
whereas for higher temperatures/low stress ($\tau_y/\tau_{y0} > 0.5$),
\begin{align}
\taus(T) = \tau_{y0} \exp
    \left(
        -\frac{1}{0.55} \frac{k T}{E_{b}}
        \ln \frac{\dot{\varepsilon}_{0}}{\dot{\varepsilon}}
    \right).
    \label{eq:tau_T}
\end{align}

In our calculations, we assume a $\dot{\varepsilon}$ of $5 \cdot 10^{-4}\,s^{-1}$ and $\dot{\varepsilon}_{0}$ of $10^4\,s^{-1}$.

In the above equations, $\tau_{y0}$ and $\Delta E_{b}$ are, respectively, the zero-temperature yield stress
and the activation barrier, given by the following expressions:

\begin{align}
    \tau_{y0} = & 0.04 \alpha^{-\frac{1}{3}} \overline{\mu}
       \Big( \dfrac{1 + \overline{\nu}}{1 - \overline{\nu}}   \Big)^\frac{4}{3} \delta^\frac{4}{3},
    \label{eqn:vc_yield} \\
     \Delta E_{b} = & 2.00 \alpha^{\frac{1}{3}} \overline{\mu} \overline{b}^3
        \Big( \dfrac{1 + \overline{\nu}}{1 - \overline{\nu}}   \Big)^\frac{2}{3} \delta^\frac{2}{3},
    \label{eqn:vc_energy}
\end{align}

where $\alpha = 1/12$, $\overline{B}$, $\overline{\mu}$ and $\overline{\nu}$ are averaged bulk, shear modulus and Poisson ratio, $b$ is the average Burger's vector and the misfit parameter is $\delta = \dfrac{4}{3 \sqrt{3}}\sqrt{\sum_i c_i (\Delta V_i)^2} / (V )$. The averaged
linear elastic properties are obtained by:

\begin{align}
    \overline{\mu} = \sqrt{\frac{1}{2} C_{44} (C_{11} - C_{12})}
\end{align}

\begin{align}
    \overline{B} = C_{11} + 2 C_{12}  
\end{align}

\begin{align}
    \overline{\nu} = \frac{3 \overline{B} - 2 \overline{\mu} }{2 (3 \overline{B} + \overline{\mu})}  
\end{align}

The misfit volumes $\Delta V_i$ are obtained from a series of CPA calculations for systems with small deviations in the concentrations from the original alloy in the same way as in our earlier work \cite{Moitzi2022}. For each elemental component, four calculations are
performed where the concentration of the respective element is changed by
$-\delta$, $-\delta/2$, $\delta/2$, and $\delta$ ($\delta\,=\,0.005$) 
from the original alloy. The concentration ratios among the remaining constituent elements are kept constant. The linear elastic constants were obtained from
volume-conserving monoclinic and orthorhombic distortions following the computational
details described in Ref.~\cite{Razumovskiy2011a,Razumovskiy2011b}. 
To extract CRSS $\taus$ from experimental data of poly-crystalline samples, we subtract an estimated 
Hall-Petch contribution from Ref.~\cite{Cordero2016} and 
divide the result by the Taylor factor of 3.09.

\subsection{First-principles evaluation of materials parameters}
\label{method_abinitio}

To determine the relevant materials parameters that affect ductility and strengthening, various \textit{ab initio} methods based on density-functional theory are used. 

Elastic constants, equilibrium volumes and concentration derivatives of equilibrium volumes are evaluated with the exact muffin-tin orbital (EMTO)
\cite{Vitos2001a} code (Lyngby version \cite{Ruban2016}) implementing a Green's function
based DFT methodology combined with CPA to perform total energy calculations. 
Parameters for screened Coulomb interactions in the CPA are obtained by
the use of the EMTO-based locally self-consistent Green’s function 
technique \cite{Abrikosov1997,Peil2012}. Accurate total energies are obtained 
within the full-charge density formalism \cite{Vitos1997}. We evaluate finite-temperature
equilibrium volumes following our previously suggested methodology \cite{Moitzi2022} with 
LDA as the reference functional for the pressure correction calculation. The elemental reference volumes are obtained by extrapolation of cryogenic lattice constant measurements \cite{Woodard1969,Shah1971,Spreadborough1959,Corruccini1961,Goldak1966,Versaci1991,Smirnov1965} to 0K. Zero point vibrations are included in the calculations by a quasi-harmonic Debye model $U = 9/8 N k_b \Theta_D(V)$ as described in Refs. \cite{Moruzzi1988,Blanco2004}.

The unstable stacking fault energies, $\gamma_\text{USF}$, and surface energies, $\gamma_\text{surf}$, are obtained using Moment Tensor Potentials (MTPs) \cite{Novikov2021,Shapeev2016,Gubaev2019,Podryabinkin2017} which are actively-learned from DFT.
The total energy of MTPs is given by a sum over all $i$ atoms in the configuration
\begin{equation}
    E_{\text{MTP}} = \sum_i \sum_{\alpha} \xi_{\alpha} B_{\alpha}(n_i),
\end{equation}
where $\xi_{\alpha}$ are linear parameters and $B_\alpha(n_i)$ basis functions that depend on the neighbourhood $n_i$ within some cutoff radius around atom $i$.
The scalar basis functions are obtained from contractions of the moment tensor descriptors
\begin{equation}
M_{\mu,\nu} (n_i) = \sum_j f_\mu ( \vert \mathbf{r}_{ij} \vert ) \underbrace{ \mathbf{r}_{ij}  \otimes \cdots \otimes  \mathbf{r}_{ij} }_{\nu\,\text{times}},
\end{equation}
where the per-atom radial functions $f_\mu ( | \mathbf{r}_{ij} | )$ are expanded in terms of Chebyshev polynomials.
The total number of basis functions contained in $E_{\text{MTP}}$ is defined by the MTP level.
For the hyperparameters of the MTP, we choose a level of 16 and a cutoff of 5.0\,\AA. 
{
The training sets are constructed using the algorithm proposed in \cite{Hodapp2021}. Below, we highlight only the most important points of the algorithm including our small modifications. For a more detailed description we refer to the original work of \citet{Hodapp2021}.

In the first step, we randomly sample a set of training candidates containing 72/96-atom surface, 54-atom bulk, and 72/96-atom unstable stacking fault configurations on a regular concentration grid within a simplex. To explore more extrapolative configurations, we also introduce slight displacements of atoms and deformations of the cells.
Choosing a grid with four points in each dimensions was found to be sufficient to predict energies over the entire composition space for the initial training set (for details, see \cite{Hodapp2021}). 
In total, our initial training set contains 146 and 377 configurations.
Next, we let active learning choose the most distinct configurations from the set of training candidates (each candidate set contains around 10000 configurations), perform static DFT calculations on them and augment the interatomic potential by retraining it with these data. The search for distinct configurations is conducted using a progressively decreasing extrapolation grade, starting from 1000 and gradually reducing to 100, 10, and finally 2. We iterate multiple times, with always newly 
generated training candidate sets and stop when no new training configurations emerge anymore that exceeds the extrapolation grade.
This procedure resulted in a training set containing 219 and 631 configurations for MoNbTi and MoNbTiTa, respectively.
Finally, we run molecular dynamics simulations at 50K, 150K, and 300K, on each of the configurations from our initial training set.
During each of the simulations, active learning is used to find new potentially extrapolative configurations.
We remark here that running MD was crucial for the robustness of our potentials.
Training only on relaxations as done in \cite{Hodapp2021} has been found to be insufficient
to ensure stability of large-scale configurations during the lattice relaxation.
Our final training set contains 381 and 720 configurations for MoNbTi and MoNbTiTa, respectively.
The errors on the final training set for total energies are found to be 6 and 19 meV/atom, and for forces -- 0.099 and 0.104 eV/\AA for the MoNbTi and MoNbTiTa potentials, respectively. A detailed performance check can be found in SM~\ref{sup_perf_metric}.
}

The final properties ({unstable} stacking fault and surface energies) are then calculated on a cell with about 300000-500000 atoms to include far-field effects and to ensure proper statistics. {We provide a detailed convergence study for the cell size in SM~\ref{sup_conv_elements}.} 

The DFT computations for the MTP training are performed with the Vienna Ab-initio Package (VASP) \cite{Kresse1993,Kresse1996a,Kresse1996b} using the PBE functional \cite{pbe96} with an energy cutoff of 400 eV. We use a Monkhorst-Pack k-point mesh with a spacing of 0.15 $\text{\AA}^{-1}$.
Moreover, we use Gaussian smearing with a smearing width of 0.08 eV to avoid potential problems of negative occupations.
The standard pseudopotentials (PP) were used for Mo, Ti, and Ta~\cite{Kresse1994,Kresse1999}. For Nb, PP including semicore p-states as valence states was used.

\subsection{Multi-objective Bayesian-optimization}
\label{method_moo}

In order to effectively find the Pareto front of our multi-objective Bayesian optimization problem, we 
have adapted a methodology developed by~\citet{Galuzio2020}. This method involves generating a 
Pareto front approximation (PFA) from an easy-to-evaluate surrogate model at 
each search step, and selecting the next evaluation point from this PFA before updating 
the surrogate model. This iterative process enables the PFA to gradually approach the true Pareto front. We 
can also narrow the search space by excluding compositions that are thermodynamically unstable or possess unfavorable properties. 

To navigate through the search space, we employ the following sequential design strategies:

1. We begin the search process by building a smooth surrogate model.
To that end, we are using Gaussian processes regression (GPR) based on the observations available from previous rounds. For the initial round, we sample a coarse regular grid in the search space. The choice of the covariance function $K(\mathbf{x}, \mathbf{x'})$ is determined using cross-validation on grid test sets. We have selected a sum of white noise 
kernels and radial basis function kernels as the best:
\begin{equation}
    K(\mathbf{x}, \mathbf{x'}) = \sigma^2 \mathbf{I} + C \exp(-\gamma \left\| \mathbf{x} - \mathbf{x'} \right\|^2),
\end{equation}
where $\sigma, C, \lambda$ are hyper-parameters estimated by maximizing the marginal likelihood on the training data.

The surrogate models for our objectives, namely the ductility index $D$ and the yield strength $\taus$, are then 
used to make predictions at unobserved areas of the search space and quantify
the uncertainty around them. This information is crucial for guiding the search towards promising areas in the search space.

2. Next, we use the non-dominated sorting genetic algorithm II (NSGAII)~\cite{Deb2022} implemented within the DEAP library \cite{DEAP_JMLR2012} in order to identify 100 PFA points $\Omega_\text{PFA}$ using our surrogate model.
A penalty function is used to handle constraints, such as limiting the search space and keeping 
it within a certain concentration range.

3. Select the next search points $\mathbf{x}^{n + 1}$ from the current PFA, $\mathbf{x}^{n + 1} = \mathbf{x}^{*}_{t_\text{next}} \in \Omega_\text{PFA}$, that are farthermost from all previous sampling points according to
\begin{equation}
    t_{\text{next}} = \text{argmax} \left( q \dfrac{d^{(i)}_{f} - \mu_f}{\sigma_f} + (1 - q) \dfrac{d^{(i)}_{x} - \mu_x}{\sigma_x} \right),
\end{equation}
where $d^{(i)}_{\{\cdot\}}$ corresponds to the smallest distance from 
the $i^\text{th}$ point in $\Omega_{PFA}$ to all other points found by the algorithm. The subscript ``f'' indicates the distance 
calculated in objective space, while the subscript ``x'' indicates the distance in the 
search space. The distances are standardized by $\mu_{\{\cdot\}}$, representing the average, 
and $\sigma_{\{\cdot\}}$, representing the standard deviation, of the respective distance measure. We choose the weighting parameter $q$ as $0.8$ to slightly favour objective space exploration.

4. Check the stopping criterion: If the maximum number of iterations has been reached or only minimal Pareto-improvement is achieved, stop the algorithm and output the Pareto front and the Pareto set. 
{ The stationarity condition on the Pareto front measures 
Pareto-improvement by determining how much the front changes with more points. If there is little change after several iterations, the loop terminates.} Otherwise, go back to 
step 3 and continue the search. The final Pareto set can be discontinuous and portion of the Pareto set can be located in different regions of the concentration space. {To identify separate portions}, 
we use multivariate Gaussian mixture models to perform automatic spatial 
clustering and label the corresponding parts at the Pareto front. { 
For MoNbTi, 36 points are required, whereas for MoNbTiTa, 143 points are needed, with 72 for set 1 and 71 for set 2.}

\subsection{Dimensionality reduction and visualization}
\label{method_tsne}
 
We employed a t-Distributed Stochastic Neighbor Embedding (t-SNE), a dimensionality reduction technique, to 
visualize the 4-dimensional concentration space in two dimensions. This projection allows us to visualize our objectives 
and gain insights into the overall landscape. The biggest advantage of t-SNE is that it can capture much of 
the high-dimensional data's local structure while also revealing global structure. Furthermore, it is a non-linear 
method that can also capture the structure of complex manifolds. However, t-SNE provides only an approximate 
preservation of local structure and distances, which leads to the emergence of an observed patchy pattern in 
our case. The corner points of the projection correspond to the elemental compounds, the edges represent binary 
compositions, and the center represents the equimolar compound. The distance from the corner points in the 
projection is approximately proportional to the effective composition. 
{The t-SNE projections displayed in Fig.~\ref{fig:fig_03} in the main text are performed from
the entire concentration space, with points}
evaluated using the Gaussian process surrogate models for the ductility index and the strength.

\subsection{Virtual bond approximation model}
\label{method_vba}

Within the VBA, to second order in inter-component interactions, the energy of an alloy system can be expressed as follows:
\begin{align}
    E(\{c_i\}, \lambda) = & \sum_{i} c_i \, w_i e^{(1)}(N_i, \lambda) + \notag \\
       & + \sum_{i j} c_i c_j \, w_{ij} e^{(2)}(N_{ij}, \lambda),
    \label{eqn:vba_base}
\end{align}
where $c_i$ are atomic fractions of components, $e^{(k)}$ are the virtual bond functions of the average $d$-valences,
$N^{(k)} \equiv \{ N_i, N_{ij} \}$, and of bandwidth parameters $w^{(k)} \equiv \{ w_i, w_{ij} \}$.
Here $N_i$ and $w_i$ refer to a respective elemental value, while second-order parameters are defined as
$N_{ij} = (N_i + N_j) / 2$ and $w_{ij} = \sqrt{w_i w_j}$,
where the latter expression is inspired by the tight-binding description of transition-metal alloys \cite{Shiba1971}.
$\lambda$ represents all other parameters, such as temperature, volume, lattice vectors, etc.
The bandwidth parameters were taken from Ref.~\cite{Andersen1985} and adjusted by a small correction depending on the row of an element.
{
Alternatively, one can fit these parameters to elemental compounds and a few binary alloys but
usually the relative values obtained this way are similar to those from the above reference.}
Note also that the expression $E(\{ c_i \}, \lambda)$ can represent either the Gibbs free energy, $G(\{ c_i \}, P, T)$, or
the Helmholtz free energy, $F(\{ c_i \}, V, T)$, where additional parameters (e.g., lattice vectors) are implied.
Considering derivatives of the thermodynamic potentials with respect to various parameters,
we can get VBA representations of any property that can be derived from the free energy.

From the Gibbs free energy, we can the derive an expression for the volume,
$V(\{ c_i \}) = \partial{G(\{ c_i \}, P)} / \partial{P}|_{P = 0}$, omitting the temperature dependence for simplicity.
Similarly, we can obtain an expression for the bulk modulus and general elastic (isothermal) constants by 
$B = V \, \partial^2{F(\{ c_i \}, V)} / \partial{V^2}$ 
and $C_{\alpha\beta} = \frac{\partial^2 F(\{ c_i \}, V)}{\partial \epsilon_\alpha \partial \epsilon_\beta}$,
where $\epsilon_{\alpha}$ is the strain in Voigt notation.
In the context of surface and USF energies, the molar energy difference,
$\Delta E = E_{\mathrm{defect}} - E_{\mathrm{bulk}}$, between the defective and bulk structures is used %
complemented by the area normal to the planar fault deduced from the predicted volume.
The final expression always assumes the form:
\begin{align}
    \mathbf{\xi}(\{c_i \}) = & \sum_i c_i \, w_i \, \mathbf{\xi}^{(1)} \big( N_i \big) + \notag \\
       & + \sum_{i j} c_i \, c_j \, w_{ij} \, \mathbf{\xi}^{(2)} \big( N_{ij} \big),
\end{align}
where $\mathbf{\xi}(\{ c_i \})$ is one of $B$, $C'$, $C_{44}$, $V$, or $\Delta E$, and $\mathbf{\xi}^{(k)}$ represent
the corresponding virtual bond functions.
To obtain the unknown $\mathbf{\xi}^{(k)}$, we perform a series of calculations 
of the associated quantity for several systems (elemental compounds and selected binary, three- and four-component alloys).
The unknown values of virtual bond functions at integer (for $\xi^{(1)}$) 
or half-integer values (for $\xi^{(2)}$) are then obtained by 
means of linear regression from the above equation,
since the values of $c_i$ and $w_i$ are known for each composition. The 
discrete function values obtained in this manner for our refractory alloys behave very well
and can be represented as values of smooth functions of valences.
In fact, we use second and third-order polynomials to accurately represent, respectively, the first order virtual bond function $\mathbf{\xi}^{(1)}$
and the second order virtual bond function $\mathbf{\xi}^{(2)}$ as a way of conveniently obtaining a value for a given $N$.
This implies that virtual bond functions can be used to consistently 
characterize the system.
An example of the virtual bond function can be found in Supplementary~\ref{sup_vba}.

The material parameters for determining the corresponding virtual bond functions have been obtained using CPA calculations.
For planar defects we first calculate \emph{unrelaxed} surface and USF energies with 
CPA for elemental compounds and binary alloys and parameterize the VBA model, as described above.
We have checked that the VBA model predicts the CPA values very well (see Supplementary~\ref{sup_vba}).
To account for relaxation effects, we simply adjust the virtual bond functions
to align with the \emph{relaxed} surface and USF energies of individual elemental compounds. 
More precisely, we scale the virtual bond functions by a factor that follows a linear relationship with the $d$-valences.
Such an approach works because fitting to CPA values already includes most of the chemical interactions
and the strongest relaxation effects have geometrical nature, which can be taken into account by the rescaling procedure of the universal virtual bond functions.

\acknowledgments{
We are grateful to William A. Curtin, Francesco Maresca, and Vsevolod Razumovskiy for fruitful discussions.
This work was supported by the Forschungsf\"orderungsgesellschaft (FFG)
project No. 878968 ``ADAMANT'', Austrian Science Fond (FWF) project No. P33491-N ``ReCALL'',
and COMET program IC-MPPE (project No. 859480).
This program is supported by the Austrian Federal Ministries for Climate Action, Environment, Energy, Mobility, Innovation and Technology (BMK) and for Digital and Economic Affairs (BMDW), represented by the Austrian research funding association (FFG), and the federal states of Styria, Upper Austria and Tyrol. All calculations in this work have been done using Vienna Scientific Cluster (VSC-5).
}

\section{Competing Interests}

The Authors declare no Competing Financial or Non-Financial Interests.

\section{Author Contributions}

F. M.: Writing - Original Draft, Data Analysis, Methodology, Software; 
L. R: Writing - Review \& Editing, Conceptualization; 
A. V. R: Writing - Review \& Editing, Software, Validation; 
M. H: Writing - Review \& Editing, Supervision, Software, Conceptualization, Methodology; 
O. P.: Writing - Review \& Editing, Supervision, Software, Project administration, Funding acquisition, Conceptualization, Methodology

{
\section{Data Availability}

The optimization results for MoNbTi and MoNbTiTa, MTP potentials including their configuration files, and the virtual bond approximation model are all available at: [\url{https://github.com/pmu2022/TradeOffRelations}].

\section{Code Availability}

The codes used to carry out this work are described and referenced in the Methods
section and are available free-of-charge with the exception of VASP. The software utilized in this study are accesible at: MTP-2 \url{https://gitlab.com/ashapeev/mlip-2}, LAMMPS \url{https://github.com/lammps/lammps}, ATAT v3.6 \url{https://www.brown.edu/Departments/Engineering/Labs/avdw/atat/}. Additionally, EMTO, ELSGF, mtp4py, and the automated workflows are accessible upon reasonable request from the authors.
}

\FloatBarrier
\clearpage
\bibliography{main}%

\begin{thebibliography}{136}%
\makeatletter
\providecommand \@ifxundefined [1]{%
 \@ifx{#1\undefined}
}%
\providecommand \@ifnum [1]{%
 \ifnum #1\expandafter \@firstoftwo
 \else \expandafter \@secondoftwo
 \fi
}%
\providecommand \@ifx [1]{%
 \ifx #1\expandafter \@firstoftwo
 \else \expandafter \@secondoftwo
 \fi
}%
\providecommand \natexlab [1]{#1}%
\providecommand \enquote  [1]{``#1''}%
\providecommand \bibnamefont  [1]{#1}%
\providecommand \bibfnamefont [1]{#1}%
\providecommand \citenamefont [1]{#1}%
\providecommand \href@noop [0]{\@secondoftwo}%
\providecommand \href [0]{\begingroup \@sanitize@url \@href}%
\providecommand \@href[1]{\@@startlink{#1}\@@href}%
\providecommand \@@href[1]{\endgroup#1\@@endlink}%
\providecommand \@sanitize@url [0]{\catcode `\\12\catcode `\$12\catcode `\&12\catcode `\#12\catcode `\^12\catcode `\_12\catcode `\%12\relax}%
\providecommand \@@startlink[1]{}%
\providecommand \@@endlink[0]{}%
\providecommand \url  [0]{\begingroup\@sanitize@url \@url }%
\providecommand \@url [1]{\endgroup\@href {#1}{\urlprefix }}%
\providecommand \urlprefix  [0]{URL }%
\providecommand \Eprint [0]{\href }%
\providecommand \doibase [0]{http://dx.doi.org/}%
\providecommand \selectlanguage [0]{\@gobble}%
\providecommand \bibinfo  [0]{\@secondoftwo}%
\providecommand \bibfield  [0]{\@secondoftwo}%
\providecommand \translation [1]{[#1]}%
\providecommand \BibitemOpen [0]{}%
\providecommand \bibitemStop [0]{}%
\providecommand \bibitemNoStop [0]{.\EOS\space}%
\providecommand \EOS [0]{\spacefactor3000\relax}%
\providecommand \BibitemShut  [1]{\csname bibitem#1\endcsname}%
\let\auto@bib@innerbib\@empty
\bibitem [{\citenamefont {Senkov}\ \emph {et~al.}(2011)\citenamefont {Senkov}, \citenamefont {Wilks}, \citenamefont {Scott},\ and\ \citenamefont {Miracle}}]{Senkov2011}%
  \BibitemOpen
  \bibfield  {author} {\bibinfo {author} {\bibfnamefont {O.}~\bibnamefont {Senkov}}, \bibinfo {author} {\bibfnamefont {G.}~\bibnamefont {Wilks}}, \bibinfo {author} {\bibfnamefont {J.}~\bibnamefont {Scott}}, \ and\ \bibinfo {author} {\bibfnamefont {D.}~\bibnamefont {Miracle}},\ }\href {\doibase https://doi.org/10.1016/j.intermet.2011.01.004} {\bibfield  {journal} {\bibinfo  {journal} {Intermetallics}\ }\textbf {\bibinfo {volume} {19}},\ \bibinfo {pages} {698} (\bibinfo {year} {2011})}\BibitemShut {NoStop}%
\bibitem [{\citenamefont {Senkov}\ \emph {et~al.}(2016)\citenamefont {Senkov}, \citenamefont {Isheim}, \citenamefont {Seidman},\ and\ \citenamefont {Pilchak}}]{Senkov2016}%
  \BibitemOpen
  \bibfield  {author} {\bibinfo {author} {\bibfnamefont {O.}~\bibnamefont {Senkov}}, \bibinfo {author} {\bibfnamefont {D.}~\bibnamefont {Isheim}}, \bibinfo {author} {\bibfnamefont {D.}~\bibnamefont {Seidman}}, \ and\ \bibinfo {author} {\bibfnamefont {A.}~\bibnamefont {Pilchak}},\ }\href {\doibase 10.3390/e18030102} {\bibfield  {journal} {\bibinfo  {journal} {Entropy}\ }\textbf {\bibinfo {volume} {18}},\ \bibinfo {pages} {102} (\bibinfo {year} {2016})}\BibitemShut {NoStop}%
\bibitem [{\citenamefont {Wei}\ \emph {et~al.}(2020)\citenamefont {Wei}, \citenamefont {Kim}, \citenamefont {Kang}, \citenamefont {Zhang}, \citenamefont {Zhang}, \citenamefont {Furuhara}, \citenamefont {Park},\ and\ \citenamefont {Tasan}}]{Wei2020}%
  \BibitemOpen
  \bibfield  {author} {\bibinfo {author} {\bibfnamefont {S.}~\bibnamefont {Wei}}, \bibinfo {author} {\bibfnamefont {S.~J.}\ \bibnamefont {Kim}}, \bibinfo {author} {\bibfnamefont {J.}~\bibnamefont {Kang}}, \bibinfo {author} {\bibfnamefont {Y.}~\bibnamefont {Zhang}}, \bibinfo {author} {\bibfnamefont {Y.}~\bibnamefont {Zhang}}, \bibinfo {author} {\bibfnamefont {T.}~\bibnamefont {Furuhara}}, \bibinfo {author} {\bibfnamefont {E.~S.}\ \bibnamefont {Park}}, \ and\ \bibinfo {author} {\bibfnamefont {C.~C.}\ \bibnamefont {Tasan}},\ }\href {\doibase 10.1038/s41563-020-0750-4} {\bibfield  {journal} {\bibinfo  {journal} {Nature Materials}\ }\textbf {\bibinfo {volume} {19}},\ \bibinfo {pages} {1175} (\bibinfo {year} {2020})}\BibitemShut {NoStop}%
\bibitem [{\citenamefont {Gao}\ \emph {et~al.}(2015)\citenamefont {Gao}, \citenamefont {Carney}, \citenamefont {Do{\u{g}}an}, \citenamefont {Jablonksi}, \citenamefont {Hawk},\ and\ \citenamefont {Alman}}]{Gao2015}%
  \BibitemOpen
  \bibfield  {author} {\bibinfo {author} {\bibfnamefont {M.~C.}\ \bibnamefont {Gao}}, \bibinfo {author} {\bibfnamefont {C.~S.}\ \bibnamefont {Carney}}, \bibinfo {author} {\bibfnamefont {{\"O}.~N.}\ \bibnamefont {Do{\u{g}}an}}, \bibinfo {author} {\bibfnamefont {P.~D.}\ \bibnamefont {Jablonksi}}, \bibinfo {author} {\bibfnamefont {J.~A.}\ \bibnamefont {Hawk}}, \ and\ \bibinfo {author} {\bibfnamefont {D.~E.}\ \bibnamefont {Alman}},\ }\href {\doibase 10.1007/s11837-015-1617-z} {\bibfield  {journal} {\bibinfo  {journal} {JOM}\ }\textbf {\bibinfo {volume} {67}},\ \bibinfo {pages} {2653} (\bibinfo {year} {2015})}\BibitemShut {NoStop}%
\bibitem [{\citenamefont {Dixit}\ \emph {et~al.}(2022)\citenamefont {Dixit}, \citenamefont {Rodriguez}, \citenamefont {Jones}, \citenamefont {Buzby}, \citenamefont {Dixit}, \citenamefont {Argibay}, \citenamefont {DelRio}, \citenamefont {Lim},\ and\ \citenamefont {Fleming}}]{Dixit2022}%
  \BibitemOpen
  \bibfield  {author} {\bibinfo {author} {\bibfnamefont {S.}~\bibnamefont {Dixit}}, \bibinfo {author} {\bibfnamefont {S.}~\bibnamefont {Rodriguez}}, \bibinfo {author} {\bibfnamefont {M.~R.}\ \bibnamefont {Jones}}, \bibinfo {author} {\bibfnamefont {P.}~\bibnamefont {Buzby}}, \bibinfo {author} {\bibfnamefont {R.}~\bibnamefont {Dixit}}, \bibinfo {author} {\bibfnamefont {N.}~\bibnamefont {Argibay}}, \bibinfo {author} {\bibfnamefont {F.~W.}\ \bibnamefont {DelRio}}, \bibinfo {author} {\bibfnamefont {H.~H.}\ \bibnamefont {Lim}}, \ and\ \bibinfo {author} {\bibfnamefont {D.}~\bibnamefont {Fleming}},\ }\href {\doibase 10.1007/s11666-022-01324-0} {\bibfield  {journal} {\bibinfo  {journal} {Journal of Thermal Spray Technology}\ }\textbf {\bibinfo {volume} {31}},\ \bibinfo {pages} {1021} (\bibinfo {year} {2022})}\BibitemShut {NoStop}%
\bibitem [{\citenamefont {Lo}\ \emph {et~al.}(2019)\citenamefont {Lo}, \citenamefont {Chang}, \citenamefont {Murakami}, \citenamefont {Yeh},\ and\ \citenamefont {Yeh}}]{Lo2019}%
  \BibitemOpen
  \bibfield  {author} {\bibinfo {author} {\bibfnamefont {K.-C.}\ \bibnamefont {Lo}}, \bibinfo {author} {\bibfnamefont {Y.-J.}\ \bibnamefont {Chang}}, \bibinfo {author} {\bibfnamefont {H.}~\bibnamefont {Murakami}}, \bibinfo {author} {\bibfnamefont {J.-W.}\ \bibnamefont {Yeh}}, \ and\ \bibinfo {author} {\bibfnamefont {A.-C.}\ \bibnamefont {Yeh}},\ }\href {\doibase 10.1038/s41598-019-43819-x} {\bibfield  {journal} {\bibinfo  {journal} {Scientific Reports}\ }\textbf {\bibinfo {volume} {9}},\ \bibinfo {pages} {7266} (\bibinfo {year} {2019})}\BibitemShut {NoStop}%
\bibitem [{\citenamefont {Xie}\ \emph {et~al.}(2022)\citenamefont {Xie}, \citenamefont {Li}, \citenamefont {Liu}, \citenamefont {Huang}, \citenamefont {He}, \citenamefont {Yu}, \citenamefont {Xiong}, \citenamefont {Wang},\ and\ \citenamefont {Hou}}]{Xie2022}%
  \BibitemOpen
  \bibfield  {author} {\bibinfo {author} {\bibfnamefont {X.}~\bibnamefont {Xie}}, \bibinfo {author} {\bibfnamefont {N.}~\bibnamefont {Li}}, \bibinfo {author} {\bibfnamefont {W.}~\bibnamefont {Liu}}, \bibinfo {author} {\bibfnamefont {S.}~\bibnamefont {Huang}}, \bibinfo {author} {\bibfnamefont {X.}~\bibnamefont {He}}, \bibinfo {author} {\bibfnamefont {Q.}~\bibnamefont {Yu}}, \bibinfo {author} {\bibfnamefont {H.}~\bibnamefont {Xiong}}, \bibinfo {author} {\bibfnamefont {E.}~\bibnamefont {Wang}}, \ and\ \bibinfo {author} {\bibfnamefont {X.}~\bibnamefont {Hou}},\ }\href {\doibase 10.1186/s10033-022-00814-0} {\bibfield  {journal} {\bibinfo  {journal} {Chinese Journal of Mechanical Engineering}\ }\textbf {\bibinfo {volume} {35}},\ \bibinfo {pages} {142} (\bibinfo {year} {2022})}\BibitemShut {NoStop}%
\bibitem [{\citenamefont {Juan}\ \emph {et~al.}(2015)\citenamefont {Juan}, \citenamefont {Tsai}, \citenamefont {Tsai}, \citenamefont {Lin}, \citenamefont {Wang}, \citenamefont {Yang}, \citenamefont {Chen}, \citenamefont {Lin},\ and\ \citenamefont {Yeh}}]{Juan2015}%
  \BibitemOpen
  \bibfield  {author} {\bibinfo {author} {\bibfnamefont {C.-C.}\ \bibnamefont {Juan}}, \bibinfo {author} {\bibfnamefont {M.-H.}\ \bibnamefont {Tsai}}, \bibinfo {author} {\bibfnamefont {C.-W.}\ \bibnamefont {Tsai}}, \bibinfo {author} {\bibfnamefont {C.-M.}\ \bibnamefont {Lin}}, \bibinfo {author} {\bibfnamefont {W.-R.}\ \bibnamefont {Wang}}, \bibinfo {author} {\bibfnamefont {C.-C.}\ \bibnamefont {Yang}}, \bibinfo {author} {\bibfnamefont {S.-K.}\ \bibnamefont {Chen}}, \bibinfo {author} {\bibfnamefont {S.-J.}\ \bibnamefont {Lin}}, \ and\ \bibinfo {author} {\bibfnamefont {J.-W.}\ \bibnamefont {Yeh}},\ }\href {\doibase https://doi.org/10.1016/j.intermet.2015.03.013} {\bibfield  {journal} {\bibinfo  {journal} {Intermetallics}\ }\textbf {\bibinfo {volume} {62}},\ \bibinfo {pages} {76} (\bibinfo {year} {2015})}\BibitemShut {NoStop}%
\bibitem [{\citenamefont {Miracle}\ and\ \citenamefont {Senkov}(2017)}]{Miracle2017}%
  \BibitemOpen
  \bibfield  {author} {\bibinfo {author} {\bibfnamefont {D.}~\bibnamefont {Miracle}}\ and\ \bibinfo {author} {\bibfnamefont {O.}~\bibnamefont {Senkov}},\ }\href {\doibase https://doi.org/10.1016/j.actamat.2016.08.081} {\bibfield  {journal} {\bibinfo  {journal} {Acta Materialia}\ }\textbf {\bibinfo {volume} {122}},\ \bibinfo {pages} {448} (\bibinfo {year} {2017})}\BibitemShut {NoStop}%
\bibitem [{\citenamefont {Zou}\ \emph {et~al.}(2014)\citenamefont {Zou}, \citenamefont {Maiti}, \citenamefont {Steurer},\ and\ \citenamefont {Spolenak}}]{Zou2014}%
  \BibitemOpen
  \bibfield  {author} {\bibinfo {author} {\bibfnamefont {Y.}~\bibnamefont {Zou}}, \bibinfo {author} {\bibfnamefont {S.}~\bibnamefont {Maiti}}, \bibinfo {author} {\bibfnamefont {W.}~\bibnamefont {Steurer}}, \ and\ \bibinfo {author} {\bibfnamefont {R.}~\bibnamefont {Spolenak}},\ }\href {\doibase https://doi.org/10.1016/j.actamat.2013.11.049} {\bibfield  {journal} {\bibinfo  {journal} {Acta Materialia}\ }\textbf {\bibinfo {volume} {65}},\ \bibinfo {pages} {85} (\bibinfo {year} {2014})}\BibitemShut {NoStop}%
\bibitem [{\citenamefont {Sheikh}\ \emph {et~al.}(2016)\citenamefont {Sheikh}, \citenamefont {Shafeie}, \citenamefont {Hu}, \citenamefont {Ahlstr{\"o}m}, \citenamefont {Persson}, \citenamefont {Vesely}, \citenamefont {Zyka}, \citenamefont {Klement},\ and\ \citenamefont {Guo}}]{Sheikh2016}%
  \BibitemOpen
  \bibfield  {author} {\bibinfo {author} {\bibfnamefont {S.}~\bibnamefont {Sheikh}}, \bibinfo {author} {\bibfnamefont {S.}~\bibnamefont {Shafeie}}, \bibinfo {author} {\bibfnamefont {Q.}~\bibnamefont {Hu}}, \bibinfo {author} {\bibfnamefont {J.}~\bibnamefont {Ahlstr{\"o}m}}, \bibinfo {author} {\bibfnamefont {C.}~\bibnamefont {Persson}}, \bibinfo {author} {\bibfnamefont {J.}~\bibnamefont {Vesely}}, \bibinfo {author} {\bibfnamefont {J.}~\bibnamefont {Zyka}}, \bibinfo {author} {\bibfnamefont {U.}~\bibnamefont {Klement}}, \ and\ \bibinfo {author} {\bibfnamefont {S.}~\bibnamefont {Guo}},\ }\href {\doibase 10.1063/1.4966659} {\bibfield  {journal} {\bibinfo  {journal} {Journal of Applied Physics}\ }\textbf {\bibinfo {volume} {120}},\ \bibinfo {pages} {164902} (\bibinfo {year} {2016})},\ \Eprint {http://arxiv.org/abs/https://doi.org/10.1063/1.4966659} {https://doi.org/10.1063/1.4966659} \BibitemShut {NoStop}%
\bibitem [{\citenamefont {Li}\ \emph {et~al.}(2022)\citenamefont {Li}, \citenamefont {Xiong}, \citenamefont {Yang}, \citenamefont {Zhang}, \citenamefont {He}, \citenamefont {Wang},\ and\ \citenamefont {Mao}}]{Li2022}%
  \BibitemOpen
  \bibfield  {author} {\bibinfo {author} {\bibfnamefont {W.}~\bibnamefont {Li}}, \bibinfo {author} {\bibfnamefont {K.}~\bibnamefont {Xiong}}, \bibinfo {author} {\bibfnamefont {L.}~\bibnamefont {Yang}}, \bibinfo {author} {\bibfnamefont {S.}~\bibnamefont {Zhang}}, \bibinfo {author} {\bibfnamefont {J.}~\bibnamefont {He}}, \bibinfo {author} {\bibfnamefont {Y.}~\bibnamefont {Wang}}, \ and\ \bibinfo {author} {\bibfnamefont {Y.}~\bibnamefont {Mao}},\ }\href {\doibase https://doi.org/10.1016/j.msea.2022.144046} {\bibfield  {journal} {\bibinfo  {journal} {Materials Science and Engineering: A}\ }\textbf {\bibinfo {volume} {856}},\ \bibinfo {pages} {144046} (\bibinfo {year} {2022})}\BibitemShut {NoStop}%
\bibitem [{\citenamefont {Luo}\ \emph {et~al.}(1991)\citenamefont {Luo}, \citenamefont {Jacobson},\ and\ \citenamefont {Shin}}]{Luo1991}%
  \BibitemOpen
  \bibfield  {author} {\bibinfo {author} {\bibfnamefont {A.}~\bibnamefont {Luo}}, \bibinfo {author} {\bibfnamefont {D.}~\bibnamefont {Jacobson}}, \ and\ \bibinfo {author} {\bibfnamefont {K.}~\bibnamefont {Shin}},\ }\href {\doibase https://doi.org/10.1016/0263-4368(91)90028-M} {\bibfield  {journal} {\bibinfo  {journal} {International Journal of Refractory Metals and Hard Materials}\ }\textbf {\bibinfo {volume} {10}},\ \bibinfo {pages} {107} (\bibinfo {year} {1991})}\BibitemShut {NoStop}%
\bibitem [{\citenamefont {Romaner}\ \emph {et~al.}(2010)\citenamefont {Romaner}, \citenamefont {Ambrosch-Draxl},\ and\ \citenamefont {Pippan}}]{Romaner2010}%
  \BibitemOpen
  \bibfield  {author} {\bibinfo {author} {\bibfnamefont {L.}~\bibnamefont {Romaner}}, \bibinfo {author} {\bibfnamefont {C.}~\bibnamefont {Ambrosch-Draxl}}, \ and\ \bibinfo {author} {\bibfnamefont {R.}~\bibnamefont {Pippan}},\ }\href {\doibase 10.1103/PhysRevLett.104.195503} {\bibfield  {journal} {\bibinfo  {journal} {Phys. Rev. Lett.}\ }\textbf {\bibinfo {volume} {104}},\ \bibinfo {pages} {195503} (\bibinfo {year} {2010})}\BibitemShut {NoStop}%
\bibitem [{\citenamefont {Geach}\ and\ \citenamefont {Hughes}(1955)}]{Geach1955}%
  \BibitemOpen
  \bibfield  {author} {\bibinfo {author} {\bibfnamefont {G.~A.}\ \bibnamefont {Geach}}\ and\ \bibinfo {author} {\bibfnamefont {J.~R.}\ \bibnamefont {Hughes}},\ }\href@noop {} {\bibfield  {journal} {\bibinfo  {journal} {Plansee Proceeding}\ ,\ \bibinfo {pages} {245}} (\bibinfo {year} {1955})},\ \bibinfo {note} {pergamon Press}\BibitemShut {NoStop}%
\bibitem [{\citenamefont {Khatamsaz}\ \emph {et~al.}(2022)\citenamefont {Khatamsaz}, \citenamefont {Vela}, \citenamefont {Singh}, \citenamefont {Johnson}, \citenamefont {Allaire},\ and\ \citenamefont {Arr{\'o}yave}}]{Khatamsaz2022}%
  \BibitemOpen
  \bibfield  {author} {\bibinfo {author} {\bibfnamefont {D.}~\bibnamefont {Khatamsaz}}, \bibinfo {author} {\bibfnamefont {B.}~\bibnamefont {Vela}}, \bibinfo {author} {\bibfnamefont {P.}~\bibnamefont {Singh}}, \bibinfo {author} {\bibfnamefont {D.~D.}\ \bibnamefont {Johnson}}, \bibinfo {author} {\bibfnamefont {D.}~\bibnamefont {Allaire}}, \ and\ \bibinfo {author} {\bibfnamefont {R.}~\bibnamefont {Arr{\'o}yave}},\ }\href {\doibase https://doi.org/10.1016/j.actamat.2022.118133} {\bibfield  {journal} {\bibinfo  {journal} {Acta Materialia}\ }\textbf {\bibinfo {volume} {236}},\ \bibinfo {pages} {118133} (\bibinfo {year} {2022})}\BibitemShut {NoStop}%
\bibitem [{\citenamefont {Gao}\ \emph {et~al.}(2023)\citenamefont {Gao}, \citenamefont {Liu}, \citenamefont {Chen}, \citenamefont {Liu}, \citenamefont {Yang},\ and\ \citenamefont {Hao}}]{Gao2023}%
  \BibitemOpen
  \bibfield  {author} {\bibinfo {author} {\bibfnamefont {Q.}~\bibnamefont {Gao}}, \bibinfo {author} {\bibfnamefont {H.}~\bibnamefont {Liu}}, \bibinfo {author} {\bibfnamefont {P.}~\bibnamefont {Chen}}, \bibinfo {author} {\bibfnamefont {X.}~\bibnamefont {Liu}}, \bibinfo {author} {\bibfnamefont {H.}~\bibnamefont {Yang}}, \ and\ \bibinfo {author} {\bibfnamefont {J.}~\bibnamefont {Hao}},\ }\href {\doibase https://doi.org/10.1016/j.optlastec.2023.109220} {\bibfield  {journal} {\bibinfo  {journal} {Optics \& Laser Technology}\ }\textbf {\bibinfo {volume} {161}},\ \bibinfo {pages} {109220} (\bibinfo {year} {2023})}\BibitemShut {NoStop}%
\bibitem [{\citenamefont {Debnath}\ \emph {et~al.}(2020)\citenamefont {Debnath}, \citenamefont {Vinoth}, \citenamefont {Mukherjee},\ and\ \citenamefont {Datta}}]{Debnath2020}%
  \BibitemOpen
  \bibfield  {author} {\bibinfo {author} {\bibfnamefont {B.}~\bibnamefont {Debnath}}, \bibinfo {author} {\bibfnamefont {A.}~\bibnamefont {Vinoth}}, \bibinfo {author} {\bibfnamefont {M.}~\bibnamefont {Mukherjee}}, \ and\ \bibinfo {author} {\bibfnamefont {S.}~\bibnamefont {Datta}},\ }\href {\doibase 10.1088/1757-899X/912/5/052021} {\bibfield  {journal} {\bibinfo  {journal} {IOP Conference Series: Materials Science and Engineering}\ }\textbf {\bibinfo {volume} {912}},\ \bibinfo {pages} {052021} (\bibinfo {year} {2020})}\BibitemShut {NoStop}%
\bibitem [{\citenamefont {Rickman}\ \emph {et~al.}(2019)\citenamefont {Rickman}, \citenamefont {Chan}, \citenamefont {Harmer}, \citenamefont {Smeltzer}, \citenamefont {Marvel}, \citenamefont {Roy},\ and\ \citenamefont {Balasubramanian}}]{Rickman2019}%
  \BibitemOpen
  \bibfield  {author} {\bibinfo {author} {\bibfnamefont {J.~M.}\ \bibnamefont {Rickman}}, \bibinfo {author} {\bibfnamefont {H.~M.}\ \bibnamefont {Chan}}, \bibinfo {author} {\bibfnamefont {M.~P.}\ \bibnamefont {Harmer}}, \bibinfo {author} {\bibfnamefont {J.~A.}\ \bibnamefont {Smeltzer}}, \bibinfo {author} {\bibfnamefont {C.~J.}\ \bibnamefont {Marvel}}, \bibinfo {author} {\bibfnamefont {A.}~\bibnamefont {Roy}}, \ and\ \bibinfo {author} {\bibfnamefont {G.}~\bibnamefont {Balasubramanian}},\ }\href {\doibase 10.1038/s41467-019-10533-1} {\bibfield  {journal} {\bibinfo  {journal} {Nature Communications}\ }\textbf {\bibinfo {volume} {10}},\ \bibinfo {pages} {2618} (\bibinfo {year} {2019})}\BibitemShut {NoStop}%
\bibitem [{\citenamefont {Ferrari}\ \emph {et~al.}(2021)\citenamefont {Ferrari}, \citenamefont {Lysogorskiy},\ and\ \citenamefont {Drautz}}]{Ferrari2021}%
  \BibitemOpen
  \bibfield  {author} {\bibinfo {author} {\bibfnamefont {A.}~\bibnamefont {Ferrari}}, \bibinfo {author} {\bibfnamefont {Y.}~\bibnamefont {Lysogorskiy}}, \ and\ \bibinfo {author} {\bibfnamefont {R.}~\bibnamefont {Drautz}},\ }\href {\doibase 10.1103/PhysRevMaterials.5.063606} {\bibfield  {journal} {\bibinfo  {journal} {Phys. Rev. Mater.}\ }\textbf {\bibinfo {volume} {5}},\ \bibinfo {pages} {063606} (\bibinfo {year} {2021})}\BibitemShut {NoStop}%
\bibitem [{\citenamefont {Ouyang}\ \emph {et~al.}(2023)\citenamefont {Ouyang}, \citenamefont {Singh}, \citenamefont {Su}, \citenamefont {Johnson}, \citenamefont {Kramer}, \citenamefont {Perepezko}, \citenamefont {Senkov}, \citenamefont {Miracle},\ and\ \citenamefont {Cui}}]{Ouyang2023}%
  \BibitemOpen
  \bibfield  {author} {\bibinfo {author} {\bibfnamefont {G.}~\bibnamefont {Ouyang}}, \bibinfo {author} {\bibfnamefont {P.}~\bibnamefont {Singh}}, \bibinfo {author} {\bibfnamefont {R.}~\bibnamefont {Su}}, \bibinfo {author} {\bibfnamefont {D.~D.}\ \bibnamefont {Johnson}}, \bibinfo {author} {\bibfnamefont {M.~J.}\ \bibnamefont {Kramer}}, \bibinfo {author} {\bibfnamefont {J.~H.}\ \bibnamefont {Perepezko}}, \bibinfo {author} {\bibfnamefont {O.~N.}\ \bibnamefont {Senkov}}, \bibinfo {author} {\bibfnamefont {D.}~\bibnamefont {Miracle}}, \ and\ \bibinfo {author} {\bibfnamefont {J.}~\bibnamefont {Cui}},\ }\href {\doibase 10.1038/s41524-023-01095-4} {\bibfield  {journal} {\bibinfo  {journal} {npj Computational Materials}\ }\textbf {\bibinfo {volume} {9}},\ \bibinfo {pages} {141} (\bibinfo {year} {2023})}\BibitemShut {NoStop}%
\bibitem [{\citenamefont {Roy}\ \emph {et~al.}(2023)\citenamefont {Roy}, \citenamefont {Hussain}, \citenamefont {Sharma}, \citenamefont {Balasubramanian}, \citenamefont {Taufique}, \citenamefont {Devanathan}, \citenamefont {Singh},\ and\ \citenamefont {Johnson}}]{Roy2023}%
  \BibitemOpen
  \bibfield  {author} {\bibinfo {author} {\bibfnamefont {A.}~\bibnamefont {Roy}}, \bibinfo {author} {\bibfnamefont {A.}~\bibnamefont {Hussain}}, \bibinfo {author} {\bibfnamefont {P.}~\bibnamefont {Sharma}}, \bibinfo {author} {\bibfnamefont {G.}~\bibnamefont {Balasubramanian}}, \bibinfo {author} {\bibfnamefont {M.}~\bibnamefont {Taufique}}, \bibinfo {author} {\bibfnamefont {R.}~\bibnamefont {Devanathan}}, \bibinfo {author} {\bibfnamefont {P.}~\bibnamefont {Singh}}, \ and\ \bibinfo {author} {\bibfnamefont {D.~D.}\ \bibnamefont {Johnson}},\ }\href {\doibase https://doi.org/10.1016/j.actamat.2023.119177} {\bibfield  {journal} {\bibinfo  {journal} {Acta Materialia}\ }\textbf {\bibinfo {volume} {257}},\ \bibinfo {pages} {119177} (\bibinfo {year} {2023})}\BibitemShut {NoStop}%
\bibitem [{\citenamefont {Singh}\ \emph {et~al.}(2023)\citenamefont {Singh}, \citenamefont {Vela}, \citenamefont {Ouyang}, \citenamefont {Argibay}, \citenamefont {Cui}, \citenamefont {Arr{\'o}yave},\ and\ \citenamefont {Johnson}}]{Singh2023}%
  \BibitemOpen
  \bibfield  {author} {\bibinfo {author} {\bibfnamefont {P.}~\bibnamefont {Singh}}, \bibinfo {author} {\bibfnamefont {B.}~\bibnamefont {Vela}}, \bibinfo {author} {\bibfnamefont {G.}~\bibnamefont {Ouyang}}, \bibinfo {author} {\bibfnamefont {N.}~\bibnamefont {Argibay}}, \bibinfo {author} {\bibfnamefont {J.}~\bibnamefont {Cui}}, \bibinfo {author} {\bibfnamefont {R.}~\bibnamefont {Arr{\'o}yave}}, \ and\ \bibinfo {author} {\bibfnamefont {D.~D.}\ \bibnamefont {Johnson}},\ }\href {\doibase https://doi.org/10.1016/j.actamat.2023.119104} {\bibfield  {journal} {\bibinfo  {journal} {Acta Materialia}\ }\textbf {\bibinfo {volume} {257}},\ \bibinfo {pages} {119104} (\bibinfo {year} {2023})}\BibitemShut {NoStop}%
\bibitem [{\citenamefont {Khatamsaz}\ \emph {et~al.}(2023{\natexlab{a}})\citenamefont {Khatamsaz}, \citenamefont {Vela}, \citenamefont {Singh}, \citenamefont {Johnson}, \citenamefont {Allaire},\ and\ \citenamefont {Arr{\'o}yave}}]{Khatamsaz2023a}%
  \BibitemOpen
  \bibfield  {author} {\bibinfo {author} {\bibfnamefont {D.}~\bibnamefont {Khatamsaz}}, \bibinfo {author} {\bibfnamefont {B.}~\bibnamefont {Vela}}, \bibinfo {author} {\bibfnamefont {P.}~\bibnamefont {Singh}}, \bibinfo {author} {\bibfnamefont {D.~D.}\ \bibnamefont {Johnson}}, \bibinfo {author} {\bibfnamefont {D.}~\bibnamefont {Allaire}}, \ and\ \bibinfo {author} {\bibfnamefont {R.}~\bibnamefont {Arr{\'o}yave}},\ }\href {\doibase 10.1038/s41524-023-01006-7} {\bibfield  {journal} {\bibinfo  {journal} {npj Computational Materials}\ }\textbf {\bibinfo {volume} {9}},\ \bibinfo {pages} {49} (\bibinfo {year} {2023}{\natexlab{a}})}\BibitemShut {NoStop}%
\bibitem [{\citenamefont {Khatamsaz}\ \emph {et~al.}(2023{\natexlab{b}})\citenamefont {Khatamsaz}, \citenamefont {Vela},\ and\ \citenamefont {Arr{\'o}yave}}]{Khatamsaz2023b}%
  \BibitemOpen
  \bibfield  {author} {\bibinfo {author} {\bibfnamefont {D.}~\bibnamefont {Khatamsaz}}, \bibinfo {author} {\bibfnamefont {B.}~\bibnamefont {Vela}}, \ and\ \bibinfo {author} {\bibfnamefont {R.}~\bibnamefont {Arr{\'o}yave}},\ }\href {\doibase https://doi.org/10.1016/j.matlet.2023.135067} {\bibfield  {journal} {\bibinfo  {journal} {Materials Letters}\ }\textbf {\bibinfo {volume} {351}},\ \bibinfo {pages} {135067} (\bibinfo {year} {2023}{\natexlab{b}})}\BibitemShut {NoStop}%
\bibitem [{\citenamefont {Solomou}\ \emph {et~al.}(2018)\citenamefont {Solomou}, \citenamefont {Zhao}, \citenamefont {Boluki}, \citenamefont {Joy}, \citenamefont {Qian}, \citenamefont {Karaman}, \citenamefont {Arr{\'o}yave},\ and\ \citenamefont {Lagoudas}}]{Solomou2018}%
  \BibitemOpen
  \bibfield  {author} {\bibinfo {author} {\bibfnamefont {A.}~\bibnamefont {Solomou}}, \bibinfo {author} {\bibfnamefont {G.}~\bibnamefont {Zhao}}, \bibinfo {author} {\bibfnamefont {S.}~\bibnamefont {Boluki}}, \bibinfo {author} {\bibfnamefont {J.~K.}\ \bibnamefont {Joy}}, \bibinfo {author} {\bibfnamefont {X.}~\bibnamefont {Qian}}, \bibinfo {author} {\bibfnamefont {I.}~\bibnamefont {Karaman}}, \bibinfo {author} {\bibfnamefont {R.}~\bibnamefont {Arr{\'o}yave}}, \ and\ \bibinfo {author} {\bibfnamefont {D.~C.}\ \bibnamefont {Lagoudas}},\ }\href {\doibase https://doi.org/10.1016/j.matdes.2018.10.014} {\bibfield  {journal} {\bibinfo  {journal} {Materials \& Design}\ }\textbf {\bibinfo {volume} {160}},\ \bibinfo {pages} {810} (\bibinfo {year} {2018})}\BibitemShut {NoStop}%
\bibitem [{\citenamefont {Hu}\ \emph {et~al.}(2021)\citenamefont {Hu}, \citenamefont {Sundar}, \citenamefont {Ogata},\ and\ \citenamefont {Qi}}]{Hu2021}%
  \BibitemOpen
  \bibfield  {author} {\bibinfo {author} {\bibfnamefont {Y.-J.}\ \bibnamefont {Hu}}, \bibinfo {author} {\bibfnamefont {A.}~\bibnamefont {Sundar}}, \bibinfo {author} {\bibfnamefont {S.}~\bibnamefont {Ogata}}, \ and\ \bibinfo {author} {\bibfnamefont {L.}~\bibnamefont {Qi}},\ }\href {\doibase https://doi.org/10.1016/j.actamat.2021.116800} {\bibfield  {journal} {\bibinfo  {journal} {Acta Materialia}\ }\textbf {\bibinfo {volume} {210}},\ \bibinfo {pages} {116800} (\bibinfo {year} {2021})}\BibitemShut {NoStop}%
\bibitem [{\citenamefont {Hart}\ \emph {et~al.}(2021)\citenamefont {Hart}, \citenamefont {Mueller}, \citenamefont {Toher},\ and\ \citenamefont {Curtarolo}}]{Hart2021}%
  \BibitemOpen
  \bibfield  {author} {\bibinfo {author} {\bibfnamefont {G.~L.~W.}\ \bibnamefont {Hart}}, \bibinfo {author} {\bibfnamefont {T.}~\bibnamefont {Mueller}}, \bibinfo {author} {\bibfnamefont {C.}~\bibnamefont {Toher}}, \ and\ \bibinfo {author} {\bibfnamefont {S.}~\bibnamefont {Curtarolo}},\ }\href {\doibase 10.1038/s41578-021-00340-w} {\bibfield  {journal} {\bibinfo  {journal} {Nature Reviews Materials}\ }\textbf {\bibinfo {volume} {6}},\ \bibinfo {pages} {730} (\bibinfo {year} {2021})}\BibitemShut {NoStop}%
\bibitem [{\citenamefont {Zhang}\ \emph {et~al.}(2020)\citenamefont {Zhang}, \citenamefont {Liu}, \citenamefont {Bi}, \citenamefont {Yin}, \citenamefont {Zhang},\ and\ \citenamefont {Eisenbach}}]{Zhang2020}%
  \BibitemOpen
  \bibfield  {author} {\bibinfo {author} {\bibfnamefont {J.}~\bibnamefont {Zhang}}, \bibinfo {author} {\bibfnamefont {X.}~\bibnamefont {Liu}}, \bibinfo {author} {\bibfnamefont {S.}~\bibnamefont {Bi}}, \bibinfo {author} {\bibfnamefont {J.}~\bibnamefont {Yin}}, \bibinfo {author} {\bibfnamefont {G.}~\bibnamefont {Zhang}}, \ and\ \bibinfo {author} {\bibfnamefont {M.}~\bibnamefont {Eisenbach}},\ }\href {\doibase https://doi.org/10.1016/j.matdes.2019.108247} {\bibfield  {journal} {\bibinfo  {journal} {Materials {\&} Design}\ }\textbf {\bibinfo {volume} {185}},\ \bibinfo {pages} {108247} (\bibinfo {year} {2020})}\BibitemShut {NoStop}%
\bibitem [{\citenamefont {Shaikh}\ \emph {et~al.}(2020)\citenamefont {Shaikh}, \citenamefont {Hariharan}, \citenamefont {Yadav},\ and\ \citenamefont {Murty}}]{Shaikh2020}%
  \BibitemOpen
  \bibfield  {author} {\bibinfo {author} {\bibfnamefont {S.~M.}\ \bibnamefont {Shaikh}}, \bibinfo {author} {\bibfnamefont {V.}~\bibnamefont {Hariharan}}, \bibinfo {author} {\bibfnamefont {S.~K.}\ \bibnamefont {Yadav}}, \ and\ \bibinfo {author} {\bibfnamefont {B.}~\bibnamefont {Murty}},\ }\href {\doibase https://doi.org/10.1016/j.intermet.2020.106926} {\bibfield  {journal} {\bibinfo  {journal} {Intermetallics}\ }\textbf {\bibinfo {volume} {127}},\ \bibinfo {pages} {106926} (\bibinfo {year} {2020})}\BibitemShut {NoStop}%
\bibitem [{\citenamefont {Gao}\ \emph {et~al.}(2017)\citenamefont {Gao}, \citenamefont {Gao}, \citenamefont {Hawk}, \citenamefont {Ouyang}, \citenamefont {Alman},\ and\ \citenamefont {Widom}}]{Gao2017}%
  \BibitemOpen
  \bibfield  {author} {\bibinfo {author} {\bibfnamefont {M.~C.}\ \bibnamefont {Gao}}, \bibinfo {author} {\bibfnamefont {P.}~\bibnamefont {Gao}}, \bibinfo {author} {\bibfnamefont {J.~A.}\ \bibnamefont {Hawk}}, \bibinfo {author} {\bibfnamefont {L.}~\bibnamefont {Ouyang}}, \bibinfo {author} {\bibfnamefont {D.~E.}\ \bibnamefont {Alman}}, \ and\ \bibinfo {author} {\bibfnamefont {M.}~\bibnamefont {Widom}},\ }\href {\doibase 10.1557/jmr.2017.366} {\bibfield  {journal} {\bibinfo  {journal} {Journal of Materials Research}\ }\textbf {\bibinfo {volume} {32}},\ \bibinfo {pages} {3627–3641} (\bibinfo {year} {2017})}\BibitemShut {NoStop}%
\bibitem [{\citenamefont {Elder}\ \emph {et~al.}(2023)\citenamefont {Elder}, \citenamefont {Berry}, \citenamefont {Bocklund}, \citenamefont {McCall}, \citenamefont {Perron},\ and\ \citenamefont {McKeown}}]{Elder2023a}%
  \BibitemOpen
  \bibfield  {author} {\bibinfo {author} {\bibfnamefont {K.~L.~M.}\ \bibnamefont {Elder}}, \bibinfo {author} {\bibfnamefont {J.}~\bibnamefont {Berry}}, \bibinfo {author} {\bibfnamefont {B.}~\bibnamefont {Bocklund}}, \bibinfo {author} {\bibfnamefont {S.~K.}\ \bibnamefont {McCall}}, \bibinfo {author} {\bibfnamefont {A.}~\bibnamefont {Perron}}, \ and\ \bibinfo {author} {\bibfnamefont {J.~T.}\ \bibnamefont {McKeown}},\ }\href {\doibase 10.1038/s41524-023-01030-7} {\bibfield  {journal} {\bibinfo  {journal} {npj Computational Materials}\ }\textbf {\bibinfo {volume} {9}},\ \bibinfo {pages} {84} (\bibinfo {year} {2023})}\BibitemShut {NoStop}%
\bibitem [{\citenamefont {Singh}\ \emph {et~al.}(2018{\natexlab{a}})\citenamefont {Singh}, \citenamefont {Smirnov},\ and\ \citenamefont {Johnson}}]{Singh2018}%
  \BibitemOpen
  \bibfield  {author} {\bibinfo {author} {\bibfnamefont {P.}~\bibnamefont {Singh}}, \bibinfo {author} {\bibfnamefont {A.~V.}\ \bibnamefont {Smirnov}}, \ and\ \bibinfo {author} {\bibfnamefont {D.~D.}\ \bibnamefont {Johnson}},\ }\href {\doibase 10.1103/PhysRevMaterials.2.055004} {\bibfield  {journal} {\bibinfo  {journal} {Phys. Rev. Mater.}\ }\textbf {\bibinfo {volume} {2}},\ \bibinfo {pages} {055004} (\bibinfo {year} {2018}{\natexlab{a}})}\BibitemShut {NoStop}%
\bibitem [{\citenamefont {Li}\ \emph {et~al.}(2020{\natexlab{a}})\citenamefont {Li}, \citenamefont {Chen}, \citenamefont {Zheng}, \citenamefont {Zuo},\ and\ \citenamefont {Ong}}]{li_complex_2020}%
  \BibitemOpen
  \bibfield  {author} {\bibinfo {author} {\bibfnamefont {X.-G.}\ \bibnamefont {Li}}, \bibinfo {author} {\bibfnamefont {C.}~\bibnamefont {Chen}}, \bibinfo {author} {\bibfnamefont {H.}~\bibnamefont {Zheng}}, \bibinfo {author} {\bibfnamefont {Y.}~\bibnamefont {Zuo}}, \ and\ \bibinfo {author} {\bibfnamefont {S.~P.}\ \bibnamefont {Ong}},\ }\href {\doibase 10.1038/s41524-020-0339-0} {\bibfield  {journal} {\bibinfo  {journal} {npj Computational Materials}\ }\textbf {\bibinfo {volume} {6}},\ \bibinfo {pages} {70} (\bibinfo {year} {2020}{\natexlab{a}})}\BibitemShut {NoStop}%
\bibitem [{\citenamefont {Song}\ \emph {et~al.}(2023)\citenamefont {Song}, \citenamefont {Zhao}, \citenamefont {Liu}, \citenamefont {Wang}, \citenamefont {Lindgren}, \citenamefont {Wang}, \citenamefont {Chen}, \citenamefont {Xu}, \citenamefont {Liang}, \citenamefont {Ying}, \citenamefont {Xu}, \citenamefont {Zhao}, \citenamefont {Shi}, \citenamefont {Wang}, \citenamefont {Lyu}, \citenamefont {Zeng}, \citenamefont {Liang}, \citenamefont {Dong}, \citenamefont {Sun}, \citenamefont {Chen}, \citenamefont {Zhang}, \citenamefont {Guo}, \citenamefont {Qian}, \citenamefont {Sun}, \citenamefont {Erhart}, \citenamefont {Ala-Nissila}, \citenamefont {Su},\ and\ \citenamefont {Fan}}]{Song2023}%
  \BibitemOpen
  \bibfield  {author} {\bibinfo {author} {\bibfnamefont {K.}~\bibnamefont {Song}}, \bibinfo {author} {\bibfnamefont {R.}~\bibnamefont {Zhao}}, \bibinfo {author} {\bibfnamefont {J.}~\bibnamefont {Liu}}, \bibinfo {author} {\bibfnamefont {Y.}~\bibnamefont {Wang}}, \bibinfo {author} {\bibfnamefont {E.}~\bibnamefont {Lindgren}}, \bibinfo {author} {\bibfnamefont {Y.}~\bibnamefont {Wang}}, \bibinfo {author} {\bibfnamefont {S.}~\bibnamefont {Chen}}, \bibinfo {author} {\bibfnamefont {K.}~\bibnamefont {Xu}}, \bibinfo {author} {\bibfnamefont {T.}~\bibnamefont {Liang}}, \bibinfo {author} {\bibfnamefont {P.}~\bibnamefont {Ying}}, \bibinfo {author} {\bibfnamefont {N.}~\bibnamefont {Xu}}, \bibinfo {author} {\bibfnamefont {Z.}~\bibnamefont {Zhao}}, \bibinfo {author} {\bibfnamefont {J.}~\bibnamefont {Shi}}, \bibinfo {author} {\bibfnamefont {J.}~\bibnamefont {Wang}}, \bibinfo {author} {\bibfnamefont {S.}~\bibnamefont {Lyu}}, \bibinfo {author} {\bibfnamefont {Z.}~\bibnamefont {Zeng}}, \bibinfo {author} {\bibfnamefont
  {S.}~\bibnamefont {Liang}}, \bibinfo {author} {\bibfnamefont {H.}~\bibnamefont {Dong}}, \bibinfo {author} {\bibfnamefont {L.}~\bibnamefont {Sun}}, \bibinfo {author} {\bibfnamefont {Y.}~\bibnamefont {Chen}}, \bibinfo {author} {\bibfnamefont {Z.}~\bibnamefont {Zhang}}, \bibinfo {author} {\bibfnamefont {W.}~\bibnamefont {Guo}}, \bibinfo {author} {\bibfnamefont {P.}~\bibnamefont {Qian}}, \bibinfo {author} {\bibfnamefont {J.}~\bibnamefont {Sun}}, \bibinfo {author} {\bibfnamefont {P.}~\bibnamefont {Erhart}}, \bibinfo {author} {\bibfnamefont {T.}~\bibnamefont {Ala-Nissila}}, \bibinfo {author} {\bibfnamefont {Y.}~\bibnamefont {Su}}, \ and\ \bibinfo {author} {\bibfnamefont {Z.}~\bibnamefont {Fan}},\ }\href@noop {} {\enquote {\bibinfo {title} {General-purpose machine-learned potential for 16 elemental metals and their alloys},}\ } (\bibinfo {year} {2023}),\ \Eprint {http://arxiv.org/abs/2311.04732} {arXiv:2311.04732 [cond-mat.mtrl-sci]} \BibitemShut {NoStop}%
\bibitem [{\citenamefont {Maresca}\ and\ \citenamefont {Curtin}(2020{\natexlab{a}})}]{Maresca2020_a}%
  \BibitemOpen
  \bibfield  {author} {\bibinfo {author} {\bibfnamefont {F.}~\bibnamefont {Maresca}}\ and\ \bibinfo {author} {\bibfnamefont {W.~A.}\ \bibnamefont {Curtin}},\ }\href {\doibase https://doi.org/10.1016/j.actamat.2019.10.007} {\bibfield  {journal} {\bibinfo  {journal} {Acta Materialia}\ }\textbf {\bibinfo {volume} {182}},\ \bibinfo {pages} {144} (\bibinfo {year} {2020}{\natexlab{a}})}\BibitemShut {NoStop}%
\bibitem [{\citenamefont {Maresca}\ and\ \citenamefont {Curtin}(2020{\natexlab{b}})}]{Maresca2020_b}%
  \BibitemOpen
  \bibfield  {author} {\bibinfo {author} {\bibfnamefont {F.}~\bibnamefont {Maresca}}\ and\ \bibinfo {author} {\bibfnamefont {W.~A.}\ \bibnamefont {Curtin}},\ }\href {\doibase https://doi.org/10.1016/j.actamat.2019.10.015} {\bibfield  {journal} {\bibinfo  {journal} {Acta Materialia}\ }\textbf {\bibinfo {volume} {182}},\ \bibinfo {pages} {235} (\bibinfo {year} {2020}{\natexlab{b}})}\BibitemShut {NoStop}%
\bibitem [{\citenamefont {Lee}\ \emph {et~al.}(2021)\citenamefont {Lee}, \citenamefont {Maresca}, \citenamefont {Feng}, \citenamefont {Chou}, \citenamefont {Ungar}, \citenamefont {Widom}, \citenamefont {An}, \citenamefont {Poplawsky}, \citenamefont {Chou}, \citenamefont {Liaw},\ and\ \citenamefont {Curtin}}]{Lee2021}%
  \BibitemOpen
  \bibfield  {author} {\bibinfo {author} {\bibfnamefont {C.}~\bibnamefont {Lee}}, \bibinfo {author} {\bibfnamefont {F.}~\bibnamefont {Maresca}}, \bibinfo {author} {\bibfnamefont {R.}~\bibnamefont {Feng}}, \bibinfo {author} {\bibfnamefont {Y.}~\bibnamefont {Chou}}, \bibinfo {author} {\bibfnamefont {T.}~\bibnamefont {Ungar}}, \bibinfo {author} {\bibfnamefont {M.}~\bibnamefont {Widom}}, \bibinfo {author} {\bibfnamefont {K.}~\bibnamefont {An}}, \bibinfo {author} {\bibfnamefont {J.~D.}\ \bibnamefont {Poplawsky}}, \bibinfo {author} {\bibfnamefont {Y.-C.}\ \bibnamefont {Chou}}, \bibinfo {author} {\bibfnamefont {P.~K.}\ \bibnamefont {Liaw}}, \ and\ \bibinfo {author} {\bibfnamefont {W.~A.}\ \bibnamefont {Curtin}},\ }\href {\doibase 10.1038/s41467-021-25807-w} {\bibfield  {journal} {\bibinfo  {journal} {Nature Communications}\ }\textbf {\bibinfo {volume} {12}},\ \bibinfo {pages} {5474} (\bibinfo {year} {2021})}\BibitemShut {NoStop}%
\bibitem [{\citenamefont {Novikov}\ \emph {et~al.}(2022)\citenamefont {Novikov}, \citenamefont {Kovalyova}, \citenamefont {Shapeev},\ and\ \citenamefont {Hodapp}}]{Novikov2022}%
  \BibitemOpen
  \bibfield  {author} {\bibinfo {author} {\bibfnamefont {I.}~\bibnamefont {Novikov}}, \bibinfo {author} {\bibfnamefont {O.}~\bibnamefont {Kovalyova}}, \bibinfo {author} {\bibfnamefont {A.}~\bibnamefont {Shapeev}}, \ and\ \bibinfo {author} {\bibfnamefont {M.}~\bibnamefont {Hodapp}},\ }\href {\doibase 10.1557/s43578-022-00783-z} {\bibfield  {journal} {\bibinfo  {journal} {Journal of Materials Research}\ }\textbf {\bibinfo {volume} {37}},\ \bibinfo {pages} {3491} (\bibinfo {year} {2022})}\BibitemShut {NoStop}%
\bibitem [{\citenamefont {Rice}\ and\ \citenamefont {Thomson}(1974)}]{Rice1974}%
  \BibitemOpen
  \bibfield  {author} {\bibinfo {author} {\bibfnamefont {J.~R.}\ \bibnamefont {Rice}}\ and\ \bibinfo {author} {\bibfnamefont {R.}~\bibnamefont {Thomson}},\ }\href {\doibase 10.1080/14786437408213555} {\bibfield  {journal} {\bibinfo  {journal} {The Philosophical Magazine: A Journal of Theoretical Experimental and Applied Physics}\ }\textbf {\bibinfo {volume} {29}},\ \bibinfo {pages} {73} (\bibinfo {year} {1974})},\ \Eprint {http://arxiv.org/abs/https://doi.org/10.1080/14786437408213555} {https://doi.org/10.1080/14786437408213555} \BibitemShut {NoStop}%
\bibitem [{\citenamefont {Mak}\ \emph {et~al.}(2021)\citenamefont {Mak}, \citenamefont {Yin},\ and\ \citenamefont {Curtin}}]{Mak2021}%
  \BibitemOpen
  \bibfield  {author} {\bibinfo {author} {\bibfnamefont {E.}~\bibnamefont {Mak}}, \bibinfo {author} {\bibfnamefont {B.}~\bibnamefont {Yin}}, \ and\ \bibinfo {author} {\bibfnamefont {W.}~\bibnamefont {Curtin}},\ }\href {\doibase https://doi.org/10.1016/j.jmps.2021.104389} {\bibfield  {journal} {\bibinfo  {journal} {Journal of the Mechanics and Physics of Solids}\ }\textbf {\bibinfo {volume} {152}},\ \bibinfo {pages} {104389} (\bibinfo {year} {2021})}\BibitemShut {NoStop}%
\bibitem [{\citenamefont {Li}\ \emph {et~al.}(2020{\natexlab{b}})\citenamefont {Li}, \citenamefont {Li}, \citenamefont {Irving}, \citenamefont {Varga}, \citenamefont {Vitos},\ and\ \citenamefont {Sch\"onecker}}]{Li2020_a}%
  \BibitemOpen
  \bibfield  {author} {\bibinfo {author} {\bibfnamefont {X.}~\bibnamefont {Li}}, \bibinfo {author} {\bibfnamefont {W.}~\bibnamefont {Li}}, \bibinfo {author} {\bibfnamefont {D.~L.}\ \bibnamefont {Irving}}, \bibinfo {author} {\bibfnamefont {L.~K.}\ \bibnamefont {Varga}}, \bibinfo {author} {\bibfnamefont {L.}~\bibnamefont {Vitos}}, \ and\ \bibinfo {author} {\bibfnamefont {S.}~\bibnamefont {Sch\"onecker}},\ }\href {\doibase https://doi.org/10.1016/j.actamat.2020.03.004} {\bibfield  {journal} {\bibinfo  {journal} {Acta Materialia}\ }\textbf {\bibinfo {volume} {189}},\ \bibinfo {pages} {174} (\bibinfo {year} {2020}{\natexlab{b}})}\BibitemShut {NoStop}%
\bibitem [{\citenamefont {van~de Walle}(2009)}]{avdw:atat2}%
  \BibitemOpen
  \bibfield  {author} {\bibinfo {author} {\bibfnamefont {A.}~\bibnamefont {van~de Walle}},\ }\href {\doibase 10.1016/j.calphad.2008.12.005} {\bibfield  {journal} {\bibinfo  {journal} {Calphad}\ }\textbf {\bibinfo {volume} {33}},\ \bibinfo {pages} {266} (\bibinfo {year} {2009})}\BibitemShut {NoStop}%
\bibitem [{\citenamefont {van~de Walle}\ and\ \citenamefont {Ceder}(2002)}]{avdw:maps}%
  \BibitemOpen
  \bibfield  {author} {\bibinfo {author} {\bibfnamefont {A.}~\bibnamefont {van~de Walle}}\ and\ \bibinfo {author} {\bibfnamefont {G.}~\bibnamefont {Ceder}},\ }\href {\doibase 10.1361/105497102770331596} {\bibfield  {journal} {\bibinfo  {journal} {J. Phase Equilib.}\ }\textbf {\bibinfo {volume} {23}},\ \bibinfo {pages} {348} (\bibinfo {year} {2002})}\BibitemShut {NoStop}%
\bibitem [{\citenamefont {Singh}\ \emph {et~al.}(2021)\citenamefont {Singh}, \citenamefont {Sharma}, \citenamefont {Singh}, \citenamefont {Balasubramanian},\ and\ \citenamefont {Johnson}}]{Singh2021}%
  \BibitemOpen
  \bibfield  {author} {\bibinfo {author} {\bibfnamefont {R.}~\bibnamefont {Singh}}, \bibinfo {author} {\bibfnamefont {A.}~\bibnamefont {Sharma}}, \bibinfo {author} {\bibfnamefont {P.}~\bibnamefont {Singh}}, \bibinfo {author} {\bibfnamefont {G.}~\bibnamefont {Balasubramanian}}, \ and\ \bibinfo {author} {\bibfnamefont {D.~D.}\ \bibnamefont {Johnson}},\ }\href {\doibase 10.1038/s43588-020-00006-7} {\bibfield  {journal} {\bibinfo  {journal} {Nature Computational Science}\ }\textbf {\bibinfo {volume} {1}},\ \bibinfo {pages} {54} (\bibinfo {year} {2021})}\BibitemShut {NoStop}%
\bibitem [{\citenamefont {Tasn\'adi}\ \emph {et~al.}(2015)\citenamefont {Tasn\'adi}, \citenamefont {Wang}, \citenamefont {Od\'en},\ and\ \citenamefont {Abrikosov}}]{Tasnadi2015}%
  \BibitemOpen
  \bibfield  {author} {\bibinfo {author} {\bibfnamefont {F.}~\bibnamefont {Tasn\'adi}}, \bibinfo {author} {\bibfnamefont {F.}~\bibnamefont {Wang}}, \bibinfo {author} {\bibfnamefont {M.}~\bibnamefont {Od\'en}}, \ and\ \bibinfo {author} {\bibfnamefont {I.~A.}\ \bibnamefont {Abrikosov}},\ }\href {\doibase https://doi.org/10.1016/j.commatsci.2015.03.030} {\bibfield  {journal} {\bibinfo  {journal} {Computational Materials Science}\ }\textbf {\bibinfo {volume} {103}},\ \bibinfo {pages} {194} (\bibinfo {year} {2015})}\BibitemShut {NoStop}%
\bibitem [{\citenamefont {Holec}\ \emph {et~al.}(2014)\citenamefont {Holec}, \citenamefont {Tasn{\'a}di}, \citenamefont {Wagner}, \citenamefont {Fri{\'a}k}, \citenamefont {Neugebauer}, \citenamefont {Mayrhofer},\ and\ \citenamefont {Keckes}}]{Holec2014}%
  \BibitemOpen
  \bibfield  {author} {\bibinfo {author} {\bibfnamefont {D.}~\bibnamefont {Holec}}, \bibinfo {author} {\bibfnamefont {F.}~\bibnamefont {Tasn{\'a}di}}, \bibinfo {author} {\bibfnamefont {P.}~\bibnamefont {Wagner}}, \bibinfo {author} {\bibfnamefont {M.}~\bibnamefont {Fri{\'a}k}}, \bibinfo {author} {\bibfnamefont {J.}~\bibnamefont {Neugebauer}}, \bibinfo {author} {\bibfnamefont {P.~H.}\ \bibnamefont {Mayrhofer}}, \ and\ \bibinfo {author} {\bibfnamefont {J.}~\bibnamefont {Keckes}},\ }\href {\doibase 10.1103/PhysRevB.90.184106} {\bibfield  {journal} {\bibinfo  {journal} {Phys. Rev. B}\ }\textbf {\bibinfo {volume} {90}},\ \bibinfo {pages} {184106} (\bibinfo {year} {2014})}\BibitemShut {NoStop}%
\bibitem [{\citenamefont {Soven}(1967)}]{Soven1967}%
  \BibitemOpen
  \bibfield  {author} {\bibinfo {author} {\bibfnamefont {P.}~\bibnamefont {Soven}},\ }\href {\doibase 10.1103/PhysRev.156.809} {\bibfield  {journal} {\bibinfo  {journal} {Physical Review}\ }\textbf {\bibinfo {volume} {156}},\ \bibinfo {pages} {809} (\bibinfo {year} {1967})}\BibitemShut {NoStop}%
\bibitem [{\citenamefont {Velick{\'{y}}}\ \emph {et~al.}(1968)\citenamefont {Velick{\'{y}}}, \citenamefont {Kirkpatrick},\ and\ \citenamefont {Ehrenreich}}]{Velicky1968}%
  \BibitemOpen
  \bibfield  {author} {\bibinfo {author} {\bibfnamefont {B.}~\bibnamefont {Velick{\'{y}}}}, \bibinfo {author} {\bibfnamefont {S.}~\bibnamefont {Kirkpatrick}}, \ and\ \bibinfo {author} {\bibfnamefont {H.}~\bibnamefont {Ehrenreich}},\ }\href {\doibase 10.1103/PhysRev.175.747} {\bibfield  {journal} {\bibinfo  {journal} {Physical Review}\ }\textbf {\bibinfo {volume} {175}},\ \bibinfo {pages} {747} (\bibinfo {year} {1968})}\BibitemShut {NoStop}%
\bibitem [{\citenamefont {Behler}\ and\ \citenamefont {Parrinello}(2007)}]{behler_generalized_2007}%
  \BibitemOpen
  \bibfield  {author} {\bibinfo {author} {\bibfnamefont {J.}~\bibnamefont {Behler}}\ and\ \bibinfo {author} {\bibfnamefont {M.}~\bibnamefont {Parrinello}},\ }\href {\doibase 10.1103/PhysRevLett.98.146401} {\bibfield  {journal} {\bibinfo  {journal} {Phys. Rev. Lett.}\ }\textbf {\bibinfo {volume} {98}},\ \bibinfo {pages} {146401} (\bibinfo {year} {2007})}\BibitemShut {NoStop}%
\bibitem [{\citenamefont {Bartok}\ and\ \citenamefont {Csanyi}(2015)}]{Bartok2015}%
  \BibitemOpen
  \bibfield  {author} {\bibinfo {author} {\bibfnamefont {A.~P.}\ \bibnamefont {Bartok}}\ and\ \bibinfo {author} {\bibfnamefont {G.}~\bibnamefont {Csanyi}},\ }\href {\doibase https://doi.org/10.1002/qua.24927} {\bibfield  {journal} {\bibinfo  {journal} {International Journal of Quantum Chemistry}\ }\textbf {\bibinfo {volume} {115}},\ \bibinfo {pages} {1051} (\bibinfo {year} {2015})},\ \Eprint {http://arxiv.org/abs/https://onlinelibrary.wiley.com/doi/pdf/10.1002/qua.24927} {https://onlinelibrary.wiley.com/doi/pdf/10.1002/qua.24927} \BibitemShut {NoStop}%
\bibitem [{\citenamefont {Thompson}\ \emph {et~al.}(2015)\citenamefont {Thompson}, \citenamefont {Swiler}, \citenamefont {Trott}, \citenamefont {Foiles},\ and\ \citenamefont {Tucker}}]{thompson_spectral_2015}%
  \BibitemOpen
  \bibfield  {author} {\bibinfo {author} {\bibfnamefont {A.}~\bibnamefont {Thompson}}, \bibinfo {author} {\bibfnamefont {L.}~\bibnamefont {Swiler}}, \bibinfo {author} {\bibfnamefont {C.}~\bibnamefont {Trott}}, \bibinfo {author} {\bibfnamefont {S.}~\bibnamefont {Foiles}}, \ and\ \bibinfo {author} {\bibfnamefont {G.}~\bibnamefont {Tucker}},\ }\href {\doibase 10.1016/j.jcp.2014.12.018} {\bibfield  {journal} {\bibinfo  {journal} {Journal of Computational Physics}\ }\textbf {\bibinfo {volume} {285}},\ \bibinfo {pages} {316} (\bibinfo {year} {2015})}\BibitemShut {NoStop}%
\bibitem [{\citenamefont {Shapeev}(2016)}]{Shapeev2016}%
  \BibitemOpen
  \bibfield  {author} {\bibinfo {author} {\bibfnamefont {A.~V.}\ \bibnamefont {Shapeev}},\ }\href {\doibase 10.1137/15M1054183} {\bibfield  {journal} {\bibinfo  {journal} {Multiscale Modeling \& Simulation}\ }\textbf {\bibinfo {volume} {14}},\ \bibinfo {pages} {1153} (\bibinfo {year} {2016})}\BibitemShut {NoStop}%
\bibitem [{\citenamefont {Byggm\"astar}\ \emph {et~al.}(2021)\citenamefont {Byggm\"astar}, \citenamefont {Nordlund},\ and\ \citenamefont {Djurabekova}}]{Byggmastar2021}%
  \BibitemOpen
  \bibfield  {author} {\bibinfo {author} {\bibfnamefont {J.}~\bibnamefont {Byggm\"astar}}, \bibinfo {author} {\bibfnamefont {K.}~\bibnamefont {Nordlund}}, \ and\ \bibinfo {author} {\bibfnamefont {F.}~\bibnamefont {Djurabekova}},\ }\href {\doibase 10.1103/PhysRevB.104.104101} {\bibfield  {journal} {\bibinfo  {journal} {Phys. Rev. B}\ }\textbf {\bibinfo {volume} {104}},\ \bibinfo {pages} {104101} (\bibinfo {year} {2021})}\BibitemShut {NoStop}%
\bibitem [{\citenamefont {Lopanitsyna}\ \emph {et~al.}(2023)\citenamefont {Lopanitsyna}, \citenamefont {Fraux}, \citenamefont {Springer}, \citenamefont {De},\ and\ \citenamefont {Ceriotti}}]{Lopanitsyna2023}%
  \BibitemOpen
  \bibfield  {author} {\bibinfo {author} {\bibfnamefont {N.}~\bibnamefont {Lopanitsyna}}, \bibinfo {author} {\bibfnamefont {G.}~\bibnamefont {Fraux}}, \bibinfo {author} {\bibfnamefont {M.~A.}\ \bibnamefont {Springer}}, \bibinfo {author} {\bibfnamefont {S.}~\bibnamefont {De}}, \ and\ \bibinfo {author} {\bibfnamefont {M.}~\bibnamefont {Ceriotti}},\ }\href {\doibase 10.1103/PhysRevMaterials.7.045802} {\bibfield  {journal} {\bibinfo  {journal} {Phys. Rev. Mater.}\ }\textbf {\bibinfo {volume} {7}},\ \bibinfo {pages} {045802} (\bibinfo {year} {2023})}\BibitemShut {NoStop}%
\bibitem [{\citenamefont {Hodapp}\ and\ \citenamefont {Shapeev}(2021)}]{Hodapp2021}%
  \BibitemOpen
  \bibfield  {author} {\bibinfo {author} {\bibfnamefont {M.}~\bibnamefont {Hodapp}}\ and\ \bibinfo {author} {\bibfnamefont {A.}~\bibnamefont {Shapeev}},\ }\href {\doibase 10.1103/PhysRevMaterials.5.113802} {\bibfield  {journal} {\bibinfo  {journal} {Phys. Rev. Mater.}\ }\textbf {\bibinfo {volume} {5}},\ \bibinfo {pages} {113802} (\bibinfo {year} {2021})}\BibitemShut {NoStop}%
\bibitem [{\citenamefont {Rice}(1992)}]{Rice1992}%
  \BibitemOpen
  \bibfield  {author} {\bibinfo {author} {\bibfnamefont {J.~R.}\ \bibnamefont {Rice}},\ }\href {\doibase https://doi.org/10.1016/S0022-5096(05)80012-2} {\bibfield  {journal} {\bibinfo  {journal} {Journal of the Mechanics and Physics of Solids}\ }\textbf {\bibinfo {volume} {40}},\ \bibinfo {pages} {239} (\bibinfo {year} {1992})}\BibitemShut {NoStop}%
\bibitem [{\citenamefont {Deb}\ \emph {et~al.}(2002)\citenamefont {Deb}, \citenamefont {Pratap}, \citenamefont {Agarwal},\ and\ \citenamefont {Meyarivan}}]{Deb2022}%
  \BibitemOpen
  \bibfield  {author} {\bibinfo {author} {\bibfnamefont {K.}~\bibnamefont {Deb}}, \bibinfo {author} {\bibfnamefont {A.}~\bibnamefont {Pratap}}, \bibinfo {author} {\bibfnamefont {S.}~\bibnamefont {Agarwal}}, \ and\ \bibinfo {author} {\bibfnamefont {T.}~\bibnamefont {Meyarivan}},\ }\href {\doibase 10.1109/4235.996017} {\bibfield  {journal} {\bibinfo  {journal} {IEEE Transactions on Evolutionary Computation}\ }\textbf {\bibinfo {volume} {6}},\ \bibinfo {pages} {182} (\bibinfo {year} {2002})}\BibitemShut {NoStop}%
\bibitem [{\citenamefont {Moitzi}\ \emph {et~al.}(2022)\citenamefont {Moitzi}, \citenamefont {Romaner}, \citenamefont {Ruban},\ and\ \citenamefont {Peil}}]{Moitzi2022}%
  \BibitemOpen
  \bibfield  {author} {\bibinfo {author} {\bibfnamefont {F.}~\bibnamefont {Moitzi}}, \bibinfo {author} {\bibfnamefont {L.}~\bibnamefont {Romaner}}, \bibinfo {author} {\bibfnamefont {A.~V.}\ \bibnamefont {Ruban}}, \ and\ \bibinfo {author} {\bibfnamefont {O.~E.}\ \bibnamefont {Peil}},\ }\href {\doibase 10.1103/PhysRevMaterials.6.103602} {\bibfield  {journal} {\bibinfo  {journal} {Phys. Rev. Mater.}\ }\textbf {\bibinfo {volume} {6}},\ \bibinfo {pages} {103602} (\bibinfo {year} {2022})}\BibitemShut {NoStop}%
\bibitem [{\citenamefont {Biermair}\ \emph {et~al.}(2023)\citenamefont {Biermair}, \citenamefont {Mendez-Martin}, \citenamefont {Razumovskiy}, \citenamefont {Moitzi},\ and\ \citenamefont {Ressel}}]{Biermair2023}%
  \BibitemOpen
  \bibfield  {author} {\bibinfo {author} {\bibfnamefont {F.}~\bibnamefont {Biermair}}, \bibinfo {author} {\bibfnamefont {F.}~\bibnamefont {Mendez-Martin}}, \bibinfo {author} {\bibfnamefont {V.~I.}\ \bibnamefont {Razumovskiy}}, \bibinfo {author} {\bibfnamefont {F.}~\bibnamefont {Moitzi}}, \ and\ \bibinfo {author} {\bibfnamefont {G.}~\bibnamefont {Ressel}},\ }\href {\doibase 10.3390/ma16072821} {\bibfield  {journal} {\bibinfo  {journal} {Materials}\ }\textbf {\bibinfo {volume} {16}},\ \bibinfo {pages} {2821} (\bibinfo {year} {2023})}\BibitemShut {NoStop}%
\bibitem [{\citenamefont {Zheng}\ \emph {et~al.}(2023)\citenamefont {Zheng}, \citenamefont {Fey}, \citenamefont {Li}, \citenamefont {Hu}, \citenamefont {Qi}, \citenamefont {Chen}, \citenamefont {Xu}, \citenamefont {Beyerlein},\ and\ \citenamefont {Ong}}]{Zheng2022}%
  \BibitemOpen
  \bibfield  {author} {\bibinfo {author} {\bibfnamefont {H.}~\bibnamefont {Zheng}}, \bibinfo {author} {\bibfnamefont {L.~T.~W.}\ \bibnamefont {Fey}}, \bibinfo {author} {\bibfnamefont {X.-G.}\ \bibnamefont {Li}}, \bibinfo {author} {\bibfnamefont {Y.-J.}\ \bibnamefont {Hu}}, \bibinfo {author} {\bibfnamefont {L.}~\bibnamefont {Qi}}, \bibinfo {author} {\bibfnamefont {C.}~\bibnamefont {Chen}}, \bibinfo {author} {\bibfnamefont {S.}~\bibnamefont {Xu}}, \bibinfo {author} {\bibfnamefont {I.~J.}\ \bibnamefont {Beyerlein}}, \ and\ \bibinfo {author} {\bibfnamefont {S.~P.}\ \bibnamefont {Ong}},\ }\href {\doibase 10.1038/s41524-023-01046-z} {\bibfield  {journal} {\bibinfo  {journal} {npj Computational Materials}\ }\textbf {\bibinfo {volume} {9}},\ \bibinfo {pages} {89} (\bibinfo {year} {2023})}\BibitemShut {NoStop}%
\bibitem [{\citenamefont {Xu}\ \emph {et~al.}(2020)\citenamefont {Xu}, \citenamefont {Hwang}, \citenamefont {Jian}, \citenamefont {Su},\ and\ \citenamefont {Beyerlein}}]{Xu2020}%
  \BibitemOpen
  \bibfield  {author} {\bibinfo {author} {\bibfnamefont {S.}~\bibnamefont {Xu}}, \bibinfo {author} {\bibfnamefont {E.}~\bibnamefont {Hwang}}, \bibinfo {author} {\bibfnamefont {W.-R.}\ \bibnamefont {Jian}}, \bibinfo {author} {\bibfnamefont {Y.}~\bibnamefont {Su}}, \ and\ \bibinfo {author} {\bibfnamefont {I.~J.}\ \bibnamefont {Beyerlein}},\ }\href {\doibase https://doi.org/10.1016/j.intermet.2020.106844} {\bibfield  {journal} {\bibinfo  {journal} {Intermetallics}\ }\textbf {\bibinfo {volume} {124}},\ \bibinfo {pages} {106844} (\bibinfo {year} {2020})}\BibitemShut {NoStop}%
\bibitem [{\citenamefont {Fritz~Körmann}\ and\ \citenamefont {Sluiter}(2017)}]{Kormann2017}%
  \BibitemOpen
  \bibfield  {author} {\bibinfo {author} {\bibfnamefont {A.~V.~R.}\ \bibnamefont {Fritz~Körmann}}\ and\ \bibinfo {author} {\bibfnamefont {M.~H.}\ \bibnamefont {Sluiter}},\ }\href {\doibase 10.1080/21663831.2016.1198837} {\bibfield  {journal} {\bibinfo  {journal} {Materials Research Letters}\ }\textbf {\bibinfo {volume} {5}},\ \bibinfo {pages} {35} (\bibinfo {year} {2017})}\BibitemShut {NoStop}%
\bibitem [{\citenamefont {Tr{\'e}glia}\ \emph {et~al.}(1999)\citenamefont {Tr{\'e}glia}, \citenamefont {Legrand}, \citenamefont {Ducastelle}, \citenamefont {Sa{\'u}l}, \citenamefont {Gallis}, \citenamefont {Meunier}, \citenamefont {Mottet},\ and\ \citenamefont {Senhaji}}]{Treglia1999}%
  \BibitemOpen
  \bibfield  {author} {\bibinfo {author} {\bibfnamefont {G.}~\bibnamefont {Tr{\'e}glia}}, \bibinfo {author} {\bibfnamefont {B.}~\bibnamefont {Legrand}}, \bibinfo {author} {\bibfnamefont {F.}~\bibnamefont {Ducastelle}}, \bibinfo {author} {\bibfnamefont {A.}~\bibnamefont {Sa{\'u}l}}, \bibinfo {author} {\bibfnamefont {C.}~\bibnamefont {Gallis}}, \bibinfo {author} {\bibfnamefont {I.}~\bibnamefont {Meunier}}, \bibinfo {author} {\bibfnamefont {C.}~\bibnamefont {Mottet}}, \ and\ \bibinfo {author} {\bibfnamefont {A.}~\bibnamefont {Senhaji}},\ }\href@noop {} {\bibfield  {journal} {\bibinfo  {journal} {Computational materials science}\ }\textbf {\bibinfo {volume} {15}},\ \bibinfo {pages} {196} (\bibinfo {year} {1999})}\BibitemShut {NoStop}%
\bibitem [{\citenamefont {Coury}\ \emph {et~al.}(2019)\citenamefont {Coury}, \citenamefont {Kaufman},\ and\ \citenamefont {Clarke}}]{Coury2019}%
  \BibitemOpen
  \bibfield  {author} {\bibinfo {author} {\bibfnamefont {F.~G.}\ \bibnamefont {Coury}}, \bibinfo {author} {\bibfnamefont {M.}~\bibnamefont {Kaufman}}, \ and\ \bibinfo {author} {\bibfnamefont {A.~J.}\ \bibnamefont {Clarke}},\ }\href {\doibase https://doi.org/10.1016/j.actamat.2019.06.006} {\bibfield  {journal} {\bibinfo  {journal} {Acta Materialia}\ }\textbf {\bibinfo {volume} {175}},\ \bibinfo {pages} {66} (\bibinfo {year} {2019})}\BibitemShut {NoStop}%
\bibitem [{\citenamefont {Senkov}\ \emph {et~al.}(2019{\natexlab{a}})\citenamefont {Senkov}, \citenamefont {Rao}, \citenamefont {Butler},\ and\ \citenamefont {Chaput}}]{Senkov2019c}%
  \BibitemOpen
  \bibfield  {author} {\bibinfo {author} {\bibfnamefont {O.}~\bibnamefont {Senkov}}, \bibinfo {author} {\bibfnamefont {S.}~\bibnamefont {Rao}}, \bibinfo {author} {\bibfnamefont {T.}~\bibnamefont {Butler}}, \ and\ \bibinfo {author} {\bibfnamefont {K.}~\bibnamefont {Chaput}},\ }\href {\doibase https://doi.org/10.1016/j.jallcom.2019.151685} {\bibfield  {journal} {\bibinfo  {journal} {Journal of Alloys and Compounds}\ }\textbf {\bibinfo {volume} {808}},\ \bibinfo {pages} {151685} (\bibinfo {year} {2019}{\natexlab{a}})}\BibitemShut {NoStop}%
\bibitem [{\citenamefont {Barzilai}\ \emph {et~al.}(2017)\citenamefont {Barzilai}, \citenamefont {Toher}, \citenamefont {Curtarolo},\ and\ \citenamefont {Levy}}]{Barzilai2017}%
  \BibitemOpen
  \bibfield  {author} {\bibinfo {author} {\bibfnamefont {S.}~\bibnamefont {Barzilai}}, \bibinfo {author} {\bibfnamefont {C.}~\bibnamefont {Toher}}, \bibinfo {author} {\bibfnamefont {S.}~\bibnamefont {Curtarolo}}, \ and\ \bibinfo {author} {\bibfnamefont {O.}~\bibnamefont {Levy}},\ }\href {\doibase 10.1103/PhysRevMaterials.1.023604} {\bibfield  {journal} {\bibinfo  {journal} {Phys. Rev. Mater.}\ }\textbf {\bibinfo {volume} {1}},\ \bibinfo {pages} {023604} (\bibinfo {year} {2017})}\BibitemShut {NoStop}%
\bibitem [{\citenamefont {B{\"o}nisch}\ \emph {et~al.}(2020)\citenamefont {B{\"o}nisch}, \citenamefont {Stoica},\ and\ \citenamefont {Calin}}]{Boenisch2020}%
  \BibitemOpen
  \bibfield  {author} {\bibinfo {author} {\bibfnamefont {M.}~\bibnamefont {B{\"o}nisch}}, \bibinfo {author} {\bibfnamefont {M.}~\bibnamefont {Stoica}}, \ and\ \bibinfo {author} {\bibfnamefont {M.}~\bibnamefont {Calin}},\ }\href {\doibase 10.1038/s41598-020-60038-x} {\bibfield  {journal} {\bibinfo  {journal} {Scientific Reports}\ }\textbf {\bibinfo {volume} {10}},\ \bibinfo {pages} {3045} (\bibinfo {year} {2020})}\BibitemShut {NoStop}%
\bibitem [{\citenamefont {Barzilai}\ \emph {et~al.}(2016)\citenamefont {Barzilai}, \citenamefont {Toher}, \citenamefont {Curtarolo},\ and\ \citenamefont {Levy}}]{Barzilai2016}%
  \BibitemOpen
  \bibfield  {author} {\bibinfo {author} {\bibfnamefont {S.}~\bibnamefont {Barzilai}}, \bibinfo {author} {\bibfnamefont {C.}~\bibnamefont {Toher}}, \bibinfo {author} {\bibfnamefont {S.}~\bibnamefont {Curtarolo}}, \ and\ \bibinfo {author} {\bibfnamefont {O.}~\bibnamefont {Levy}},\ }\href {\doibase https://doi.org/10.1016/j.actamat.2016.08.053} {\bibfield  {journal} {\bibinfo  {journal} {Acta Materialia}\ }\textbf {\bibinfo {volume} {120}},\ \bibinfo {pages} {255} (\bibinfo {year} {2016})}\BibitemShut {NoStop}%
\bibitem [{\citenamefont {Senkov}\ \emph {et~al.}(2021)\citenamefont {Senkov}, \citenamefont {Miracle},\ and\ \citenamefont {Rao}}]{Senkov20211}%
  \BibitemOpen
  \bibfield  {author} {\bibinfo {author} {\bibfnamefont {O.}~\bibnamefont {Senkov}}, \bibinfo {author} {\bibfnamefont {D.}~\bibnamefont {Miracle}}, \ and\ \bibinfo {author} {\bibfnamefont {S.}~\bibnamefont {Rao}},\ }\href {\doibase https://doi.org/10.1016/j.msea.2021.141512} {\bibfield  {journal} {\bibinfo  {journal} {Materials Science and Engineering: A}\ }\textbf {\bibinfo {volume} {820}},\ \bibinfo {pages} {141512} (\bibinfo {year} {2021})}\BibitemShut {NoStop}%
\bibitem [{\citenamefont {Senkov}\ \emph {et~al.}(2019{\natexlab{b}})\citenamefont {Senkov}, \citenamefont {Gorsse},\ and\ \citenamefont {Miracle}}]{Senkov2019}%
  \BibitemOpen
  \bibfield  {author} {\bibinfo {author} {\bibfnamefont {O.}~\bibnamefont {Senkov}}, \bibinfo {author} {\bibfnamefont {S.}~\bibnamefont {Gorsse}}, \ and\ \bibinfo {author} {\bibfnamefont {D.}~\bibnamefont {Miracle}},\ }\href {\doibase https://doi.org/10.1016/j.actamat.2019.06.032} {\bibfield  {journal} {\bibinfo  {journal} {Acta Materialia}\ }\textbf {\bibinfo {volume} {175}},\ \bibinfo {pages} {394} (\bibinfo {year} {2019}{\natexlab{b}})}\BibitemShut {NoStop}%
\bibitem [{\citenamefont {Rao}\ \emph {et~al.}(2019)\citenamefont {Rao}, \citenamefont {Antillon}, \citenamefont {Woodward}, \citenamefont {Akdim}, \citenamefont {Parthasarathy},\ and\ \citenamefont {Senkov}}]{Rao2019}%
  \BibitemOpen
  \bibfield  {author} {\bibinfo {author} {\bibfnamefont {S.}~\bibnamefont {Rao}}, \bibinfo {author} {\bibfnamefont {E.}~\bibnamefont {Antillon}}, \bibinfo {author} {\bibfnamefont {C.}~\bibnamefont {Woodward}}, \bibinfo {author} {\bibfnamefont {B.}~\bibnamefont {Akdim}}, \bibinfo {author} {\bibfnamefont {T.}~\bibnamefont {Parthasarathy}}, \ and\ \bibinfo {author} {\bibfnamefont {O.}~\bibnamefont {Senkov}},\ }\href {\doibase https://doi.org/10.1016/j.scriptamat.2019.02.012} {\bibfield  {journal} {\bibinfo  {journal} {Scripta Materialia}\ }\textbf {\bibinfo {volume} {165}},\ \bibinfo {pages} {103} (\bibinfo {year} {2019})}\BibitemShut {NoStop}%
\bibitem [{\citenamefont {Rao}\ \emph {et~al.}(2021)\citenamefont {Rao}, \citenamefont {Woodward}, \citenamefont {Akdim}, \citenamefont {Senkov},\ and\ \citenamefont {Miracle}}]{Rao2021}%
  \BibitemOpen
  \bibfield  {author} {\bibinfo {author} {\bibfnamefont {S.}~\bibnamefont {Rao}}, \bibinfo {author} {\bibfnamefont {C.}~\bibnamefont {Woodward}}, \bibinfo {author} {\bibfnamefont {B.}~\bibnamefont {Akdim}}, \bibinfo {author} {\bibfnamefont {O.}~\bibnamefont {Senkov}}, \ and\ \bibinfo {author} {\bibfnamefont {D.}~\bibnamefont {Miracle}},\ }\href {\doibase https://doi.org/10.1016/j.actamat.2021.116758} {\bibfield  {journal} {\bibinfo  {journal} {Acta Materialia}\ }\textbf {\bibinfo {volume} {209}},\ \bibinfo {pages} {116758} (\bibinfo {year} {2021})}\BibitemShut {NoStop}%
\bibitem [{\citenamefont {Xu}\ \emph {et~al.}(2022)\citenamefont {Xu}, \citenamefont {Chavoshi},\ and\ \citenamefont {Su}}]{Xu2022}%
  \BibitemOpen
  \bibfield  {author} {\bibinfo {author} {\bibfnamefont {S.}~\bibnamefont {Xu}}, \bibinfo {author} {\bibfnamefont {S.~Z.}\ \bibnamefont {Chavoshi}}, \ and\ \bibinfo {author} {\bibfnamefont {Y.}~\bibnamefont {Su}},\ }\href {\doibase https://doi.org/10.1016/j.commatsci.2021.110942} {\bibfield  {journal} {\bibinfo  {journal} {Computational Materials Science}\ }\textbf {\bibinfo {volume} {202}},\ \bibinfo {pages} {110942} (\bibinfo {year} {2022})}\BibitemShut {NoStop}%
\bibitem [{\citenamefont {Pasini}\ \emph {et~al.}(2022)\citenamefont {Pasini}, \citenamefont {Zhang}, \citenamefont {Reeve},\ and\ \citenamefont {Choi}}]{Lupopasini2022}%
  \BibitemOpen
  \bibfield  {author} {\bibinfo {author} {\bibfnamefont {M.~L.}\ \bibnamefont {Pasini}}, \bibinfo {author} {\bibfnamefont {P.}~\bibnamefont {Zhang}}, \bibinfo {author} {\bibfnamefont {S.~T.}\ \bibnamefont {Reeve}}, \ and\ \bibinfo {author} {\bibfnamefont {J.~Y.}\ \bibnamefont {Choi}},\ }\href {\doibase 10.1088/2632-2153/ac6a51} {\bibfield  {journal} {\bibinfo  {journal} {Machine Learning: Science and Technology}\ }\textbf {\bibinfo {volume} {3}},\ \bibinfo {pages} {025007} (\bibinfo {year} {2022})}\BibitemShut {NoStop}%
\bibitem [{\citenamefont {{Lupo Pasini}}\ \emph {et~al.}(2023)\citenamefont {{Lupo Pasini}}, \citenamefont {Jung},\ and\ \citenamefont {Irle}}]{Lupopasini2023}%
  \BibitemOpen
  \bibfield  {author} {\bibinfo {author} {\bibfnamefont {M.}~\bibnamefont {{Lupo Pasini}}}, \bibinfo {author} {\bibfnamefont {G.~S.}\ \bibnamefont {Jung}}, \ and\ \bibinfo {author} {\bibfnamefont {S.}~\bibnamefont {Irle}},\ }\href {\doibase https://doi.org/10.1016/j.commatsci.2023.112141} {\bibfield  {journal} {\bibinfo  {journal} {Computational Materials Science}\ }\textbf {\bibinfo {volume} {224}},\ \bibinfo {pages} {112141} (\bibinfo {year} {2023})}\BibitemShut {NoStop}%
\bibitem [{\citenamefont {Nyshadham}\ \emph {et~al.}(2019)\citenamefont {Nyshadham}, \citenamefont {Rupp}, \citenamefont {Bekker}, \citenamefont {Shapeev}, \citenamefont {Mueller}, \citenamefont {Rosenbrock}, \citenamefont {Cs{\'a}nyi}, \citenamefont {Wingate},\ and\ \citenamefont {Hart}}]{Nyshadham2019}%
  \BibitemOpen
  \bibfield  {author} {\bibinfo {author} {\bibfnamefont {C.}~\bibnamefont {Nyshadham}}, \bibinfo {author} {\bibfnamefont {M.}~\bibnamefont {Rupp}}, \bibinfo {author} {\bibfnamefont {B.}~\bibnamefont {Bekker}}, \bibinfo {author} {\bibfnamefont {A.~V.}\ \bibnamefont {Shapeev}}, \bibinfo {author} {\bibfnamefont {T.}~\bibnamefont {Mueller}}, \bibinfo {author} {\bibfnamefont {C.~W.}\ \bibnamefont {Rosenbrock}}, \bibinfo {author} {\bibfnamefont {G.}~\bibnamefont {Cs{\'a}nyi}}, \bibinfo {author} {\bibfnamefont {D.~W.}\ \bibnamefont {Wingate}}, \ and\ \bibinfo {author} {\bibfnamefont {G.~L.~W.}\ \bibnamefont {Hart}},\ }\href {\doibase 10.1038/s41524-019-0189-9} {\bibfield  {journal} {\bibinfo  {journal} {npj Computational Materials}\ }\textbf {\bibinfo {volume} {5}},\ \bibinfo {pages} {51} (\bibinfo {year} {2019})}\BibitemShut {NoStop}%
\bibitem [{\citenamefont {Liu}\ \emph {et~al.}(2021)\citenamefont {Liu}, \citenamefont {Zhang}, \citenamefont {Yin}, \citenamefont {Bi}, \citenamefont {Eisenbach},\ and\ \citenamefont {Wang}}]{Liu2021}%
  \BibitemOpen
  \bibfield  {author} {\bibinfo {author} {\bibfnamefont {X.}~\bibnamefont {Liu}}, \bibinfo {author} {\bibfnamefont {J.}~\bibnamefont {Zhang}}, \bibinfo {author} {\bibfnamefont {J.}~\bibnamefont {Yin}}, \bibinfo {author} {\bibfnamefont {S.}~\bibnamefont {Bi}}, \bibinfo {author} {\bibfnamefont {M.}~\bibnamefont {Eisenbach}}, \ and\ \bibinfo {author} {\bibfnamefont {Y.}~\bibnamefont {Wang}},\ }\href {\doibase https://doi.org/10.1016/j.commatsci.2020.110135} {\bibfield  {journal} {\bibinfo  {journal} {Computational Materials Science}\ }\textbf {\bibinfo {volume} {187}},\ \bibinfo {pages} {110135} (\bibinfo {year} {2021})}\BibitemShut {NoStop}%
\bibitem [{\citenamefont {Huang}\ \emph {et~al.}(2024)\citenamefont {Huang}, \citenamefont {Zheng}, \citenamefont {Xu},\ and\ \citenamefont {Fu}}]{Huang2024}%
  \BibitemOpen
  \bibfield  {author} {\bibinfo {author} {\bibfnamefont {X.}~\bibnamefont {Huang}}, \bibinfo {author} {\bibfnamefont {L.}~\bibnamefont {Zheng}}, \bibinfo {author} {\bibfnamefont {H.}~\bibnamefont {Xu}}, \ and\ \bibinfo {author} {\bibfnamefont {H.}~\bibnamefont {Fu}},\ }\href {\doibase https://doi.org/10.1016/j.matdes.2024.112797} {\bibfield  {journal} {\bibinfo  {journal} {Materials \& Design}\ }\textbf {\bibinfo {volume} {239}},\ \bibinfo {pages} {112797} (\bibinfo {year} {2024})}\BibitemShut {NoStop}%
\bibitem [{\citenamefont {Yang}\ and\ \citenamefont {Qi}(2018)}]{Yang2018}%
  \BibitemOpen
  \bibfield  {author} {\bibinfo {author} {\bibfnamefont {C.}~\bibnamefont {Yang}}\ and\ \bibinfo {author} {\bibfnamefont {L.}~\bibnamefont {Qi}},\ }\href {\doibase 10.1103/PhysRevB.97.014107} {\bibfield  {journal} {\bibinfo  {journal} {Phys. Rev. B}\ }\textbf {\bibinfo {volume} {97}},\ \bibinfo {pages} {014107} (\bibinfo {year} {2018})}\BibitemShut {NoStop}%
\bibitem [{\citenamefont {Ruban}\ \emph {et~al.}(1998)\citenamefont {Ruban}, \citenamefont {Skriver},\ and\ \citenamefont {N\o{}rskov}}]{Ruban1998}%
  \BibitemOpen
  \bibfield  {author} {\bibinfo {author} {\bibfnamefont {A.~V.}\ \bibnamefont {Ruban}}, \bibinfo {author} {\bibfnamefont {H.~L.}\ \bibnamefont {Skriver}}, \ and\ \bibinfo {author} {\bibfnamefont {J.~K.}\ \bibnamefont {N\o{}rskov}},\ }\href {\doibase 10.1103/PhysRevLett.80.1240} {\bibfield  {journal} {\bibinfo  {journal} {Phys. Rev. Lett.}\ }\textbf {\bibinfo {volume} {80}},\ \bibinfo {pages} {1240} (\bibinfo {year} {1998})}\BibitemShut {NoStop}%
\bibitem [{\citenamefont {Pettifor}\ \emph {et~al.}(1995)\citenamefont {Pettifor} \emph {et~al.}}]{Pettifor1995}%
  \BibitemOpen
  \bibfield  {author} {\bibinfo {author} {\bibfnamefont {D.~G.}\ \bibnamefont {Pettifor}} \emph {et~al.},\ }\href@noop {} {\emph {\bibinfo {title} {Bonding and structure of molecules and solids}}}\ (\bibinfo  {publisher} {Oxford university press},\ \bibinfo {year} {1995})\BibitemShut {NoStop}%
\bibitem [{\citenamefont {Ferrari}\ \emph {et~al.}(2018)\citenamefont {Ferrari}, \citenamefont {Paulsen}, \citenamefont {Frenzel}, \citenamefont {Rogal}, \citenamefont {Eggeler},\ and\ \citenamefont {Drautz}}]{Ferrari2018}%
  \BibitemOpen
  \bibfield  {author} {\bibinfo {author} {\bibfnamefont {A.}~\bibnamefont {Ferrari}}, \bibinfo {author} {\bibfnamefont {A.}~\bibnamefont {Paulsen}}, \bibinfo {author} {\bibfnamefont {J.}~\bibnamefont {Frenzel}}, \bibinfo {author} {\bibfnamefont {J.}~\bibnamefont {Rogal}}, \bibinfo {author} {\bibfnamefont {G.}~\bibnamefont {Eggeler}}, \ and\ \bibinfo {author} {\bibfnamefont {R.}~\bibnamefont {Drautz}},\ }\href {\doibase 10.1103/PhysRevMaterials.2.073609} {\bibfield  {journal} {\bibinfo  {journal} {Phys. Rev. Mater.}\ }\textbf {\bibinfo {volume} {2}},\ \bibinfo {pages} {073609} (\bibinfo {year} {2018})}\BibitemShut {NoStop}%
\bibitem [{\citenamefont {Wu}\ \emph {et~al.}(2014)\citenamefont {Wu}, \citenamefont {Xu}, \citenamefont {Guo}, \citenamefont {Su}, \citenamefont {Du},\ and\ \citenamefont {Zhang}}]{Wu2014a}%
  \BibitemOpen
  \bibfield  {author} {\bibinfo {author} {\bibfnamefont {Y.}~\bibnamefont {Wu}}, \bibinfo {author} {\bibfnamefont {C.}~\bibnamefont {Xu}}, \bibinfo {author} {\bibfnamefont {J.}~\bibnamefont {Guo}}, \bibinfo {author} {\bibfnamefont {Q.}~\bibnamefont {Su}}, \bibinfo {author} {\bibfnamefont {G.}~\bibnamefont {Du}}, \ and\ \bibinfo {author} {\bibfnamefont {J.}~\bibnamefont {Zhang}},\ }\href {\doibase https://doi.org/10.1016/j.matlet.2014.09.044} {\bibfield  {journal} {\bibinfo  {journal} {Materials Letters}\ }\textbf {\bibinfo {volume} {137}},\ \bibinfo {pages} {277} (\bibinfo {year} {2014})}\BibitemShut {NoStop}%
\bibitem [{\citenamefont {Wu}\ \emph {et~al.}(2015)\citenamefont {Wu}, \citenamefont {Cai}, \citenamefont {Chen}, \citenamefont {Wang}, \citenamefont {Si}, \citenamefont {Wang}, \citenamefont {Wang},\ and\ \citenamefont {Hui}}]{Wu2015a}%
  \BibitemOpen
  \bibfield  {author} {\bibinfo {author} {\bibfnamefont {Y.}~\bibnamefont {Wu}}, \bibinfo {author} {\bibfnamefont {Y.}~\bibnamefont {Cai}}, \bibinfo {author} {\bibfnamefont {X.}~\bibnamefont {Chen}}, \bibinfo {author} {\bibfnamefont {T.}~\bibnamefont {Wang}}, \bibinfo {author} {\bibfnamefont {J.}~\bibnamefont {Si}}, \bibinfo {author} {\bibfnamefont {L.}~\bibnamefont {Wang}}, \bibinfo {author} {\bibfnamefont {Y.}~\bibnamefont {Wang}}, \ and\ \bibinfo {author} {\bibfnamefont {X.}~\bibnamefont {Hui}},\ }\href {\doibase https://doi.org/10.1016/j.matdes.2015.06.072} {\bibfield  {journal} {\bibinfo  {journal} {Materials \& Design}\ }\textbf {\bibinfo {volume} {83}},\ \bibinfo {pages} {651} (\bibinfo {year} {2015})}\BibitemShut {NoStop}%
\bibitem [{\citenamefont {Fazakas}\ \emph {et~al.}(2014)\citenamefont {Fazakas}, \citenamefont {Zadorozhnyy}, \citenamefont {Varga}, \citenamefont {Inoue}, \citenamefont {Louzguine-Luzgin}, \citenamefont {Tian},\ and\ \citenamefont {Vitos}}]{Fazakas2014a}%
  \BibitemOpen
  \bibfield  {author} {\bibinfo {author} {\bibfnamefont {A.}~\bibnamefont {Fazakas}}, \bibinfo {author} {\bibfnamefont {V.}~\bibnamefont {Zadorozhnyy}}, \bibinfo {author} {\bibfnamefont {L.}~\bibnamefont {Varga}}, \bibinfo {author} {\bibfnamefont {A.}~\bibnamefont {Inoue}}, \bibinfo {author} {\bibfnamefont {D.}~\bibnamefont {Louzguine-Luzgin}}, \bibinfo {author} {\bibfnamefont {F.}~\bibnamefont {Tian}}, \ and\ \bibinfo {author} {\bibfnamefont {L.}~\bibnamefont {Vitos}},\ }\href {\doibase https://doi.org/10.1016/j.ijrmhm.2014.07.009} {\bibfield  {journal} {\bibinfo  {journal} {International Journal of Refractory Metals and Hard Materials}\ }\textbf {\bibinfo {volume} {47}},\ \bibinfo {pages} {131} (\bibinfo {year} {2014})}\BibitemShut {NoStop}%
\bibitem [{\citenamefont {Xiong}\ \emph {et~al.}(2023)\citenamefont {Xiong}, \citenamefont {Guo}, \citenamefont {Zhan}, \citenamefont {Liu},\ and\ \citenamefont {Cao}}]{Xiong2023a}%
  \BibitemOpen
  \bibfield  {author} {\bibinfo {author} {\bibfnamefont {W.}~\bibnamefont {Xiong}}, \bibinfo {author} {\bibfnamefont {A.~X.}\ \bibnamefont {Guo}}, \bibinfo {author} {\bibfnamefont {S.}~\bibnamefont {Zhan}}, \bibinfo {author} {\bibfnamefont {C.-T.}\ \bibnamefont {Liu}}, \ and\ \bibinfo {author} {\bibfnamefont {S.~C.}\ \bibnamefont {Cao}},\ }\href {\doibase https://doi.org/10.1016/j.jmst.2022.08.046} {\bibfield  {journal} {\bibinfo  {journal} {Journal of Materials Science \& Technology}\ }\textbf {\bibinfo {volume} {142}},\ \bibinfo {pages} {196} (\bibinfo {year} {2023})}\BibitemShut {NoStop}%
\bibitem [{\citenamefont {Kotykhov}\ \emph {et~al.}(2023)\citenamefont {Kotykhov}, \citenamefont {Gubaev}, \citenamefont {Hodapp}, \citenamefont {Tantardini}, \citenamefont {Shapeev},\ and\ \citenamefont {Novikov}}]{kotykhov_constrained_2023}%
  \BibitemOpen
  \bibfield  {author} {\bibinfo {author} {\bibfnamefont {A.~S.}\ \bibnamefont {Kotykhov}}, \bibinfo {author} {\bibfnamefont {K.}~\bibnamefont {Gubaev}}, \bibinfo {author} {\bibfnamefont {M.}~\bibnamefont {Hodapp}}, \bibinfo {author} {\bibfnamefont {C.}~\bibnamefont {Tantardini}}, \bibinfo {author} {\bibfnamefont {A.~V.}\ \bibnamefont {Shapeev}}, \ and\ \bibinfo {author} {\bibfnamefont {I.~S.}\ \bibnamefont {Novikov}},\ }\href {\doibase 10.1038/s41598-023-46951-x} {\bibfield  {journal} {\bibinfo  {journal} {Scientific Reports}\ }\textbf {\bibinfo {volume} {13}},\ \bibinfo {pages} {19728} (\bibinfo {year} {2023})}\BibitemShut {NoStop}%
\bibitem [{\citenamefont {Hodapp}\ and\ \citenamefont {Shapeev}(2020)}]{hodapp_operando_2020}%
  \BibitemOpen
  \bibfield  {author} {\bibinfo {author} {\bibfnamefont {M.}~\bibnamefont {Hodapp}}\ and\ \bibinfo {author} {\bibfnamefont {A.}~\bibnamefont {Shapeev}},\ }\href {\doibase 10.1088/2632-2153/aba373} {\bibfield  {journal} {\bibinfo  {journal} {Mach. Learn.: Sci. Technol.}\ }\textbf {\bibinfo {volume} {1}},\ \bibinfo {pages} {045005} (\bibinfo {year} {2020})}\BibitemShut {NoStop}%
\bibitem [{\citenamefont {Andric}\ and\ \citenamefont {Curtin}(2018)}]{Andric2019}%
  \BibitemOpen
  \bibfield  {author} {\bibinfo {author} {\bibfnamefont {P.}~\bibnamefont {Andric}}\ and\ \bibinfo {author} {\bibfnamefont {W.~A.}\ \bibnamefont {Curtin}},\ }\href {\doibase 10.1088/1361-651X/aae40c} {\bibfield  {journal} {\bibinfo  {journal} {Modelling and Simulation in Materials Science and Engineering}\ }\textbf {\bibinfo {volume} {27}},\ \bibinfo {pages} {013001} (\bibinfo {year} {2018})}\BibitemShut {NoStop}%
\bibitem [{\citenamefont {Tyson}\ \emph {et~al.}(1973)\citenamefont {Tyson}, \citenamefont {Ayres},\ and\ \citenamefont {Stein}}]{Tyson1973}%
  \BibitemOpen
  \bibfield  {author} {\bibinfo {author} {\bibfnamefont {W.}~\bibnamefont {Tyson}}, \bibinfo {author} {\bibfnamefont {R.}~\bibnamefont {Ayres}}, \ and\ \bibinfo {author} {\bibfnamefont {D.}~\bibnamefont {Stein}},\ }\href {\doibase https://doi.org/10.1016/0001-6160(73)90071-0} {\bibfield  {journal} {\bibinfo  {journal} {Acta Metallurgica}\ }\textbf {\bibinfo {volume} {21}},\ \bibinfo {pages} {621} (\bibinfo {year} {1973})}\BibitemShut {NoStop}%
\bibitem [{\citenamefont {Vitek}(1968)}]{Vitek1968}%
  \BibitemOpen
  \bibfield  {author} {\bibinfo {author} {\bibfnamefont {V.}~\bibnamefont {Vitek}},\ }\href {\doibase 10.1080/14786436808227500} {\bibfield  {journal} {\bibinfo  {journal} {The Philosophical Magazine: A Journal of Theoretical Experimental and Applied Physics}\ }\textbf {\bibinfo {volume} {18}},\ \bibinfo {pages} {773} (\bibinfo {year} {1968})},\ \Eprint {http://arxiv.org/abs/https://doi.org/10.1080/14786436808227500} {https://doi.org/10.1080/14786436808227500} \BibitemShut {NoStop}%
\bibitem [{\citenamefont {Ting}\ and\ \citenamefont {Ting}(1996)}]{Ting1996}%
  \BibitemOpen
  \bibfield  {author} {\bibinfo {author} {\bibfnamefont {T.~C.-t.}\ \bibnamefont {Ting}}\ and\ \bibinfo {author} {\bibfnamefont {T.~C.-t.}\ \bibnamefont {Ting}},\ }\href@noop {} {\emph {\bibinfo {title} {Anisotropic elasticity: theory and applications}}},\ \bibinfo {number} {45}\ (\bibinfo  {publisher} {Oxford University Press on Demand},\ \bibinfo {year} {1996})\BibitemShut {NoStop}%
\bibitem [{\citenamefont {Razumovskiy}\ \emph {et~al.}(2011{\natexlab{a}})\citenamefont {Razumovskiy}, \citenamefont {Ruban},\ and\ \citenamefont {Korzhavyi}}]{Razumovskiy2011a}%
  \BibitemOpen
  \bibfield  {author} {\bibinfo {author} {\bibfnamefont {V.~I.}\ \bibnamefont {Razumovskiy}}, \bibinfo {author} {\bibfnamefont {A.~V.}\ \bibnamefont {Ruban}}, \ and\ \bibinfo {author} {\bibfnamefont {P.~A.}\ \bibnamefont {Korzhavyi}},\ }\href {\doibase 10.1103/PhysRevB.84.024106} {\bibfield  {journal} {\bibinfo  {journal} {Phys. Rev. B}\ }\textbf {\bibinfo {volume} {84}},\ \bibinfo {pages} {024106} (\bibinfo {year} {2011}{\natexlab{a}})}\BibitemShut {NoStop}%
\bibitem [{\citenamefont {Razumovskiy}\ \emph {et~al.}(2011{\natexlab{b}})\citenamefont {Razumovskiy}, \citenamefont {Ruban},\ and\ \citenamefont {Korzhavyi}}]{Razumovskiy2011b}%
  \BibitemOpen
  \bibfield  {author} {\bibinfo {author} {\bibfnamefont {V.~I.}\ \bibnamefont {Razumovskiy}}, \bibinfo {author} {\bibfnamefont {A.~V.}\ \bibnamefont {Ruban}}, \ and\ \bibinfo {author} {\bibfnamefont {P.~A.}\ \bibnamefont {Korzhavyi}},\ }\href {\doibase 10.1103/PhysRevLett.107.205504} {\bibfield  {journal} {\bibinfo  {journal} {Phys. Rev. Lett.}\ }\textbf {\bibinfo {volume} {107}},\ \bibinfo {pages} {205504} (\bibinfo {year} {2011}{\natexlab{b}})}\BibitemShut {NoStop}%
\bibitem [{\citenamefont {Cordero}\ \emph {et~al.}(2016)\citenamefont {Cordero}, \citenamefont {Knight},\ and\ \citenamefont {Schuh}}]{Cordero2016}%
  \BibitemOpen
  \bibfield  {author} {\bibinfo {author} {\bibfnamefont {Z.~C.}\ \bibnamefont {Cordero}}, \bibinfo {author} {\bibfnamefont {B.~E.}\ \bibnamefont {Knight}}, \ and\ \bibinfo {author} {\bibfnamefont {C.~A.}\ \bibnamefont {Schuh}},\ }\href {\doibase 10.1080/09506608.2016.1191808} {\bibfield  {journal} {\bibinfo  {journal} {International Materials Reviews}\ }\textbf {\bibinfo {volume} {61}},\ \bibinfo {pages} {495} (\bibinfo {year} {2016})},\ \Eprint {http://arxiv.org/abs/https://doi.org/10.1080/09506608.2016.1191808} {https://doi.org/10.1080/09506608.2016.1191808} \BibitemShut {NoStop}%
\bibitem [{\citenamefont {Vitos}(2001)}]{Vitos2001a}%
  \BibitemOpen
  \bibfield  {author} {\bibinfo {author} {\bibfnamefont {L.}~\bibnamefont {Vitos}},\ }\href {\doibase 10.1103/PhysRevB.64.014107} {\bibfield  {journal} {\bibinfo  {journal} {Physical Review B}\ }\textbf {\bibinfo {volume} {64}},\ \bibinfo {pages} {014107} (\bibinfo {year} {2001})}\BibitemShut {NoStop}%
\bibitem [{\citenamefont {Ruban}\ and\ \citenamefont {Dehghani}(2016)}]{Ruban2016}%
  \BibitemOpen
  \bibfield  {author} {\bibinfo {author} {\bibfnamefont {A.~V.}\ \bibnamefont {Ruban}}\ and\ \bibinfo {author} {\bibfnamefont {M.}~\bibnamefont {Dehghani}},\ }\href {\doibase https://doi.org/10.1103/PhysRevB.94.104111} {\bibfield  {journal} {\bibinfo  {journal} {Physical Review B}\ }\textbf {\bibinfo {volume} {94}},\ \bibinfo {pages} {104111} (\bibinfo {year} {2016})}\BibitemShut {NoStop}%
\bibitem [{\citenamefont {Abrikosov}\ \emph {et~al.}(1997)\citenamefont {Abrikosov}, \citenamefont {Simak}, \citenamefont {Johansson}, \citenamefont {Ruban},\ and\ \citenamefont {Skriver}}]{Abrikosov1997}%
  \BibitemOpen
  \bibfield  {author} {\bibinfo {author} {\bibfnamefont {I.~A.}\ \bibnamefont {Abrikosov}}, \bibinfo {author} {\bibfnamefont {S.~I.}\ \bibnamefont {Simak}}, \bibinfo {author} {\bibfnamefont {B.}~\bibnamefont {Johansson}}, \bibinfo {author} {\bibfnamefont {A.~V.}\ \bibnamefont {Ruban}}, \ and\ \bibinfo {author} {\bibfnamefont {H.~L.}\ \bibnamefont {Skriver}},\ }\href {\doibase 10.1103/PhysRevB.56.9319} {\bibfield  {journal} {\bibinfo  {journal} {Phys. Rev. B}\ }\textbf {\bibinfo {volume} {56}},\ \bibinfo {pages} {9319} (\bibinfo {year} {1997})}\BibitemShut {NoStop}%
\bibitem [{\citenamefont {Peil}\ \emph {et~al.}(2012)\citenamefont {Peil}, \citenamefont {Ruban},\ and\ \citenamefont {Johansson}}]{Peil2012}%
  \BibitemOpen
  \bibfield  {author} {\bibinfo {author} {\bibfnamefont {O.~E.}\ \bibnamefont {Peil}}, \bibinfo {author} {\bibfnamefont {A.~V.}\ \bibnamefont {Ruban}}, \ and\ \bibinfo {author} {\bibfnamefont {B.}~\bibnamefont {Johansson}},\ }\href {\doibase 10.1103/PhysRevB.85.165140} {\bibfield  {journal} {\bibinfo  {journal} {Phys. Rev. B}\ }\textbf {\bibinfo {volume} {85}},\ \bibinfo {pages} {165140} (\bibinfo {year} {2012})}\BibitemShut {NoStop}%
\bibitem [{\citenamefont {Vitos}\ \emph {et~al.}(1997)\citenamefont {Vitos}, \citenamefont {Koll\'ar},\ and\ \citenamefont {Skriver}}]{Vitos1997}%
  \BibitemOpen
  \bibfield  {author} {\bibinfo {author} {\bibfnamefont {L.}~\bibnamefont {Vitos}}, \bibinfo {author} {\bibfnamefont {J.}~\bibnamefont {Koll\'ar}}, \ and\ \bibinfo {author} {\bibfnamefont {H.~L.}\ \bibnamefont {Skriver}},\ }\href {\doibase 10.1103/PhysRevB.55.13521} {\bibfield  {journal} {\bibinfo  {journal} {Physical Review B}\ }\textbf {\bibinfo {volume} {55}},\ \bibinfo {pages} {13521} (\bibinfo {year} {1997})}\BibitemShut {NoStop}%
\bibitem [{\citenamefont {Woodard}(1969)}]{Woodard1969}%
  \BibitemOpen
  \bibfield  {author} {\bibinfo {author} {\bibfnamefont {C.~L.}\ \bibnamefont {Woodard}},\ }\emph {\bibinfo {title} {X-ray determination of lattice parameters and thermal expansion coefficients of aluminum, silver and molybdenum at cryogenic temperatures}},\ \href@noop {} {\bibinfo {type} {{PhD} dissertation}},\ \bibinfo  {school} {University of Missouri--Rolla} (\bibinfo {year} {1969})\BibitemShut {NoStop}%
\bibitem [{\citenamefont {Shah}\ and\ \citenamefont {Straumanis}(1971)}]{Shah1971}%
  \BibitemOpen
  \bibfield  {author} {\bibinfo {author} {\bibfnamefont {J.~S.}\ \bibnamefont {Shah}}\ and\ \bibinfo {author} {\bibfnamefont {M.~E.}\ \bibnamefont {Straumanis}},\ }\href {\doibase 10.1063/1.1660727} {\bibfield  {journal} {\bibinfo  {journal} {Journal of Applied Physics}\ }\textbf {\bibinfo {volume} {42}},\ \bibinfo {pages} {3288} (\bibinfo {year} {1971})},\ \Eprint {http://arxiv.org/abs/https://doi.org/10.1063/1.1660727} {https://doi.org/10.1063/1.1660727} \BibitemShut {NoStop}%
\bibitem [{\citenamefont {Spreadborough}\ and\ \citenamefont {Christian}(1959)}]{Spreadborough1959}%
  \BibitemOpen
  \bibfield  {author} {\bibinfo {author} {\bibfnamefont {J.}~\bibnamefont {Spreadborough}}\ and\ \bibinfo {author} {\bibfnamefont {J.~W.}\ \bibnamefont {Christian}},\ }\href {\doibase 10.1088/0370-1328/74/5/314} {\bibfield  {journal} {\bibinfo  {journal} {Proceedings of the Physical Society}\ }\textbf {\bibinfo {volume} {74}},\ \bibinfo {pages} {609} (\bibinfo {year} {1959})}\BibitemShut {NoStop}%
\bibitem [{\citenamefont {Corruccini}\ and\ \citenamefont {Gniewek}(1961)}]{Corruccini1961}%
  \BibitemOpen
  \bibfield  {author} {\bibinfo {author} {\bibfnamefont {R.}~\bibnamefont {Corruccini}}\ and\ \bibinfo {author} {\bibfnamefont {J.}~\bibnamefont {Gniewek}},\ }\href {https://books.google.at/books?id=kmt4MNc6J5EC} {\emph {\bibinfo {title} {Thermal Expansion of Technical Solids at Low Temperatures: A Compilation from the Literature}}},\ Monograph 29 Series\ (\bibinfo  {publisher} {U.S. Department of Commerce, National Bureau of Standards},\ \bibinfo {year} {1961})\BibitemShut {NoStop}%
\bibitem [{\citenamefont {Goldak}\ \emph {et~al.}(1966)\citenamefont {Goldak}, \citenamefont {Lloyd},\ and\ \citenamefont {Barrett}}]{Goldak1966}%
  \BibitemOpen
  \bibfield  {author} {\bibinfo {author} {\bibfnamefont {J.}~\bibnamefont {Goldak}}, \bibinfo {author} {\bibfnamefont {L.~T.}\ \bibnamefont {Lloyd}}, \ and\ \bibinfo {author} {\bibfnamefont {C.~S.}\ \bibnamefont {Barrett}},\ }\href {\doibase 10.1103/PhysRev.144.478} {\bibfield  {journal} {\bibinfo  {journal} {Phys. Rev.}\ }\textbf {\bibinfo {volume} {144}},\ \bibinfo {pages} {478} (\bibinfo {year} {1966})}\BibitemShut {NoStop}%
\bibitem [{\citenamefont {Versaci}\ and\ \citenamefont {Ipohorski}(1991)}]{Versaci1991}%
  \BibitemOpen
  \bibfield  {author} {\bibinfo {author} {\bibfnamefont {R.~A.}\ \bibnamefont {Versaci}}\ and\ \bibinfo {author} {\bibfnamefont {M.}~\bibnamefont {Ipohorski}},\ }\href {http://inis.iaea.org/search/search.aspx?orig\%5Fq=RN:23041495} {\emph {\bibinfo {title} {Temperature dependence of lattice parameters of alpha-zirconium}}},\ \bibinfo {type} {Tech. Rep.}\ (\bibinfo  {institution} {Cnea--500},\ \bibinfo {year} {1991})\BibitemShut {NoStop}%
\bibitem [{\citenamefont {Smirnov}\ and\ \citenamefont {Finkel}(1965)}]{Smirnov1965}%
  \BibitemOpen
  \bibfield  {author} {\bibinfo {author} {\bibfnamefont {Y.~N.}\ \bibnamefont {Smirnov}}\ and\ \bibinfo {author} {\bibfnamefont {V.~A.}\ \bibnamefont {Finkel}},\ }\href {https://www.osti.gov/biblio/4586965} {\bibfield  {journal} {\bibinfo  {journal} {Zhurnal Eksperimental'noi i Teoreticheskoi Fiziki (U.S.S.R.) For English translation see Sov. Phys. - JETP (Engl. Transl.)}\ }\textbf {\bibinfo {volume} {49}} (\bibinfo {year} {1965})}\BibitemShut {NoStop}%
\bibitem [{\citenamefont {Moruzzi}\ \emph {et~al.}(1988)\citenamefont {Moruzzi}, \citenamefont {Janak},\ and\ \citenamefont {Schwarz}}]{Moruzzi1988}%
  \BibitemOpen
  \bibfield  {author} {\bibinfo {author} {\bibfnamefont {V.~L.}\ \bibnamefont {Moruzzi}}, \bibinfo {author} {\bibfnamefont {J.~F.}\ \bibnamefont {Janak}}, \ and\ \bibinfo {author} {\bibfnamefont {K.}~\bibnamefont {Schwarz}},\ }\href {\doibase 10.1103/PhysRevB.37.790} {\bibfield  {journal} {\bibinfo  {journal} {Phys. Rev. B}\ }\textbf {\bibinfo {volume} {37}},\ \bibinfo {pages} {790} (\bibinfo {year} {1988})}\BibitemShut {NoStop}%
\bibitem [{\citenamefont {Blanco}\ \emph {et~al.}(2004)\citenamefont {Blanco}, \citenamefont {Francisco},\ and\ \citenamefont {Luaa}}]{Blanco2004}%
  \BibitemOpen
  \bibfield  {author} {\bibinfo {author} {\bibfnamefont {M.}~\bibnamefont {Blanco}}, \bibinfo {author} {\bibfnamefont {E.}~\bibnamefont {Francisco}}, \ and\ \bibinfo {author} {\bibfnamefont {V.}~\bibnamefont {Luaa}},\ }\href {\doibase https://doi.org/10.1016/j.comphy.2003.12.001} {\bibfield  {journal} {\bibinfo  {journal} {Computer Physics Communications}\ }\textbf {\bibinfo {volume} {158}},\ \bibinfo {pages} {57} (\bibinfo {year} {2004})}\BibitemShut {NoStop}%
\bibitem [{\citenamefont {Novikov}\ \emph {et~al.}(2021)\citenamefont {Novikov}, \citenamefont {Gubaev}, \citenamefont {Podryabinkin},\ and\ \citenamefont {Shapeev}}]{Novikov2021}%
  \BibitemOpen
  \bibfield  {author} {\bibinfo {author} {\bibfnamefont {I.~S.}\ \bibnamefont {Novikov}}, \bibinfo {author} {\bibfnamefont {K.}~\bibnamefont {Gubaev}}, \bibinfo {author} {\bibfnamefont {E.~V.}\ \bibnamefont {Podryabinkin}}, \ and\ \bibinfo {author} {\bibfnamefont {A.~V.}\ \bibnamefont {Shapeev}},\ }\href {\doibase 10.1088/2632-2153/abc9fe} {\bibfield  {journal} {\bibinfo  {journal} {Machine Learning: Science and Technology}\ }\textbf {\bibinfo {volume} {2}},\ \bibinfo {pages} {025002} (\bibinfo {year} {2021})}\BibitemShut {NoStop}%
\bibitem [{\citenamefont {Gubaev}\ \emph {et~al.}(2019)\citenamefont {Gubaev}, \citenamefont {Podryabinkin}, \citenamefont {Hart},\ and\ \citenamefont {Shapeev}}]{Gubaev2019}%
  \BibitemOpen
  \bibfield  {author} {\bibinfo {author} {\bibfnamefont {K.}~\bibnamefont {Gubaev}}, \bibinfo {author} {\bibfnamefont {E.~V.}\ \bibnamefont {Podryabinkin}}, \bibinfo {author} {\bibfnamefont {G.~L.}\ \bibnamefont {Hart}}, \ and\ \bibinfo {author} {\bibfnamefont {A.~V.}\ \bibnamefont {Shapeev}},\ }\href {\doibase 10.1016/j.commatsci.2018.09.031} {\bibfield  {journal} {\bibinfo  {journal} {Computational Materials Science}\ }\textbf {\bibinfo {volume} {156}},\ \bibinfo {pages} {148} (\bibinfo {year} {2019})}\BibitemShut {NoStop}%
\bibitem [{\citenamefont {Podryabinkin}\ and\ \citenamefont {Shapeev}(2017)}]{Podryabinkin2017}%
  \BibitemOpen
  \bibfield  {author} {\bibinfo {author} {\bibfnamefont {E.~V.}\ \bibnamefont {Podryabinkin}}\ and\ \bibinfo {author} {\bibfnamefont {A.~V.}\ \bibnamefont {Shapeev}},\ }\href {\doibase 10.1016/j.commatsci.2017.08.031} {\bibfield  {journal} {\bibinfo  {journal} {Computational Materials Science}\ }\textbf {\bibinfo {volume} {140}},\ \bibinfo {pages} {171} (\bibinfo {year} {2017})}\BibitemShut {NoStop}%
\bibitem [{\citenamefont {Kresse}\ and\ \citenamefont {Hafner}(1993)}]{Kresse1993}%
  \BibitemOpen
  \bibfield  {author} {\bibinfo {author} {\bibfnamefont {G.}~\bibnamefont {Kresse}}\ and\ \bibinfo {author} {\bibfnamefont {J.}~\bibnamefont {Hafner}},\ }\href {\doibase 10.1103/PhysRevB.47.558} {\bibfield  {journal} {\bibinfo  {journal} {Phys. Rev. B}\ }\textbf {\bibinfo {volume} {47}},\ \bibinfo {pages} {558} (\bibinfo {year} {1993})}\BibitemShut {NoStop}%
\bibitem [{\citenamefont {Kresse}\ and\ \citenamefont {Furthm\"uller}(1996{\natexlab{a}})}]{Kresse1996a}%
  \BibitemOpen
  \bibfield  {author} {\bibinfo {author} {\bibfnamefont {G.}~\bibnamefont {Kresse}}\ and\ \bibinfo {author} {\bibfnamefont {J.}~\bibnamefont {Furthm\"uller}},\ }\href {\doibase 10.1103/PhysRevB.54.11169} {\bibfield  {journal} {\bibinfo  {journal} {Phys. Rev. B}\ }\textbf {\bibinfo {volume} {54}},\ \bibinfo {pages} {11169} (\bibinfo {year} {1996}{\natexlab{a}})}\BibitemShut {NoStop}%
\bibitem [{\citenamefont {Kresse}\ and\ \citenamefont {Furthm\"uller}(1996{\natexlab{b}})}]{Kresse1996b}%
  \BibitemOpen
  \bibfield  {author} {\bibinfo {author} {\bibfnamefont {G.}~\bibnamefont {Kresse}}\ and\ \bibinfo {author} {\bibfnamefont {J.}~\bibnamefont {Furthm\"uller}},\ }\href {\doibase https://doi.org/10.1016/0927-0256(96)00008-0} {\bibfield  {journal} {\bibinfo  {journal} {Computational Materials Science}\ }\textbf {\bibinfo {volume} {6}},\ \bibinfo {pages} {15} (\bibinfo {year} {1996}{\natexlab{b}})}\BibitemShut {NoStop}%
\bibitem [{\citenamefont {Perdew}\ \emph {et~al.}(1996)\citenamefont {Perdew}, \citenamefont {Burke},\ and\ \citenamefont {Ernzerhof}}]{pbe96}%
  \BibitemOpen
  \bibfield  {author} {\bibinfo {author} {\bibfnamefont {J.~P.}\ \bibnamefont {Perdew}}, \bibinfo {author} {\bibfnamefont {K.}~\bibnamefont {Burke}}, \ and\ \bibinfo {author} {\bibfnamefont {M.}~\bibnamefont {Ernzerhof}},\ }\href {\doibase https://doi.org/10.1103/PhysRevLett.77.3865} {\bibfield  {journal} {\bibinfo  {journal} {Physical review letters}\ }\textbf {\bibinfo {volume} {77}},\ \bibinfo {pages} {3865} (\bibinfo {year} {1996})}\BibitemShut {NoStop}%
\bibitem [{\citenamefont {Kresse}\ and\ \citenamefont {Hafner}(1994)}]{Kresse1994}%
  \BibitemOpen
  \bibfield  {author} {\bibinfo {author} {\bibfnamefont {G.}~\bibnamefont {Kresse}}\ and\ \bibinfo {author} {\bibfnamefont {J.}~\bibnamefont {Hafner}},\ }\href {\doibase 10.1088/0953-8984/6/40/015} {\bibfield  {journal} {\bibinfo  {journal} {Journal of Physics: Condensed Matter}\ }\textbf {\bibinfo {volume} {6}},\ \bibinfo {pages} {8245} (\bibinfo {year} {1994})}\BibitemShut {NoStop}%
\bibitem [{\citenamefont {Kresse}\ and\ \citenamefont {Joubert}(1999)}]{Kresse1999}%
  \BibitemOpen
  \bibfield  {author} {\bibinfo {author} {\bibfnamefont {G.}~\bibnamefont {Kresse}}\ and\ \bibinfo {author} {\bibfnamefont {D.}~\bibnamefont {Joubert}},\ }\href {\doibase 10.1103/PhysRevB.59.1758} {\bibfield  {journal} {\bibinfo  {journal} {Phys. Rev. B}\ }\textbf {\bibinfo {volume} {59}},\ \bibinfo {pages} {1758} (\bibinfo {year} {1999})}\BibitemShut {NoStop}%
\bibitem [{\citenamefont {Galuzio}\ \emph {et~al.}(2020)\citenamefont {Galuzio}, \citenamefont {{de Vasconcelos Segundo}}, \citenamefont {dos Santos~Coelho},\ and\ \citenamefont {Mariani}}]{Galuzio2020}%
  \BibitemOpen
  \bibfield  {author} {\bibinfo {author} {\bibfnamefont {P.~P.}\ \bibnamefont {Galuzio}}, \bibinfo {author} {\bibfnamefont {E.~H.}\ \bibnamefont {{de Vasconcelos Segundo}}}, \bibinfo {author} {\bibfnamefont {L.}~\bibnamefont {dos Santos~Coelho}}, \ and\ \bibinfo {author} {\bibfnamefont {V.~C.}\ \bibnamefont {Mariani}},\ }\href {\doibase https://doi.org/10.1016/j.softx.2020.100520} {\bibfield  {journal} {\bibinfo  {journal} {SoftwareX}\ }\textbf {\bibinfo {volume} {12}},\ \bibinfo {pages} {100520} (\bibinfo {year} {2020})}\BibitemShut {NoStop}%
\bibitem [{\citenamefont {Fortin}\ \emph {et~al.}(2012)\citenamefont {Fortin}, \citenamefont {{De Rainville}}, \citenamefont {Gardner}, \citenamefont {Parizeau},\ and\ \citenamefont {Gagn\'e}}]{DEAP_JMLR2012}%
  \BibitemOpen
  \bibfield  {author} {\bibinfo {author} {\bibfnamefont {F.-A.}\ \bibnamefont {Fortin}}, \bibinfo {author} {\bibfnamefont {F.-M.}\ \bibnamefont {{De Rainville}}}, \bibinfo {author} {\bibfnamefont {M.-A.}\ \bibnamefont {Gardner}}, \bibinfo {author} {\bibfnamefont {M.}~\bibnamefont {Parizeau}}, \ and\ \bibinfo {author} {\bibfnamefont {C.}~\bibnamefont {Gagn\'e}},\ }\href@noop {} {\bibfield  {journal} {\bibinfo  {journal} {Journal of Machine Learning Research}\ }\textbf {\bibinfo {volume} {13}},\ \bibinfo {pages} {2171} (\bibinfo {year} {2012})}\BibitemShut {NoStop}%
\bibitem [{\citenamefont {Shiba}(1971)}]{Shiba1971}%
  \BibitemOpen
  \bibfield  {author} {\bibinfo {author} {\bibfnamefont {H.}~\bibnamefont {Shiba}},\ }\href@noop {} {\bibfield  {journal} {\bibinfo  {journal} {Progress of Theoretical Physics}\ }\textbf {\bibinfo {volume} {46}},\ \bibinfo {pages} {77} (\bibinfo {year} {1971})}\BibitemShut {NoStop}%
\bibitem [{\citenamefont {Andersen}\ \emph {et~al.}(1985)\citenamefont {Andersen}, \citenamefont {Jepsen},\ and\ \citenamefont {Gl{\"o}tzel}}]{Andersen1985}%
  \BibitemOpen
  \bibfield  {author} {\bibinfo {author} {\bibfnamefont {O.}~\bibnamefont {Andersen}}, \bibinfo {author} {\bibfnamefont {O.}~\bibnamefont {Jepsen}}, \ and\ \bibinfo {author} {\bibfnamefont {O.}~\bibnamefont {Gl{\"o}tzel}},\ }\href@noop {} {\bibfield  {journal} {\bibinfo  {journal} {Proceedings of the International School of Physics, Course LXXXIX, Varenna}\ } (\bibinfo {year} {1985})}\BibitemShut {NoStop}%
\bibitem [{\citenamefont {Singh}\ \emph {et~al.}(2018{\natexlab{b}})\citenamefont {Singh}, \citenamefont {Sharma}, \citenamefont {Smirnov}, \citenamefont {Diallo}, \citenamefont {Ray}, \citenamefont {Balasubramanian},\ and\ \citenamefont {Johnson}}]{Singh2018b}%
  \BibitemOpen
  \bibfield  {author} {\bibinfo {author} {\bibfnamefont {P.}~\bibnamefont {Singh}}, \bibinfo {author} {\bibfnamefont {A.}~\bibnamefont {Sharma}}, \bibinfo {author} {\bibfnamefont {A.~V.}\ \bibnamefont {Smirnov}}, \bibinfo {author} {\bibfnamefont {M.~S.}\ \bibnamefont {Diallo}}, \bibinfo {author} {\bibfnamefont {P.~K.}\ \bibnamefont {Ray}}, \bibinfo {author} {\bibfnamefont {G.}~\bibnamefont {Balasubramanian}}, \ and\ \bibinfo {author} {\bibfnamefont {D.~D.}\ \bibnamefont {Johnson}},\ }\href {\doibase 10.1038/s41524-018-0072-0} {\bibfield  {journal} {\bibinfo  {journal} {npj Computational Materials}\ }\textbf {\bibinfo {volume} {4}},\ \bibinfo {pages} {16} (\bibinfo {year} {2018}{\natexlab{b}})}\BibitemShut {NoStop}%
\bibitem [{\citenamefont {Inoue}\ \emph {et~al.}(1993)\citenamefont {Inoue}, \citenamefont {Zhang},\ and\ \citenamefont {Masumoto}}]{Inoune1993}%
  \BibitemOpen
  \bibfield  {author} {\bibinfo {author} {\bibfnamefont {A.}~\bibnamefont {Inoue}}, \bibinfo {author} {\bibfnamefont {T.}~\bibnamefont {Zhang}}, \ and\ \bibinfo {author} {\bibfnamefont {T.}~\bibnamefont {Masumoto}},\ }\href {\doibase https://doi.org/10.1016/0022-3093(93)90030-2} {\bibfield  {journal} {\bibinfo  {journal} {Journal of Non-Crystalline Solids}\ }\textbf {\bibinfo {volume} {156-158}},\ \bibinfo {pages} {598} (\bibinfo {year} {1993})}\BibitemShut {NoStop}%
\bibitem [{\citenamefont {Kalali}\ \emph {et~al.}(2024)\citenamefont {Kalali}, \citenamefont {Kanchi}, \citenamefont {Phani}, \citenamefont {Rao}, \citenamefont {Murty},\ and\ \citenamefont {Rajulapati}}]{Kalali2024}%
  \BibitemOpen
  \bibfield  {author} {\bibinfo {author} {\bibfnamefont {D.~G.}\ \bibnamefont {Kalali}}, \bibinfo {author} {\bibfnamefont {A.}~\bibnamefont {Kanchi}}, \bibinfo {author} {\bibfnamefont {P.~S.}\ \bibnamefont {Phani}}, \bibinfo {author} {\bibfnamefont {K.~B.~S.}\ \bibnamefont {Rao}}, \bibinfo {author} {\bibfnamefont {S.~N.}\ \bibnamefont {Murty}}, \ and\ \bibinfo {author} {\bibfnamefont {K.~V.}\ \bibnamefont {Rajulapati}},\ }\href {\doibase https://doi.org/10.1016/j.ijrmhm.2023.106487} {\bibfield  {journal} {\bibinfo  {journal} {International Journal of Refractory Metals and Hard Materials}\ }\textbf {\bibinfo {volume} {118}},\ \bibinfo {pages} {106487} (\bibinfo {year} {2024})}\BibitemShut {NoStop}%
\bibitem [{\citenamefont {Startt}\ \emph {et~al.}(2022)\citenamefont {Startt}, \citenamefont {Kustas}, \citenamefont {Pegues}, \citenamefont {Yang},\ and\ \citenamefont {Dingreville}}]{Start2022}%
  \BibitemOpen
  \bibfield  {author} {\bibinfo {author} {\bibfnamefont {J.}~\bibnamefont {Startt}}, \bibinfo {author} {\bibfnamefont {A.}~\bibnamefont {Kustas}}, \bibinfo {author} {\bibfnamefont {J.}~\bibnamefont {Pegues}}, \bibinfo {author} {\bibfnamefont {P.}~\bibnamefont {Yang}}, \ and\ \bibinfo {author} {\bibfnamefont {R.}~\bibnamefont {Dingreville}},\ }\href {\doibase https://doi.org/10.1016/j.matdes.2021.110311} {\bibfield  {journal} {\bibinfo  {journal} {Materials \& Design}\ }\textbf {\bibinfo {volume} {213}},\ \bibinfo {pages} {110311} (\bibinfo {year} {2022})}\BibitemShut {NoStop}%
\bibitem [{\citenamefont {Zhang}\ \emph {et~al.}(2022)\citenamefont {Zhang}, \citenamefont {Tang}, \citenamefont {Wen}, \citenamefont {Obaied}, \citenamefont {Roslyakova},\ and\ \citenamefont {Zhang}}]{Zhang2022c}%
  \BibitemOpen
  \bibfield  {author} {\bibinfo {author} {\bibfnamefont {E.}~\bibnamefont {Zhang}}, \bibinfo {author} {\bibfnamefont {Y.}~\bibnamefont {Tang}}, \bibinfo {author} {\bibfnamefont {M.}~\bibnamefont {Wen}}, \bibinfo {author} {\bibfnamefont {A.}~\bibnamefont {Obaied}}, \bibinfo {author} {\bibfnamefont {I.}~\bibnamefont {Roslyakova}}, \ and\ \bibinfo {author} {\bibfnamefont {L.}~\bibnamefont {Zhang}},\ }\href {\doibase https://doi.org/10.1016/j.ijrmhm.2022.105780} {\bibfield  {journal} {\bibinfo  {journal} {International Journal of Refractory Metals and Hard Materials}\ }\textbf {\bibinfo {volume} {103}},\ \bibinfo {pages} {105780} (\bibinfo {year} {2022})}\BibitemShut {NoStop}%
\bibitem [{\citenamefont {Tseng}\ \emph {et~al.}(2019)\citenamefont {Tseng}, \citenamefont {Juan}, \citenamefont {Tso}, \citenamefont {Chen}, \citenamefont {Tsai},\ and\ \citenamefont {Yeh}}]{Tseng2019}%
  \BibitemOpen
  \bibfield  {author} {\bibinfo {author} {\bibfnamefont {K.-K.}\ \bibnamefont {Tseng}}, \bibinfo {author} {\bibfnamefont {C.-C.}\ \bibnamefont {Juan}}, \bibinfo {author} {\bibfnamefont {S.}~\bibnamefont {Tso}}, \bibinfo {author} {\bibfnamefont {H.-C.}\ \bibnamefont {Chen}}, \bibinfo {author} {\bibfnamefont {C.-W.}\ \bibnamefont {Tsai}}, \ and\ \bibinfo {author} {\bibfnamefont {J.-W.}\ \bibnamefont {Yeh}},\ }\href {\doibase 10.3390/e21010015} {\bibfield  {journal} {\bibinfo  {journal} {Entropy}\ }\textbf {\bibinfo {volume} {21}} (\bibinfo {year} {2019}),\ 10.3390/e21010015}\BibitemShut {NoStop}%
\bibitem [{\citenamefont {Zhang}\ \emph {et~al.}(2023)\citenamefont {Zhang}, \citenamefont {Wei}, \citenamefont {Xie},\ and\ \citenamefont {Xu}}]{Zhang2023b}%
  \BibitemOpen
  \bibfield  {author} {\bibinfo {author} {\bibfnamefont {Y.}~\bibnamefont {Zhang}}, \bibinfo {author} {\bibfnamefont {Q.}~\bibnamefont {Wei}}, \bibinfo {author} {\bibfnamefont {P.}~\bibnamefont {Xie}}, \ and\ \bibinfo {author} {\bibfnamefont {X.}~\bibnamefont {Xu}},\ }\href {\doibase 10.1080/21663831.2022.2133977} {\bibfield  {journal} {\bibinfo  {journal} {Materials Research Letters}\ }\textbf {\bibinfo {volume} {11}},\ \bibinfo {pages} {169} (\bibinfo {year} {2023})}\BibitemShut {NoStop}%
\bibitem [{\citenamefont {Dobbelstein}\ \emph {et~al.}(2019)\citenamefont {Dobbelstein}, \citenamefont {Gurevich}, \citenamefont {George}, \citenamefont {Ostendorf},\ and\ \citenamefont {Laplanche}}]{Dobbelstein2019}%
  \BibitemOpen
  \bibfield  {author} {\bibinfo {author} {\bibfnamefont {H.}~\bibnamefont {Dobbelstein}}, \bibinfo {author} {\bibfnamefont {E.~L.}\ \bibnamefont {Gurevich}}, \bibinfo {author} {\bibfnamefont {E.~P.}\ \bibnamefont {George}}, \bibinfo {author} {\bibfnamefont {A.}~\bibnamefont {Ostendorf}}, \ and\ \bibinfo {author} {\bibfnamefont {G.}~\bibnamefont {Laplanche}},\ }\href {\doibase https://doi.org/10.1016/j.addma.2018.10.042} {\bibfield  {journal} {\bibinfo  {journal} {Additive Manufacturing}\ }\textbf {\bibinfo {volume} {25}},\ \bibinfo {pages} {252} (\bibinfo {year} {2019})}\BibitemShut {NoStop}%
\bibitem [{\citenamefont {Song}\ \emph {et~al.}(2017)\citenamefont {Song}, \citenamefont {Tian}, \citenamefont {Hu}, \citenamefont {Vitos}, \citenamefont {Wang}, \citenamefont {Shen},\ and\ \citenamefont {Chen}}]{Song2017}%
  \BibitemOpen
  \bibfield  {author} {\bibinfo {author} {\bibfnamefont {H.}~\bibnamefont {Song}}, \bibinfo {author} {\bibfnamefont {F.}~\bibnamefont {Tian}}, \bibinfo {author} {\bibfnamefont {Q.-M.}\ \bibnamefont {Hu}}, \bibinfo {author} {\bibfnamefont {L.}~\bibnamefont {Vitos}}, \bibinfo {author} {\bibfnamefont {Y.}~\bibnamefont {Wang}}, \bibinfo {author} {\bibfnamefont {J.}~\bibnamefont {Shen}}, \ and\ \bibinfo {author} {\bibfnamefont {N.}~\bibnamefont {Chen}},\ }\href {\doibase 10.1103/PhysRevMaterials.1.023404} {\bibfield  {journal} {\bibinfo  {journal} {Phys. Rev. Mater.}\ }\textbf {\bibinfo {volume} {1}},\ \bibinfo {pages} {023404} (\bibinfo {year} {2017})}\BibitemShut {NoStop}%
\bibitem [{\citenamefont {Tian}\ \emph {et~al.}(2015)\citenamefont {Tian}, \citenamefont {Hu}, \citenamefont {Yang}, \citenamefont {Zhao}, \citenamefont {Johansson},\ and\ \citenamefont {Vitos}}]{Tian2015}%
  \BibitemOpen
  \bibfield  {author} {\bibinfo {author} {\bibfnamefont {L.-Y.}\ \bibnamefont {Tian}}, \bibinfo {author} {\bibfnamefont {Q.-M.}\ \bibnamefont {Hu}}, \bibinfo {author} {\bibfnamefont {R.}~\bibnamefont {Yang}}, \bibinfo {author} {\bibfnamefont {J.}~\bibnamefont {Zhao}}, \bibinfo {author} {\bibfnamefont {B.}~\bibnamefont {Johansson}}, \ and\ \bibinfo {author} {\bibfnamefont {L.}~\bibnamefont {Vitos}},\ }\href {\doibase 10.1088/0953-8984/27/31/315702} {\bibfield  {journal} {\bibinfo  {journal} {Journal of Physics: Condensed Matter}\ }\textbf {\bibinfo {volume} {27}},\ \bibinfo {pages} {315702} (\bibinfo {year} {2015})}\BibitemShut {NoStop}%
\bibitem [{\citenamefont {Baruffi}\ \emph {et~al.}(2022)\citenamefont {Baruffi}, \citenamefont {Maresca},\ and\ \citenamefont {Curtin}}]{Baruffi2022}%
  \BibitemOpen
  \bibfield  {author} {\bibinfo {author} {\bibfnamefont {C.}~\bibnamefont {Baruffi}}, \bibinfo {author} {\bibfnamefont {F.}~\bibnamefont {Maresca}}, \ and\ \bibinfo {author} {\bibfnamefont {W.~A.}\ \bibnamefont {Curtin}},\ }\href {\doibase 10.1557/s43579-022-00278-2} {\bibfield  {journal} {\bibinfo  {journal} {MRS Communications}\ }\textbf {\bibinfo {volume} {12}},\ \bibinfo {pages} {1111} (\bibinfo {year} {2022})}\BibitemShut {NoStop}%
\bibitem [{\citenamefont {Melia}\ \emph {et~al.}(2020)\citenamefont {Melia}, \citenamefont {Whetten}, \citenamefont {Puckett}, \citenamefont {Jones}, \citenamefont {Heiden}, \citenamefont {Argibay},\ and\ \citenamefont {Kustas}}]{Melia2020}%
  \BibitemOpen
  \bibfield  {author} {\bibinfo {author} {\bibfnamefont {M.~A.}\ \bibnamefont {Melia}}, \bibinfo {author} {\bibfnamefont {S.~R.}\ \bibnamefont {Whetten}}, \bibinfo {author} {\bibfnamefont {R.}~\bibnamefont {Puckett}}, \bibinfo {author} {\bibfnamefont {M.}~\bibnamefont {Jones}}, \bibinfo {author} {\bibfnamefont {M.~J.}\ \bibnamefont {Heiden}}, \bibinfo {author} {\bibfnamefont {N.}~\bibnamefont {Argibay}}, \ and\ \bibinfo {author} {\bibfnamefont {A.~B.}\ \bibnamefont {Kustas}},\ }\href {\doibase https://doi.org/10.1016/j.apmt.2020.100560} {\bibfield  {journal} {\bibinfo  {journal} {Applied Materials Today}\ }\textbf {\bibinfo {volume} {19}},\ \bibinfo {pages} {100560} (\bibinfo {year} {2020})}\BibitemShut {NoStop}%
\bibitem [{\citenamefont {Larsen}\ \emph {et~al.}(2016)\citenamefont {Larsen}, \citenamefont {Schmidt},\ and\ \citenamefont {Schi{\o}tz}}]{Larsen2016}%
  \BibitemOpen
  \bibfield  {author} {\bibinfo {author} {\bibfnamefont {P.~M.}\ \bibnamefont {Larsen}}, \bibinfo {author} {\bibfnamefont {S.}~\bibnamefont {Schmidt}}, \ and\ \bibinfo {author} {\bibfnamefont {J.}~\bibnamefont {Schi{\o}tz}},\ }\href {\doibase 10.1088/0965-0393/24/5/055007} {\bibfield  {journal} {\bibinfo  {journal} {Modelling and Simulation in Materials Science and Engineering}\ }\textbf {\bibinfo {volume} {24}},\ \bibinfo {pages} {055007} (\bibinfo {year} {2016})}\BibitemShut {NoStop}%
\end{thebibliography}%

\clearpage

\begin{widetext}
    \begin{center}
        \textbf{\large Supplemental Materials: Ab initio framework for deciphering trade-off relationships in multi-component alloys}
    \end{center}
\end{widetext}

\setcounter{section}{0}
\setcounter{equation}{0}
\setcounter{figure}{0}
\setcounter{table}{0}
\setcounter{page}{1}

\makeatletter

\renewcommand{\theequation}{S\arabic{equation}}
\renewcommand{\thefigure}{S\arabic{figure}}

\renewcommand{\thesection}{S\arabic{section}}
\renewcommand{\thetable}{S\arabic{table}}
\makeatother

{

\section{Validation of phase stability of solute solution}
\label{sup_stab_ss}

\begin{figure}
   \centering
   \includegraphics{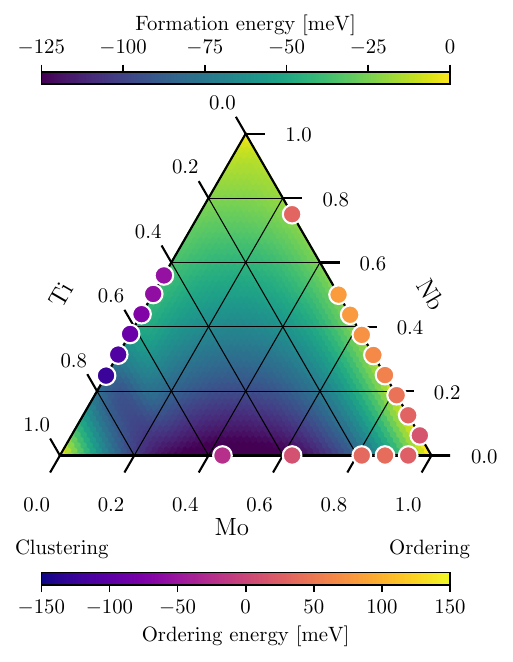}
   \caption{Ternary plot showing a heatmap of the formation energies of the MoNbTi solid solution.
   Dots denote binary compositions corresponding to intermetallic compounds on the convex hull from Ref.~\citet{Zheng2022},
   with the fill color indicating the respective ordering energies.
   }
   \label{fig:appendix_formation_monbti}
\end{figure}

\begin{figure*}
   \centering
   \includegraphics{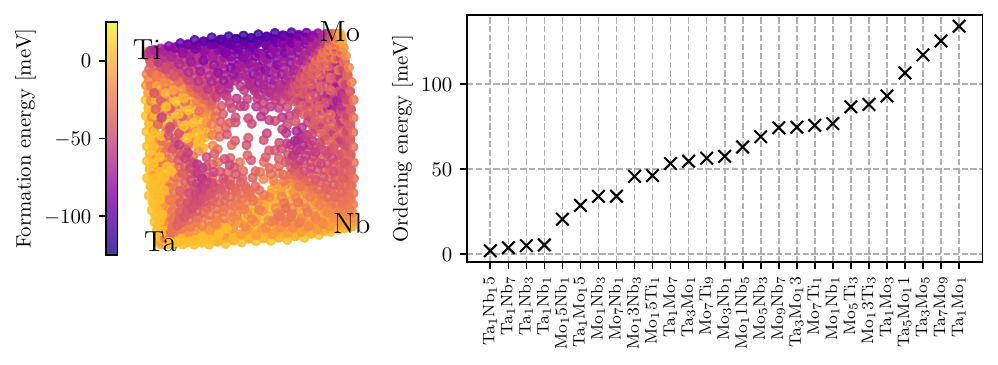}
   \caption{(upper panel) t-SNE projection of the formation energies of the MoNbTiTa system and (lower panel) positive ordering energies binaries.
   }
   \label{fig:appendix_formation_monbtita}
\end{figure*}

To assess the stability and formation of solute solutions and the occurrence of competing intermetallic phases, various approaches can be employed.

One approach involves CALPHAD utilizing thermodynamic databases, albeit constrained by the availability and also accessibility of such databases. Alternatively, there are also fully ab initio-based approaches to evaluate phase stability, although they tend to be computationally too expensive
and complex to be utilized inside an automated workflow. Accurate and reliable prediction of phase transformations can be a non-trivial task even for binary systems.

A more straightforward strategy involves assessing simpler quantities such as the formation energies of solid solutions and ordered phases, complemented by experimental observations to establish criteria for estimating phase stability.

We employ the formation energies criterion from Ref.~\cite{Singh2018b} in our work to evaluate operational stability of our alloy and to set \emph{a priori} boundaries of the design space. This criterion relies on the observation that multiple phases form when $E_f\,>\,70\,\textrm{meV}$ and metastable metallic glasses form or ordered phases appear
when $E_f\,<\,-150\,\textrm{meV}$ \cite{Inoune1993}. In our case, the formation energies of the entire system fall within these boundaries, as illustrated in Fig. \ref{fig:appendix_formation_monbti} and Fig. \ref{fig:appendix_formation_monbtita}.

Furthermore, we also take into account the ordering energies for binaries on or near the convex hull of formation energies. The ordered phases were obtained from supplementary materials of \citet{Zheng2022} and \href{materialsproject.org}{materialsproject.org}. The ordering energies are determined by subtracting the formation energy of the ordered intermetallic phase from the formation energy of the solid solution,
$E_{\textrm{ord}} = E^{(\textrm{SS})}_{f,BCC} - E^{(\textrm{I})}_{f,BCC}$.

In Fig.~\ref{fig:appendix_formation_monbti}, it is evident that only the MoNb binary side exhibits a significant ordering energy, peaking at around 80 meV. In the four-component system, slightly higher ordering energies are observed at the TaMo binary side, reaching 120 meV. However, this corresponds only to a small region close to the boundary of the concentration space, which is irrelevant for the optimization. Moreover,
in a multicomponent alloy, nucleation of an ordered phase consisting of two elements can be hindered by interactions between these two elements and other components.

Most importantly, experimental evidence strongly supports the assumption that MoNbTi and MoNbTiTa form solid solutions. Mo, Nb, and Ta are fully miscible at typical annealing temperatures and readily form a solid solution. Various experimental studies have demonstrated that the addition of Ti to the MoNbTa system also leads to the formation of a BCC solid solution \cite{Kalali2024,Start2022,Zhang2022c}.

}

\clearpage

\section{Validation of mechanical parameters}

\subsection{Validation of Ductility calculations}
\label{sup_valid_ductilty}

\begin{figure}
   \centering
   \includegraphics{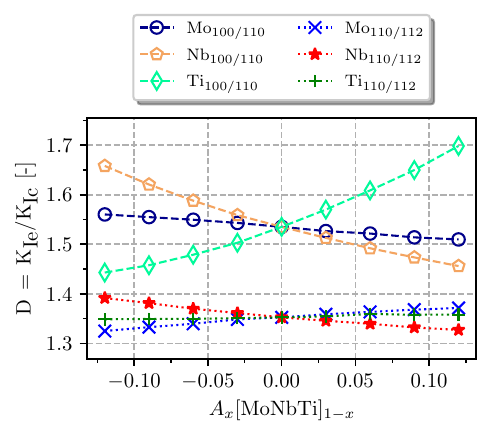}
   \caption{Comparison of the ductility index $D$ for the different fracture systems $100/110$ and $110/112$ (denoted by subscript) upon changing the concentration of one component in the equimolar MoNbTi alloy according to $A_x[\text{MoNbTi}]_{1 - x}$, where $x$ is the molar fraction added to the equimolar alloy.
   }
   \label{fig:figure2}
\end{figure}

Experimental verification of the ductility model for a variety of refractory MPEAs has been performed in Ref.~\cite{Mak2021}. Following the same approach, we compute the ductility index, $D$, as a minimum of two values, $D_{100/110}$ and $D_{110/112}$, corresponding to two surface orientations, $\{100\}$ and $\{110\}$,
and two stacking fault orientations, $\{110\}$ and $\{112\}$, respectively. USF, $\gamma_{USF}$, and the surface energy, $\gamma_{surf}$, is obtained from molecular statics of fully relaxed structures using MTPs.
Unlike the SQS approach, which requires full structure relaxation calculations for each property, crystallographic orientation, and composition in a cell with typically 800-1000 atoms for a three-component alloy, our active-learning approach generates 381 and 720 \emph{static} configurations with 54-96 atoms/configuration to fit an accurate potential in the entire composition space of MoNbTi and MoNbTiTa, respectively. 

Using the USF and surface energies, along with the linear 
elastic parameters from CPA, we evaluate parameters
$D_{100/110}$ and $D_{110/112}$ and ultimately the ductility 
index, $D$. Our values of the ductility parameters, 
$D_{100/110} = 1.53$ and $D_{110/112} = 1.35$, in the equimolar MoNbTi,
corresponding to $x = 0$ in Fig.~\ref{fig:figure2},
are somewhat different then the ones (1.29 and 1.59, respectively) 
reported in Mak et al.~\cite{Mak2021}. This can be attributed to 
supercell size convergence issues associated with SQS surface energy calculations.
In Fig.~\ref{fig:figure2}, we examine $D$ for the MoNbTi alloy as a function of varying concentrations of each component
$A \in \{ \text{Mo},\text{Nb},\text{Ti} \}$ in the two fracture systems, $100/110$ and $110/112$.
The concentration of $A$ is varied, keeping the proportion between the rest 
of the alloy equal, i.e., the general
formula of a modified alloy is $A_{x}[\textrm{MoNbTi}]_{1 - x}$. Notably, 
the $110/112$ fracture system is the dominating 
one across all cases, as it has the lowest $D$ index.

Increasing Nb content enhances ductility, while Mo and Ti additions show no positive effect. Nb's softening properties are experimentally documented in Refs.~\cite{Tseng2019,Zhang2023b,Dobbelstein2019}.

\subsection{Validation of solid solution strengthening calculations}
\label{sup_valid_sss}

\begin{figure}
   \centering
   \includegraphics{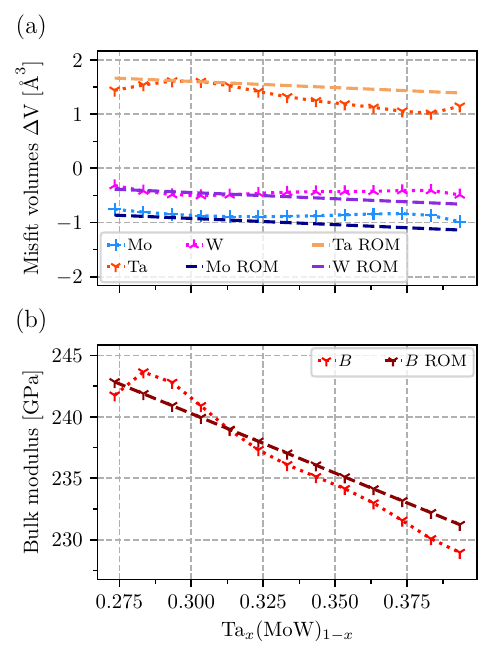}
   \caption{Comparison of directly calculated and values from the rule-of-mixture for (a) misfit volumes and (b) bulk modulus of a MoTaW alloy with respect to Ta variation.}
   \label{fig:appendix_5}
\end{figure}

~\citet{Maresca2020_b} suggested that in concentrated alloys, especially at high 
temperatures, solid solution strengthening can be described based on the alloy's average volume, 
linear elastic moduli, and misfit volumes of components. In particular, the determination of linear elastic constants and concentration derivatives of equilibrium volumes becomes challenging when employing supercell methods such as Special Quasirandom Structures (SQS). The presence of broken symmetry in SQS introduces direction-induced errors in elastic constants that are challenging to alleviate~\cite{Holec2014}. Additionally, obtaining concentration derivatives poses difficulties, as SQS is not effective for arbitrary concentrations. All of these parameters can be obtained robustly within CPA~\cite{Song2017,Tian2015, Moitzi2022} because local atomic relaxations have a minor effect.

Given the computational efficiency of CPA it eliminates the need 
for extrapolative models for misfit volumes, distinguishing it from prior studies~\cite{Elder2023a,Khatamsaz2022}. While 
misfit volumes often exhibit a linear dependence on concentration in refractory alloys, our research identifies specific alloys where this relationship deviates notably. Fig.~\ref{fig:appendix_5} illustrates the misfit volumes and bulk modulus that were directly calculated alongside those determined using ROM for variations of Ta within the MoTaW alloy. Similarly, misfit volumes and bulk modulus shows quite significant deviations form the linear trends. Especially around 0.28 the bulk modulus abruptly changes, which can be attributed to topological transitions. In Fig.~\ref{fig:paper_compare_alloy}b, we find good agreement between our CPA results and those obtained from an embedded-atom potential for $A$-atom in larger cells, as detailed in Refs.~\cite{Rao2019,Rao2021,Xu2022}.

To verify the final accuracy of the strengthening calculations, we compared the 
accuracy of our CRSS values calculated at 
300 K to room-temperature (RT) experimental data from 
tensile tests~\cite{Senkov2019,Senkov20211}, 
as well as compression tests reported in Ref.~\cite{Coury2019}.
The poly-crystalline samples used in the experiments 
were reported to be single-phase.
To extract the CRSS from these tests, we subtract the estimated 
Hall-Petch contribution of 20 MPa~\cite{Cordero2016} and 
divide the result by the Taylor factor of 3.09. In comparison to the 
experimentally measured total yield strengths, 
the strengthening contribution of the grain boundary is small.

However, for NbTaTi, the deviations are quite large. Our 
theoretical $\taus$ value for NbTaTi is approximately 88 MPa, 
which is about half of the experimental value obtained from tensile tests, 
which yields a CRSS of around 168 MPa. To elucidate the nature of this 
discrepancy, we have performed a comprehensive analysis of solid solution 
strengthening in five additional 3- and 4-component MPEAs: NbTiZr, 
MoNbTi, CrMoTaTi, MoNbTaW and MoTaW. Among the alloys tested, NbTiZr, MoNbTi, and MoNbTaW exhibit similar $\taus$ values of around 300 MPa in experiments,
whereas our theoretical approach yields 293, 195, and 234 MPa, respectively. The 4-component CrMoTaTi alloy, which exhibits a yield strength of 621 MPa in experiments, displays close agreement with our theoretical prediction of 577 MPa.

By comparing Fig.~\ref{fig:paper_compare_alloy}a with Fig.~\ref{fig:paper_compare_alloy}c, one 
can see that the theoretical predictions aligns closely with 
experimental data when the average misfit $\delta$ is high, as 
exemplified by NbTiZr and CrMoTaTi. All other alloys have smaller 
misfit deltas and their theoretical predictions consistently underestimate 
strength. Furthermore, it seems that the relative error is reduce for stiffer 
alloys with larger elastic moduli. 

Calculating CRSS in bcc crystals is challenging due to multiple active glide 
mechanisms. According to~\citet{Maresca2020_a,Maresca2020_b}, the solid solution strength of 
our multi-component refractory metal alloys is to large extend 
influenced by interactions between edge dislocations and solutes, with a 
lesser contribution from solute-screw dislocation interactions. Building 
on that strengthening theory, \citet{Baruffi2022} discovered 
that as the misfit parameter $\delta$ or elastic moduli increase, the dominance of edge-controlled 
contributions becomes more pronounced, eventually determining the strength completely. Within 
the used analytic strengthening model, we, however, only consider the edge-controlled strengthening. The 
contributions arising from screw dislocation are omitted, because their contributions 
are very challenging to calculate. The negative example of NbTaTi has a relatively small 
misfit parameter $\delta$ and is also soft. As a result, we underestimate the strength because the
screw-controlled contribution is significant and can actually not be neglected. In 
contrast, NbTiZr and CrMoTaTi are well-predicted because the screw contribution can be effectively 
disregarded. We want to emphasize, however, that as per Maresca's findings \cite{Maresca2020_a,Maresca2020_b}, the edge-controlled contribution tends to become increasingly 
dominant at elevated temperatures. Given that 
refractory alloys are primarily intended for high-temperature applications, this observation strengthens the validity of using a model for strengthening that is solely based on edge contributions for all alloys.

Even though we only consider a portion of the strengthening contribution, the findings presented in Fig.~\ref{fig:paper_compare_alloy}c demonstrate a high level of consistency between the experimentally determined CRSS and our calculated values. Overall, we tend to underestimate the strength of all the alloys tested, as expected, because we ignore the screw contributions. This also implies that, we can use the values for a conservative estimate of strength. In general, the relative changes in strength that occur during alloying are accurately reproduced. We can distinguish a weaker and a stronger alloy. For example, hardness screening of Nb in MoNbTaW revealed only a modest strength increase of about 10\% \cite{Melia2020} from equimolar MoNbTaW to MoTaW (no reliable yield strength measurements are available for MoTaW). Our methodology predicts a similar magnitude of increase, from 234 MPa to 257 MPa.

\clearpage

\section{Validation of material parameters}

\subsection{Convergence of local relaxation on surface and stacking fault energy}
\label{sup_valid_mat1}

Interatomic potentials have the advantage of allowing for larger cell sizes compared to SQS methods with. 
Using large cells is especially important when studying far-reaching elastic effects that cannot be accurately captured in small cells or need some complex compensation procedures. In the case of surface energies, the defect structure was created by 
introducing a vacuum of 20 \AA. We checked that 
fixing atomic positions of certain deep internal layers had no effect. For both surface orientation, we create orthogonal cells. 
In case of USF, we created a defect structure with two stacking faults in cell by simply shifting half of the atoms in $\frac{1}{4}a[111]$. We keep the cell orthogonal. We only allow for atomic relaxation normal to the fault plant and do not change the out-of-plane lattice dimension. By using large enough structures with more than 100 layers, we can account for inelastic displacement associated with the fault. This is in contrast to ~\cite{Mak2021, Andric2019}.

In order to assess the necessary cell size for the calculation of 
surface energies, we conducted convergence tests. Fig.~\ref{fig:figure5} displays the surface energies 
for the ${100}$ orientation as a function of the number of layers perpendicular to the fault. Convergence varied by alloy, and MoNbTi exhibited the slowest convergence. Approximately 100 layers were sufficient for convergence 
for the more sparsely-packed ${100}$ layer in the case of MoNbTi. Only about 50 ${110}$ layers are required to achieve convergence. The other alloys, MoNb and MoNbTa, require 8-10 layers to achieve convergence. 

On the other hand, the necessary cell size for unstable stacking fault converges much faster, even after only a few layers. We want to stress that we don't perform optimization of the neighboring cluster vector. Our big cell will contain statistical clustering of atoms of a specific kind. We assume that by randomly distributing the atoms in the cell in a sufficiently large cell size, we can generate a truely random alloy. For comparison, we also performed reference calculations with 4x4x10 SQS cells with 160 atoms with cluster vectors closely resembling those of truly random alloys. Except for MoNbTi, the energies of SQS cells are numerically equivalent to those of large cells. Numerical errors related to the interatomic potential can be assumed to be below 5\%~\cite{Hodapp2021}. 

\subsection{Effects of local relaxation on surface and stacking fault energy}
\label{sup_valid_mat2}

\begin{figure}
   \centering
   \includegraphics{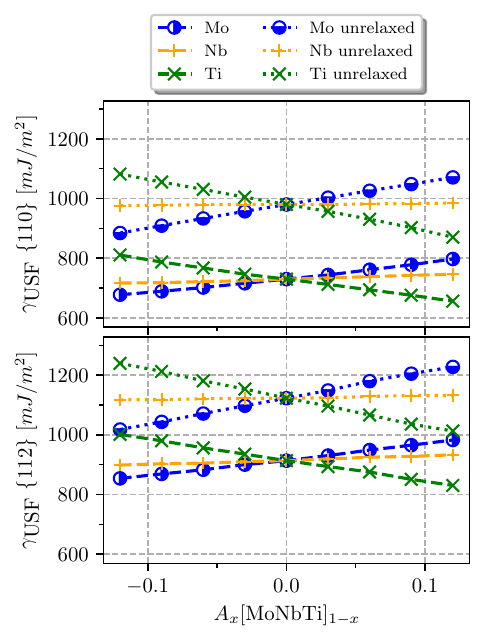}
   \caption{Comparison stacking fault energy $\{110\}$ and $\{112\}$ with and without local relaxations upon changing the concentration of one component in the equimolar MoNbTi alloy according to $x\,\text{Y} \rightarrow \text{MoNbTi}$, where $x$ is the molar fraction of $\text{Y}$ added to the equimolar alloy.}
   \label{fig:figure3}
\end{figure}

\begin{figure}
   \centering
   \includegraphics{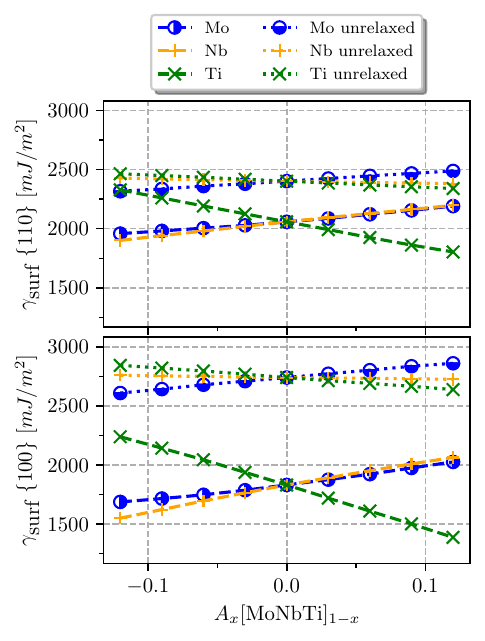}
   \caption{Comparison surface energy $\{110\}$ and $\{100\}$ with and without local relaxations upon changing the concentration of one component in the equimolar MoNbTi alloy according to $x\,\text{Y} \rightarrow \text{MoNbTi}$, where $x$ is the molar fraction of $\text{Y}$ added to the equimolar alloy.}
   \label{fig:figure4}
\end{figure}

\begin{figure}
   \centering
   \includegraphics{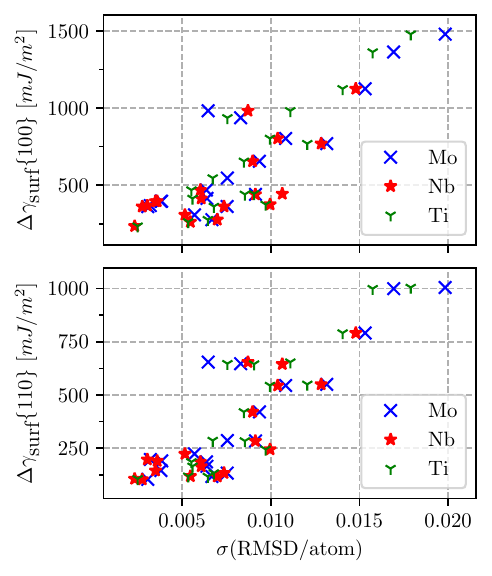}
   \caption{Energy difference of unrelaxed and relaxed surface energies and standard deviation of RMSD value from polyhedra template matching for different compositions of MoNbTi.}
   \label{fig:figure14}
\end{figure}

In order to investigate the impact of local relaxation on surface and unstable stacking fault energies, we present a 
comparison of different compositions of the MoNbTi alloy 
in Figs.~\ref{fig:figure3} and \ref{fig:figure4}. Each line shows the change in one 
alloy component in relation to the equimolar composition, while keeping the ratios of the 
remaining components constant. Neglecting the effects of local relaxation on USF $\gamma_\text{USF}$ yields a consistent 
increase of about 25 \% for both orientations. The overall trends of $\gamma_\text{USF}$ remain unchanged upon changes in composition. For surface energies $\gamma_\text{surf}$, the impact of relaxation is significant and varies depending on the element in question. Without considering relaxation effects, surface energies are largely unaffected by the addition of Nb. However, when relaxation effects are taken into account, the addition of Nb causes a significant increase in $\gamma_\text{surf}$. Similarly, changes in Ti concentration lead to more pronounced effects on $\gamma_\text{surf}$. On the other hand, for Mo, there is an almost constant shift between the relaxed and unrelaxed cases.

To show the origin of this behaviour, we used the Polyhedral Template Matching (PTM)~\cite{Larsen2016} method to 
identify the local crystalline structure in order to relate this relaxation contribution to local structural changes. 
It also provides for each atom in the structure a RMSD value, which is a measure of the spatial 
deviation from the ideal local structure. The distribution of the RMSD value for each atomic species in the structure is Gaussian-like. Especially Ti showed higher local distortions compared to all other components in all alloys 
investigated. We also found that the mean and standard deviation of the RMSD correlate to the energy difference 
between unrelaxed and relaxed surface energies. The higher the standard deviation 
of these RMSD values, as seen in Fig.~\ref{fig:figure14}, the more significant the 
relaxation contribution will be. The overall effect of this local structural changes, seems to be less relevant for USF.

{

\subsection{Performance metric for MTPs}
\label{sup_perf_metric}

The surface and unstable stacking fault energies are relatively small and sensitive quantities. Even minor errors can have significantly impact on the final results. Therefore, simply providing errors on forces and total energies calculated with MTP for target structures
may be inadequate to demonstrate the accuracy of the potential for 
a particular property. To verify accuracy of $\gamma_\text{surf}$ and $\gamma_\text{USF}$, we generate random structures of surface and unstable stacking fault configurations, along with their respective bulk reference configurations. Atomic species are placed randomly on lattice sites for various alloy compositions and the atomic positions and the cell volumes are slightly randomly perturbed.
Atomic relaxation is then performed using our MTPs.
The reference energies are then computed on the relaxed structures using DFT.
Across the test set, the mean absolute difference between the MTP and DFT forces is found to be 0.0402 eV/\AA.
Additionally, we provide a parity plot in Fig.~\ref{fig:appendix_metric} to compare the MTP and DFT surface and unstable stacking fault energies.
Overall, there is a high degree of correlation between MTP and DFT proving the accuracy of our MTPs.

\begin{figure}
   \centering
   \includegraphics{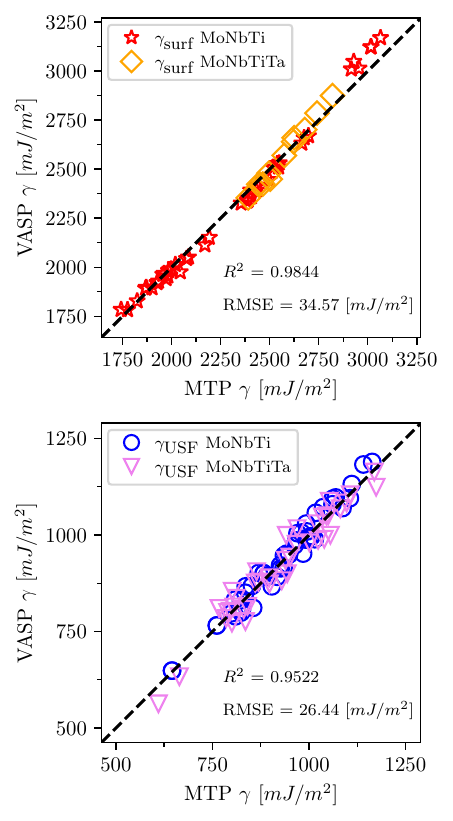}
   \caption{Parity plot of surface and unstable fault energies obtained from MTP and directly from VASP DFT code.
   }
   \label{fig:appendix_metric}
\end{figure}

\subsection{Influence of the elemental distribution}
\label{sup_conv_elements}

\begin{figure}
   \centering
   \includegraphics{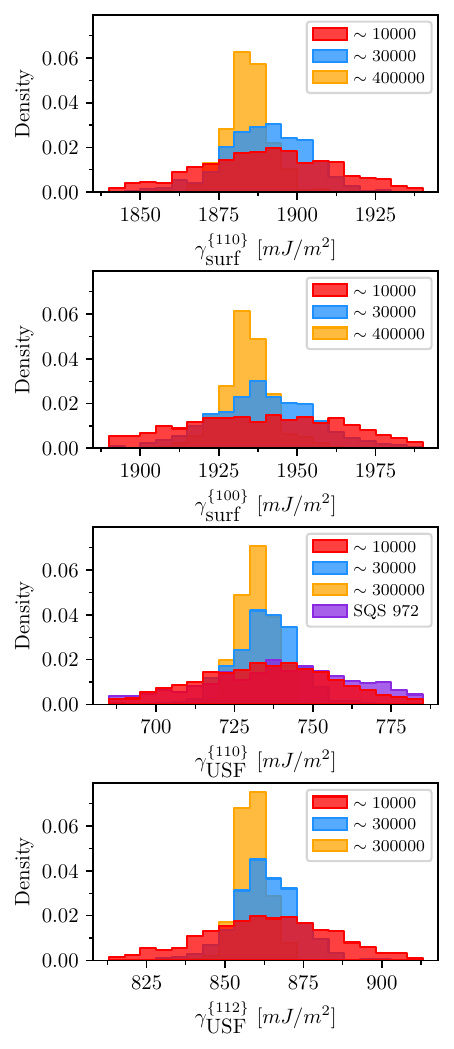}
   \caption{Histograms of surface $\gamma_{\mathrm{surf}}$ and unstable stacking fault energies $\gamma_{\mathrm{USF}}$ for purely random configurations of different number of atoms and SQS generated cells obtained from MTPs of a near equimolar MoNbTi alloy.
   }
   \label{fig:appendix_spread}
\end{figure}

In order to estimate the influence of the elemental distribution, we investigate the distributions of surface $\gamma_{\mathrm{surf}}$ and unstable stacking fault energies $\gamma_{\mathrm{USF}}$ using atomic configurations with different distributions of atomic species. To that end, we generated several hundred configurations for MoNbTi with fully random atomic distribution on the lattice sites, and using SQS. For SQS, we optimized the pair correlations of the first four shells, and the first triplet cluster correlations.
We perform the optimization of the atomic configurations on the bulk structures and neglect 2-dimensional correlation at the defects.
Fig. \ref{fig:appendix_spread} illustrates the expected narrowing of energy spread with increasing cell sizes. The standard deviation in the energies is smaller for the unstable stacking faults than for surface energies. Furthermore, the energy distribution of the SQS configuration for the stacking faults show a similar standard deviation to that of substantial bigger cell of around 10000 atoms with a purely random atom distribution. This result is in agreement with the distributions 
reported in the supplementary material of \cite{Zheng2022}, who also used MTPs.

\subsection{Error propagation}
\label{sup_error_propagation}

Two primary error sources arise: first, from representing a truly random alloy within a finite cell, accounting for variations in atomic configurations; and second, from approximations made by the potential.

From Fig.~\ref{fig:appendix_spread} we deduce that the distributions 
for the surface and unstable stacking fault energies due to different atomic configurations 
are close to Gaussian distribution. Hence, we can perform an analytic Gaussian error propagation to estimate the final error on the D parameter,
\begin{align*}
    D &= \dfrac{K_{Ie}}{K_{Ic}} \\ 
      &= \dfrac{\sqrt{\gamma_\text{USF} \,o(\doubleunderline{C}, \theta, \phi) \, \lambda_{22 (\doubleunderline{C})}} }
      {{F_{12}(\doubleunderline{C},\theta) \cos(\phi) \sqrt{2 \gamma_\text{surf} }} } \\
      &= \chi \sqrt{\dfrac{\gamma_\text{USF}}{\gamma_\text{surf} }}.
\end{align*}

Let us denote the errors as $\delta D$, $\delta \gamma_\text{USF}$, and $\delta \gamma_\text{surf}$.
Since $\chi$ is computed from CPA values, it is not subjected to statistical errors and can be treated as a constant.
Assuming statistical independence of $\delta \gamma_\text{USF}$ and $\delta \gamma_\text{surf}$, we have
\begin{align*}
    \left( \delta D \right)^2 &= \left( \frac{\partial D}{\partial \gamma_\text{USF}} \delta \gamma_\text{USF} \right)^2 + \left( \frac{\partial D}{\partial \gamma_\text{surf}} \delta \gamma_\text{surf} \right)^2
    ,
\end{align*}
with the partial derivatives,
\begin{align*}
    \frac{\partial D}{\partial \gamma_\text{USF}} &= \frac{\chi}{2\sqrt{\gamma_\text{surf} \gamma_\text{USF}}}, \\
    \frac{\partial D}{\partial \gamma_\text{surf}} &= -\frac{\chi}{2\sqrt{\gamma_\text{surf} \gamma_\text{USF}}} \times \frac{\gamma_\text{USF}}{\gamma_\text{surf}}
    .
\end{align*}
So, the error in $D$ can be written as
\begin{align*}
    \delta D &= \frac{\chi}{2\sqrt{\gamma_\text{surf} \gamma_\text{USF}}} \sqrt{\left( \delta \gamma_\text{USF} \right)^2 + \left( \frac{\gamma_\text{USF}}{\gamma_\text{surf}} \delta \gamma_\text{surf} \right)^2}
    .
\end{align*}

Using the expectation values and standard deviations for the largest cell size, we obtain a value for $D$ of $1.35 \pm 0.004$.
Hence, we conclude that taking a large enough cell size is sufficient to practically \emph{eliminate any influence of the elemental distribution on $D$}.

Obtaining the second type of error, originating from the potential, is more challenging due to limitations imposed by the size of the test set. If one were to consider the Root Mean Square Error (RMSE) values from Fig.~\ref{fig:appendix_metric} as error estimates and assume complete independence of the errors, a straightforward calculation would yield an error of $\delta D = 0.02$. This value is approximately five times larger than the influence of the atomic contribution. However, it's essential to acknowledge that the errors stemming from the potentials of $\gamma_\text{USF}$ and $\gamma_\text{surf}$ are, in fact, correlated.

To account for this correlation we add the covariance between $\gamma_{\text{USF}}$ and $\gamma_{\text{surf}}$ in error propagation:
\begin{align*}
\left( \delta D \right)^2 &= \left(\frac{\partial D}{\partial \gamma_{\text{USF}}}\right)^2 
              \left(\delta \gamma_\text{USF}\right)^2 +  
              \left(\frac{\partial D}{\partial \gamma_{\text{surf}}}\right)^2 
              \left(\delta \gamma_\text{surf}\right)^2 \\ 
          & + 2 \frac{\partial D}{\partial \gamma_{\text{USF}}} \frac{\partial D}{\partial \gamma_{\text{surf}}} \text{cov}(\gamma_{\text{USF}}, \gamma_{\text{surf}})
\end{align*}

Incorporating the covariance, we find $\delta D = 0.015$. It is important to note that this error represents a conservative upper bound.
For instance, it is clear from Fig.~\ref{fig:appendix_metric} that the largest contribution to the average error in $\gamma_\text{surf}$ comes from the region of very high surface energies. Since alloys in this region are also characterized by large $\gamma_\text{USF}$
they are expected to be very brittle and hence far from the Pareto front.
Consequently, the overall error is small enough to accurately calculate the 
$D$-parameter and its concentration dependency. Moreover, smaller deviations 
encountered during the Pareto front search are further smoothed out through the appropriate choice of the Kernel for the Gaussian process regression and appropriate sampling multiple configuration.

}

\clearpage

\FloatBarrier

\section{Validation of VBA models}
\label{sup_vba}

\FloatBarrier

\begin{figure}
   \centering
    \includegraphics{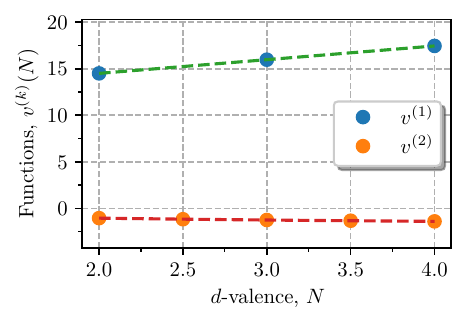}
   \caption{Virtual bond functions $v^{(1)}$ and $v^{(2)}$ fitted for the volumes. 
   }
   \label{fig:paper_misfit_overview1}
\end{figure}

\subsection{Validation of the virtual bond functions}

Fig.~\ref{fig:paper_misfit_overview1} shows virtual bond functions, $v^{(1)}$ and $v^{(2)}$, fitted for the volumes. They are smooth function of the valences and are fitted with second order polynomials. The misfit volume is directly derived from the volume expansion as follows:

\begin{equation}
    \frac{\partial V}{\partial c_i} = v^{(1)}(N_i) w_i + \sum_j 2 c_i w_{ij} v^{(2)}(N_{ij})
\end{equation}

\begin{equation}
    \Delta V_i = \frac{\partial V}{\partial c_i} + \sum_j c_j \frac{\partial V}{\partial c_j} 
\end{equation}

The volume expansion is primarily influenced by the first-order parameter function, \(v^{(1)}\). The second-order parameter function \(v^{(2)}\) can lead to non-trivial contributions to the equilibrium volume and its concentration derivatives in multi-component alloys. These functions' absolute values are determined by the scaling of the row-dependent prefactors and the band-width factor itself. The row-dependent prefactor for the bandwidth parameters is optimized for the training data of each individual property. The virtual bond function are not necessarily linear for other properties.

\subsection{Material parameter from VBA}

\begin{figure}
   \centering
   \includegraphics{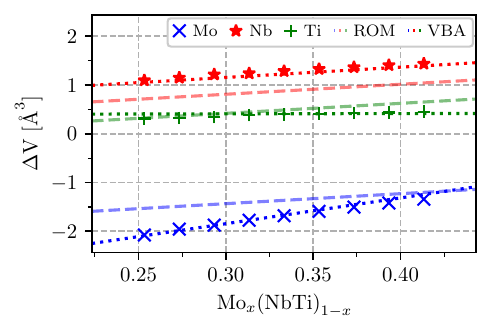}
   \caption{Comparison of misfit volumes in MoNbTi with varying concentrations of Mo obtaind from direct calculation ($\star,\times,+$), rule-of-mixture (ROM) and virtual bond approximation (VBA).
   }
   \label{fig:paper_misfit_volumes_changes}
\end{figure}

Fig.~\ref{fig:paper_misfit_volumes_changes} displays a comparison between direct and VBA calculations of misfit volumes,
where the concentration of Mo in MoNbTi is varied away from the equimolar value.
The figure illustrates that the VBA model is capable of capturing the trends in the 
concentration dependence very well, in contrast to the simple rule-of-mixture
that shows significant deviations, especially for decreasing fraction of Mo.
This is especially true for alloy systems such as MoNbTi, where relative errors between the VBA model and the rule-of-mixture model are noticeable already for the equimolar system.

However, in the case of the elastic constant, specifically $C_{44}$, noticeable deviations become apparent when employing the rule-of-mixture approach (as illustrated in Fig.~\ref{fig:paper_misfit_elastic_combined}a). While the rule-of-mixture does manage to capture the general trend, the discrepancy increase significantly, particularly for the 4-component systems and MoTaW, where it can reach 40 GPa. In contrast, the VBA model showcases good agreement with direct calculations for the majority of the alloys, exhibiting only minor discrepancies for NbTiZr and NbTaTi.

So far we have showed results for quantities that can 
be directly obtained from CPA calculations.
However, we cannot apply the same approach to the surface and USF energies 
because of the appreciable effect of atomic
relaxations. To avoid the need of heavy SQS calculations or of training MTPs for several test alloys
we take a simpler route.
First, we calculate \emph{unrelaxed} surface and USF energies with 
CPA for elemental compounds and binary alloys and parameterize the VBA model, as described above. We have checked that the VBA model predicts the CPA values very well.
To incorporate relaxation effects, we simply adjust the virtual bond functions 
to align with \emph{relaxed} surface and USF energies of individual elemental compounds. 
More precisely, we scale the virtual bond functions by a factor that follows a linear relationship with the d-valences. %
Such an approach works because fitting to CPA values already includes most of the chemical interactions
and the strongest relaxation effects have geometrical nature, which can be taken into account by the rescaling procedure of the universal virtual bond functions.

\begin{figure}
   \centering
\includegraphics{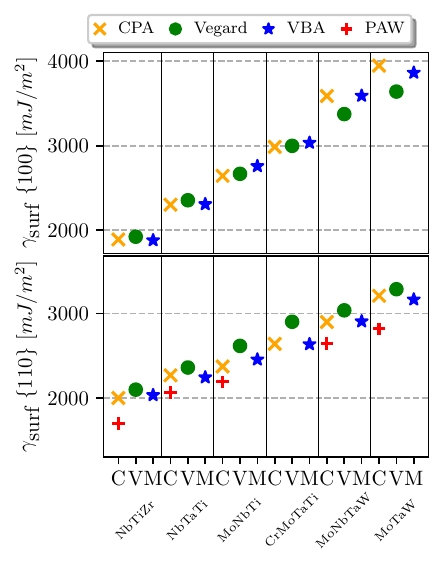}
   \caption{Comparison of surface energies
            obtained from Vegard’s law (V),
            from virtual bond approximation (M) 
            and calculated directly with CPA and PAW (C). PAW is relaxed. CPA unrelaxed. Virtual bond function are not rescaled.
   }
   \label{fig:figure15}
\end{figure}

\begin{figure}
   \centering
   \includegraphics{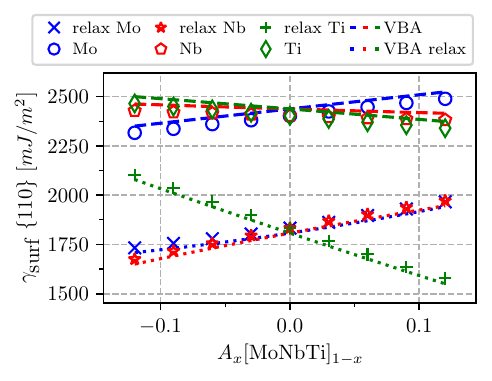}
   \caption{Comparison of directly calculated and VBA-model surface energy (${110}$) with and without local relaxations upon concentration changes in equimolar MoNbTi alloy, as
            $A_x[\text{MoNbTi}]_{1 - x}$, with $x$ as the molar fraction of $A$.}
   \label{fig:paper_surf_rescale}
\end{figure}

\begin{figure}
   \centering
\includegraphics{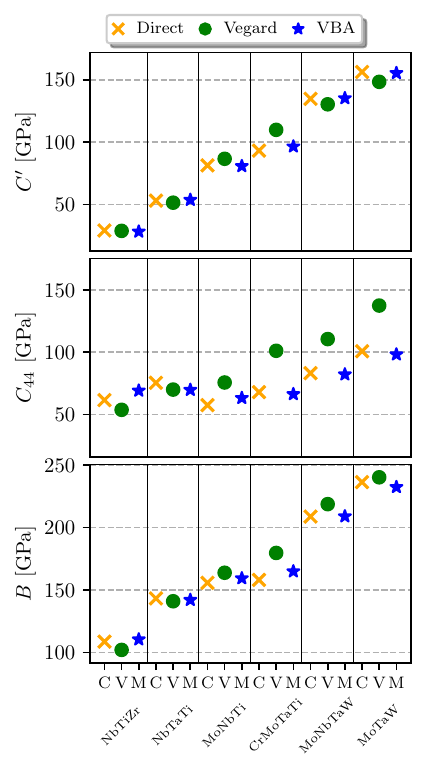}
   \caption{Comparison of linear elastic parameters ($B$, $C'$, $C_{44}$) 
            obtained from Vegard’s law (V),
            from virtual bond approximation (M) 
            and calculated directly (C).
   }
   \label{fig:figure12}
\end{figure}

In Fig. \ref{fig:figure15}, we compare directly computed surface energies from CPA (without relaxation) and PAW (with relaxations) to our model's estimates using ROM and virtual bond approximations. The energies of surfaces and the unstable stacking fault are effectively characterized by VBA model, which outperforms the simpler ROM. Notably, the VBA model significantly enhances the precision of surface energy predictions, underscoring its efficacy in capturing intricate energy behaviors. 

{
The corresponding results for the $\{110\}$ surface energy varying with the concentration of individual
components
are shown in Fig.~\ref{fig:paper_surf_rescale}, from which one can see that the simple rescaling is sufficient to take into account the
concentration dependence of relaxation effects.}

In Fig. \ref{fig:figure12}, we extend our comparative analysis to the elastic constants, following a similar approach. Notably, when examining the elastic constant $C_{44}$, it becomes evident that the VBA demonstrates a substantial enhancement compared to the traditional ROM. This improvement underscores the efficacy and accuracy of VBA in capturing essential material properties beyond what can be achieved through conventional methods.

\clearpage

{\section{Tabulated Data}}
\label{sup_tab_data}

Here, we provide data for a broad range of alloys combining values calculated in this work with literature data from
Refs.~\cite{Singh2023,Wu2014a,Wu2015a,Fazakas2014a,Xiong2023a}.

\begin{table*}[htbp]
{
\caption{Comparison of the fracture strain and $D$ parameter calculated from the VBA model. Fracture strain values are from Ref.~\cite{Singh2023}.}
  \begin{tabular}{lrr}
    \toprule
    Composition & $D$ & Fracture strain [\%] \\
    \midrule
    TiZrV$_{0.3}$Nb & 1.15 & 45.00 \\
    TiZrV$_{0.3}$NbMo$_{0.1}$ & 1.17 & 45.00 \\
    TiZrV$_{0.3}$NbMo & 1.28 & 43.00 \\
    TiZrVNbMo$_{0.3}$ & 1.26 & 42.00 \\
    ZrHfNbTa & 1.30 & 34.00 \\
    TiZrNbMo & 1.28 & 33.00 \\
    TiZrVNbMo$_{0.5}$ & 1.29 & 32.00 \\
    TiZrVNbMo$_{0.7}$ & 1.30 & 32.00 \\
    TiZrVNbMo & 1.32 & 32.00 \\
    TiZrV$_{0.25}$NbMo & 1.30 & 30.00 \\
    TiZrVNbMo$_{1.3}$ & 1.34 & 30.00 \\
    TiVNbTaMo & 1.48 & 30.00 \\
    TiZrHfVNb & 1.19 & 29.60 \\
    TiZrV$_{0.75}$NbMo & 1.32 & 29.00 \\
    Ti$_{1.5}$ZrHfNbMo & 1.20 & 28.98 \\
    TiZrV$_{0.5}$NbMo & 1.31 & 28.00 \\
    TiZrV$_{0.3}$NbMo$_{0.7}$ & 1.27 & 26.60 \\
    TiZrVNbMo & 1.33 & 26.00 \\
    TiVNbMo & 1.39 & 25.62 \\
    TiZrV$_{0.3}$NbMo & 1.30 & 25.00 \\
    TiZrHf$_{0.5}$NbMo$_{0.5}$ & 1.20 & 24.61 \\
    TiZrV$_3$NbMo & 1.37 & 24.00 \\
    TiZrHfNb$_{1.5}$Mo & 1.28 & 23.97 \\
    TiZrV$_2$NbMo & 1.35 & 23.00 \\
    TiZrV$_{1.5}$NbMo & 1.35 & 20.00 \\
    TiVNbTaW & 1.49 & 20.00 \\
    TiZrV$_{0.3}$NbMo$_{1.3}$ & 1.33 & 20.00 \\
    TiZr$_{0.5}$HfNbMo & 1.30 & 18.02 \\
    TiZrHf$_{1.5}$NbMo & 1.23 & 16.83 \\
    TiZr$_{1.5}$HfNbMo & 1.22 & 16.09 \\
    TiNbTaMoW & 1.51 & 14.10 \\
    TiZrHfNb$_{0.5}$Mo & 1.22 & 13.02 \\
    TiZrHf$_{0.5}$NbMo & 1.29 & 12.09 \\
    Ti$_{0.5}$ZrHfNbMo & 1.30 & 12.08 \\
    TiZrHfNbTaMo & 1.37 & 12.00 \\
    NbTaVW & 1.53 & 12.00 \\
    TiNbTaMoW & 1.51 & 11.50 \\
    TiZrHfNbMo$_{1.5}$ & 1.32 & 10.83 \\
    TiVNbTaMoW & 1.51 & 10.60 \\
    TiZrHfNbMo & 1.26 & 10.20 \\
    TiZrHfNbMo & 1.26 & 10.12 \\
    VNbTaMoW & 1.53 & 8.80 \\
    Ti$_{0.75}$NbTaMoW & 1.52 & 8.40 \\
    TiZrV$_{0.3}$NbMo$_{1.5}$ & 1.36 & 8.00 \\
    Ti$_{0.5}$NbTaMoW & 1.53 & 5.90 \\
    NbTaMoW & 1.53 & 2.60 \\
    Ti$_{0.25}$NbTaMoW & 1.53 & 2.50 \\
    NbTaMoW & 1.53 & 2.10 \\
    NbTaMoW & 1.53 & 1.90 \\
    VNbTaMoW & 1.54 & 1.70 \\
    VNbTaMoW & 1.54 & 1.70 \\
    \bottomrule
  \end{tabular}
  }
\end{table*}

\begin{table*}[htbp]
{
\caption{Comparison of $\tau_y$ calculated with the VBA model to experimental ones. Experimental values are from Refs.~\cite{Wu2014a,Wu2015a,Fazakas2014a,Xiong2023a}.}
\begin{tabular}{lrr}
\toprule
Composition & $\tau_{\text{y Expt.}}$ & $\tau_{\text{y Pred.}}$ \\
\midrule
    TiZrNbV & 322.28 & 227.65 \\
    TiZrNbVMo$_{0.3}$ & 382.15 & 284.91 \\
    TiZrNbVMo$_{0.5}$ & 441.70 & 312.91 \\
    TiZrNbVMo$_{0.7}$ & 517.10 & 334.93 \\
    TiZrNbVMo & 521.31 & 359.43 \\
    TiZrNbVMo$_{1.3}$ & 449.14 & 376.28 \\
    TiZrNbVMo$_{1.5}$ & 483.77 & 384.29 \\
    TiZrNbVMo$_{1.7}$ & 497.36 & 390.25 \\
    TiZrNbVMo$_{2.0}$ & 536.20 & 396.08 \\
    TiZrNbV$_{0.3}$ & 245.26 & 202.24 \\
    TiZrNbV$_{0.3}$Mo$_{0.1}$ & 266.62 & 236.27 \\
    TiZrNbV$_{0.3}$Mo$_{0.3}$ & 389.60 & 291.20 \\
    TiZrNbV$_{0.3}$Mo$_{0.5}$ & 386.04 & 332.78 \\
    TiZrNbV$_{0.3}$Mo$_{0.7}$ & 429.72 & 364.46 \\
    TiZrNbV$_{0.3}$Mo & 435.87 & 398.41 \\
    TiZrNbV$_{0.3}$Mo$_{1.3}$ & 483.77 & 420.65 \\
    TiZrNbV$_{0.3}$Mo$_{1.5}$ & 475.03 & 430.72 \\
    NbTiZr & 280.53 & 177.43 \\
    NbTiZrV & 324.22 & 227.65 \\
    NbTiZrVMo & 453.67 & 359.43 \\
    TiZrHfNbV & 401.89 & 185.11 \\
    TiZrHfNbCr & 545.91 & 485.38 \\
    NbTaTi & 190.00 & 32.75 \\
    MoNbTi & 300.00 & 206.73 \\
    CrMoTaTi & 605.00 & 443.13 \\
    MoNbTaW & 300.00 & 226.90 \\
\bottomrule
\end{tabular}
}
\end{table*}

\end{document}